\documentclass[aps,superscriptaddress,showpacs,nofootinbib]{revtex4}
\usepackage{graphics}
\usepackage[export]{adjustbox}
\usepackage{amsmath}
	
\usepackage{tcolorbox}
\graphicspath{{},{Logos/}}
\usepackage{graphicx,color}

\usepackage{bm}
\usepackage{amsfonts}
\usepackage{amsmath}
\usepackage{amssymb}
\usepackage{float}
\usepackage{hyperref}
\usepackage{subfigure}
\usepackage{slashed}
\usepackage{tikz}
\usetikzlibrary{arrows,shapes}
\usetikzlibrary{trees}
\usetikzlibrary{matrix,arrows} 				
\usetikzlibrary{positioning}				
\usetikzlibrary{calc,through}				
\usetikzlibrary{decorations.pathreplacing}  
\usepackage{pgffor}							

\usetikzlibrary{decorations.pathmorphing}	
\usetikzlibrary{decorations.markings}
\tikzset{
	vector/.style={decorate, decoration={snake}, draw},
	provector/.style={decorate, decoration={snake,amplitude=2.5pt}, draw},
	antivector/.style={decorate, decoration={snake,amplitude=-2.5pt}, draw},
	fermion/.style={draw=black, postaction={decorate},
		decoration={markings,mark=at position .55 with {\arrow[draw=black]{>}}}},
	fermionbar/.style={draw=black, postaction={decorate},
		decoration={markings,mark=at position .55 with {\arrow[draw=black]{<}}}},
	fermionnoarrow/.style={draw=black},
	gluon/.style={decorate, draw=black,
		decoration={coil,amplitude=4pt, segment length=5pt}},
	scalar/.style={dashed,draw=black, postaction={decorate},
		decoration={markings,mark=at position .55 with {\arrow[draw=black]{>}}}},
	scalarbar/.style={dashed,draw=black, postaction={decorate},
		decoration={markings,mark=at position .55 with {\arrow[draw=black]{<}}}},
	scalarnoarrow/.style={dashed,draw=black},
	electron/.style={draw=black, postaction={decorate},
		decoration={markings,mark=at position .55 with {\arrow[draw=black]{>}}}},
	bigvector/.style={decorate, decoration={snake,amplitude=4pt}, draw},
}

\tikzstyle{block} = [draw, rectangle, 
minimum height=3em, minimum width=6em]

\newcommand{\be}{\begin{equation}}
\newcommand{\ee}{\end{equation}}
\newcommand{\bea}{\begin{eqnarray}}
\newcommand{\eea}{\end{eqnarray}}

\newcommand{\dphi}{{\delta \phi}}
\newcommand{\tev}{{\rm TeV}}
\newcommand{\gev}{{\rm GeV}}

\newcommand{\yeff}{{y_{eff}}}

\begin{document}
	
\title{Supersymmetric $\nu$-Inflaton Dark Matter}

\author{Mar Bastero-Gil} \email{mbg@ugr.es} \affiliation{Departamento
  de F\'{\i}sica Te\'orica y del Cosmos, Universidad de Granada,
  Granada-18071, Spain}

\author{Ant\'onio Torres Manso} \email{atmanso@correo.ugr.es} \affiliation{Departamento
  de F\'{\i}sica Te\'orica y del Cosmos, Universidad de Granada,
  Granada-18071, Spain}

\begin{abstract}
  We present the supersymmetric extension of the unified model for inflation and Dark Matter studied in Ref. \cite{Manso:2018cba}. The scenario is based on the incomplete decay of the inflaton field into right-handed (s)neutrino pairs. By imposing a discrete interchange symmetry on the inflaton and the right-handed (s)neutrinos, one can ensure the stability of the inflaton field at the global minimum  today, while still allowing it to partially decay and reheat the Universe after inflation. Compatibility of inflationary predictions, BBN bounds and obtaining the right DM abundance for the inflaton Dark Matter candidate typically requires large values of its coupling to the neutrino sector, and we use supersymmetry to protect the inflaton from potentially dangerous large radiative corrections which may spoil the required flatness of its potential. In addition, the inflaton will decay now predominantly into sneutrinos during reheating, which in turn give rise both to the thermal bath made of Standard Model particles, and inflaton particles. We have performed a thorough analysis of the reheating process following the evolution of all the partners involved, identifying the different regimes in the parameter space for the final Dark Matter candidate. This as usual can be a WIMP-like inflaton particle or an oscillating condensate, but we find a novel regime for a FIMP-like candidate.  

\end{abstract}

\pacs{98.80.Cq, 11.10.Wx, 14.80.Bn, 14.80.Va}

\date{\today}

\maketitle


\section{Introduction}
Inflation and dark matter are two of the most important conundra in cosmology and particle physics. Dark matter (DM) particles are required to account for the observed galaxy rotation curves, the large scale structure in the Universe and weak-lensing observations \cite{Tyson:1998vp,Dahle:2007wf,Paczynski:1985jf}. As a consequence, these DM particles are expected to form a stable nonrelativistic and nonluminous fluid (for a review see \cite{Taoso:2007qk}). Similarly, the inflation paradigm \cite{Guth:1980zm,Linde:1981mu,Albrecht:1982wi}, developed to solve the flatness and horizon problems, with the additional support from an almost perfect isotropy of the Cosmic Microwave Background (CMB), has in its simplest representation a single inflaton scalar field, which is set to be weakly interacting and neutral. These requirements are key to ensure the required flatness of the associated scalar potential, easily spoiled by radiative corrections. Although inflation and DM are crucial in the modern cosmological paradigm, these cannot be encompassed within the Standard Model (SM) of particle physics. Thus, they are still to be embodied in a consistent particle framework. Here, we will entertain the possibility of having the same field to describe both the accelerated expansion in the early Universe and the matter dark sector present in the Universe energy budget. 

Scalar fields are extremely versatile, and depending on the kinematical regime and the shape of their potential, may mimic fluids with different equations of state. For instance, if $\dot{\phi}^2/2\ll V(\phi)$, in a regime of a slowly varying field, a scalar potential acts as an effective cosmological constant. Furthermore, while oscillating about the minimum of its potential, if given by a power law $V(\phi)=\phi^n$, one has $\langle\dot{\phi}^2\rangle=n \langle V(\phi)\rangle $ such that $p_\phi=\frac{n-2}{n+2}\rho_\phi $. Thus for $n=4$ the inflaton would redshift as radiation where as for $n=2$, a quadratic potential, the inflaton behaves as nonrelativistic matter. These regimes, slow-roll and oscillating phase,  are usually integrated in an inflationary model, allowing to study the evolution until late times. The main concern in unifying the two phenomena lies in the compatibility with the standard cosmological evolution, specially with the transition into a SM radiation Universe before the light element production in the so called Big Bang Nucleosynthesis (BBN)  \cite{Fields:2014uja}. Typically this requires the inflaton decaying early enough (before BBN) during the oscillating phase. However, we could still have an incomplete inflaton decay and ensure an effective reheating, while at the same time, having a stable inflaton remnant to be the DM candidate at late times. This scenario was already mentioned when developing the theory of (p)reheating in Ref.\cite{Kofman:1997yn}, and different mechanisms to block the inflaton decay at late times after reheating the Universe were further studied in \cite{Liddle:2006qz, Panotopoulos:2007ri,Cardenas:2007xh, Liddle:2008bm, Bose:2009kc, DeSantiago:2011qb, Bastero-Gil:2015lga, Daido:2017wwb, Daido:2017tbr}, the focus being in obtaining a remnant inflaton condensate as dark matter.  But it  was also soon realized that inflaton particles could be recreated later from the thermal bath and behave as a weakly interacting massive particle (WIMP), their abundance controlled by  the standard freeze-out mechanism for dark matter \cite{Lerner:2009xg, Okada:2010jd,  delaMacorra:2012sb, Khoze:2013uia, Kahlhoefer:2015jma, Choubey:2017hsq, Hooper:2018buz, Borah:2018rca, Manso:2018cba}; alternatively, for very weak couplings, it could also lead to a feebly interacting massive particle (FIMP) never coupled to the cosmic plasma, and the freeze-in mechanism for DM \cite{Tenkanen:2016twd}. 

A viable mechanism to ensure the incomplete decay  might come through a right-handed neutrino portal\footnote{For a recent thorough analysis on neutrino portals to FIMP DM  see \cite{Cosme:2020mck}}. By taking the mechanism developed in \cite{Bastero-Gil:2015lga}, where the inflaton is only allowed to decay into two fermions by imposing a discrete symmetry, one gets a consistent DM candidate when the decay into these fermions becomes kinematically forbidden. In \cite{Manso:2018cba}, by identifying the two fermions to be two of the three right-handed neutrinos, this mechanism allowed simultaneously to describe, using the same scalar field both for inflation and cold dark matter, (a) the generation of light neutrino masses through the seesaw mechanism, and (b) the observed cosmological baryon asymmetry via leptogenesis after a thermal production of the third and lighter right-handed neutrino. However in this model, referred to as the $\nu$IDM model, there are regimes with large Yukawa couplings that may lead to significant radiative corrections to the inflaton self coupling, which can come into conflict with the small values required by the amplitude of the primordial spectrum of curvature perturbations. In this work we address the problem of the radiative corrections by extending the model in \cite{Manso:2018cba} to its supersymmetric version, the S-$\nu$IDM. An interesting and alternative scenario may come through a realization in the context of Warm Inflation, as recently developed in \cite{Levy:2020zfo}.

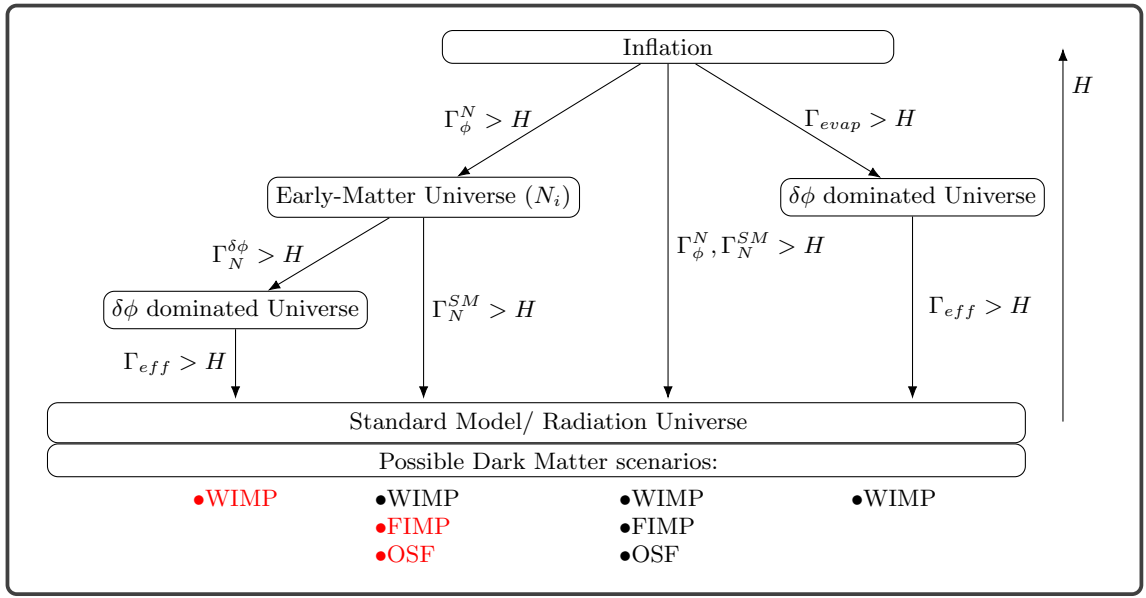
\begin{figure}[H]
	
	\centering
	\usetikzlibrary{shapes,snakes,decorations,arrows,calc,arrows.meta,fit,positioning}
	\usetikzlibrary{decorations.markings}
	
	\tikzset{
		-Latex,auto,node distance =1 cm and 1 cm,semithick,
		state/.style ={ellipse, draw, minimum width = 0.7 cm},
		point/.style = {circle, draw, inner sep=0.04cm,fill,node contents={}},
		bidirected/.style={Latex-Latex,dashed},
		el/.style = {inner sep=2pt, align=left, sloped}
	}
	\tcbox[colback=white]{
		\begin{tikzpicture}
		\node[state,shape=rectangle, rounded corners,minimum width = 6cm ] (I) at (1.75,2) {Inflation};
		\node (RI) at (1.75,-2.8) {};
	
		\node[state,shape=rectangle, rounded corners] (phi) at (5,0) {$\delta\phi$ dominated Universe};
		\node (Rphi) at (5,-2.8) {};
		
		\node[state,shape=rectangle, rounded corners] (E M) at (-1.5,0) {Early-Matter Universe ($N_i$)};
		\node (RN) at (-1.5,-2.8) {};
		
		\node[state,shape=rectangle, rounded corners] (phi2) at (-4,-1.5) {$\delta\phi$ dominated Universe};
		\node (Rphi2) at (-4,-2.8) {};
		
		\node[state,shape=rectangle, rounded corners,minimum width =13cm] (R) at (0,-3) {\quad Standard Model/ Radiation Universe};
		\node[state,shape=rectangle, rounded corners,minimum width =13cm] (DM) at (0,-3.5) {\quad Possible Dark Matter scenarios:};

		\path (I) edge node[above, left=0.1] {$\Gamma^N_\phi>H$} (E M);
		\path (E M) edge node[right] {$\Gamma^{SM}_N>H$}  (RN);
		\path (E M) edge node[above,left =0.2] {$\Gamma^{\delta\phi}_N>H$} (phi2);
		\path (phi2) edge node[below, left] {$\Gamma_{eff}>H$} (Rphi2);
		\path (I) edge node[above, right=0.1] {$\Gamma_{evap}>H$}  (phi);
		\path (phi)  edge node[above, right=0.1] {$\Gamma_{eff}>H$} (Rphi);
		\path (I)  edge node[below=0.2,right] {$\Gamma^N_\phi,\Gamma^{SM}_N>H$}  (RI);
		
		\node (Hi) at (7,2.1) {};
		\node (Hf) at (7,-3.1) {};
		\path (Hf) edge node[above=2,right] {$H$}  (Hi);
		
		\node (DM1) at (-4,-4) {\textcolor{red}{\textbullet WIMP} };
		\node[text width=4em] (DM2) at (-1.5,-4.37) { \textbullet WIMP  \textcolor{red}{ \textbullet FIMP \textbullet OSF}  };
		\node[text width=4em ] (DM3) at (1.75,-4.37) { \textbullet WIMP  {\textbullet FIMP} \textbullet OSF };
		\node (DM4) at (4.75,-4) { \textbullet WIMP  };
		\end{tikzpicture} }
	\caption{ S-$\nu$IDM paths to Standard Cosmology. $H$ denotes the Hubble parameter, which is a time decreasing function from inflation onwards. Interaction/decay rates of the species $\alpha$ into $\beta$ are denoted by $\Gamma_\alpha^\beta$; $N$ denotes collectively neutrinos/sneutrinos, $SM$ light SM particles, $\delta  \phi$ refers to inflaton particles, while $\phi$ is the inflaton condensate. $\Gamma^T_{evap}$ is the inflaton evaporation rate due to scattering with the thermal bath, and $\Gamma_{eff}$ the total interaction rate for inflaton particles. We have marked in red those scenarios excluded by BBN or DM abundance constraints, to be discussed in section \ref{sec6}.} \label{fig:Evolution}
	
\end{figure}

To provide a supersymmetric extension of the model, we embed the inflationary description within supergravity (SUGRA), by following the superconformal inflationary $\alpha$-Attractors models \cite{Kallosh:2013yoa}, where we take a non-minimal K\"ahler potential and a superpotential, both compatible with the required discrete symmetries. The inflaton scalar field follows the common slow-roll description, and at the end of inflation it starts to oscillate about a quartic potential, leading to the onset of reheating. We analyze this period following the Boltzmann equations. Reheating is achieved through the incomplete decay or evaporation interactions. Compatibility with an early-matter era, and a period of dominant excited inflaton particles exist in the parameter space, as well as a direct transition from inflation into a standard cosmological Universe, as summarized in Fig. \ref{fig:Evolution}. The remnant inflaton field, at this stage under a quadratic potential, may in principle survive as an oscillating scalar field (OSF), as a FIMP, or as the typical WIMP. In the last two scenarios the inflaton particles will evaporate through scatterings with other particles, and we find examples of all these scenarios to be compatible with all the constraints of the model.

Supersymmetric (SUSY) models with R-parity, besides providing a solution for the hierarchy problem of the SM, do naturally incorporate a candidate for DM, the lightest supersymmetric particle. Depending on the SUGRA model and the pattern of SUSY breaking, this can be either a neutralino, a sneutrino, or even the gravitino. However, the combination of current colliders SUSY searches, direct and indirect DM detection experiments, and in some cases cosmological considerations, set severe constraints on these possibilities \cite{SUSYDM}. For example, direct detection DM limits only allow the sneutrino to be a subdominant DM component \cite{sneutrinoDM}. Heavy neutralinos with masses larger than $O(1)$ TeV are still a viable candidate, but that would imply pushing the SUSY breaking scale to higher values, loosing the main motivation as an explanation to the hierarchy problem (and reducing the prospect to detect SUSY at the LHC). In phenomenological studies of the supersymmetric SM, the so called pMSSM, bino-like neutralinos could still be lighter than half the Higgs mass, and in some cases again be a subdominant DM component \cite{lowmassLSP}. Keeping a sector of the SUSY spectrum, including the lightest supersymmetric particle (LSP), close to the EW scale has the advantage that they could still be searched for (and eventually excluded) at the LHC, even if it cannot fully account for the DM relic density. 

Gravitinos on the other hand are constrained from cosmological considerations, even when they are not the LSP. One must avoid having too many gravitinos decaying at around the time of BBN, otherwise we destroy the agreement between theory and observations of the light element abundances \cite{gravitino1,gravitino2,gravitino3}. Even when they are the LSP, they can be copiously produced and overclose the universe. Avoiding therefore an overproduction of gravitinos sets an upper limit on the reheating temperature after inflation, depending on their mass. Our results will be in general consistent with heavy gravitinos with masses larger than $O(10-100)$ TeV and reheating temperatures $T_{RH} \lesssim 10^{12}$ GeV. 

Therefore, it is worth to explore SUSY multi-component DM models, keeping the original motivation for low scale SUSY models as the solution to the hierarchy problem, plus some other sector beyond the MSSM that could account or complement the DM abundance, like in Refs. \cite{axinoneutralino1,axinoneutralino2}. Our proposal is framed within this general context. In particular, we minimally extend the model to account for the inflationary period, the DM component and neutrino masses. We will focus on the inflationary and reheating period, and set the parameter space available to have the inflaton as the full DM component, assuming that the LSP is subdominant. We will not explore here the SUSY breaking mechanism and SUSY spectrum, and generically we will assume a SUSY scale $M_{SUSY} \simeq O(1\, \tev)$. When exploring the parameter space we will impose a lower bound on the reheating temperature of the order of the EW scale, such that SUSY particles behaves as relativistic species during reheating.


This work is organized as follows. We start by reviewing the $\nu$IDM model by introducing the model symmetries and their consequences. In section \ref{sec3} we describe inflation, while in \ref{sec4} we detail the reheating interactions and the construction of the Boltzmann equations.  We explore in detail the possible evolutions after inflation until the end of reheating in section \ref{sec5}, as well as the DM candidates in section \ref{sec6}, for each parametric regime. Finally, in section \ref{sec7} we summarize and discuss the main conclusions.
\\

\section{$\nu$IDM Review \cite{Manso:2018cba}}
Consider a single real scalar field, the inflaton field $\phi$, with a potential energy $V(\phi)$. The inflaton is allowed to interact through a Yukawa coupling with two right-handed neutrinos, $N_1$ and $N_2$, under the discrete symmetry $C_{2}\subset\mathbb{Z}_{2}\times S_{2}$, which acts as 
\begin{equation}
\phi  \leftrightarrow-\phi~, \qquad 
N_{1} \leftrightarrow N_{2}~.
\end{equation} 
A third right-handed neutrino is included to match the number of fermion generations in the SM, although it will not interact with the inflaton. The right-handed neutrinos are SM singlets and Weyl fermions, allowing for the important Majorana masses and the coupling with the Higgs and lepton doublets through Yukawa interactions. These lead to the seesaw mechanism, which generates the observed light neutrino masses \cite{neutrinomass}.

\begin{figure}[H]
	
	\centering 
	
	\begin{tikzpicture}[line width=1 pt, scale=1,baseline=0]
	
	\draw[fermionbar] (0:0)--(-40:1);
	\draw[fermion] (0:0)--(40:1);
	\draw[scalar] (180:1)--(0:0);
	
	\draw[fermionbar,blue] (-40:1)--(40:1);
	\draw[scalar,blue] (-40:1)--(2,-0.64);
	\draw[scalar,blue] (40:1)--(2,0.64);
	
	\end{tikzpicture}
	\hspace{1cm}
	\begin{tikzpicture}[line width=1 pt, scale=1,baseline=0]
	
	\draw[fermionbar] (0:0)--(-40:1);
	\draw[fermion] (0:0)--(40:1);
	\draw[scalar] (180:1)--(0:0);
	
	\draw[scalarbar,blue] (-40:1)--(40:1);
	\draw[fermion,blue] (-40:1)--(2,-0.64);
	\draw[fermion,blue] (40:1)--(2,0.64);
	
	\end{tikzpicture}
	\hspace{1cm}
	\begin{tikzpicture}[line width=1 pt, scale=1,baseline=-0.3]
	
	\draw[fermion] (0:0)--(-40:1);
	\draw[fermion] (0:0)--(40:1);
	\draw[scalar] (180:1)--(0:0);
	
	\draw[scalar,blue] (-40:1)--(-5:1.7);
	\draw[scalar,blue] (40:1)--(5:1.7);
	\draw[fermion,blue] (-40:1)--(-30:2);
	\draw[fermion,blue] (40:1)--(30:2);...
	
	\end{tikzpicture}
	
	\caption{ Examples of inflaton decay channels forbidden by the discrete interchange symmetry. For clarity, we represent the light Higgs and neutrino fields in blue and the inflaton and right-handed neutrinos in black. } \label{fig:FDecaysIDM}
\end{figure}
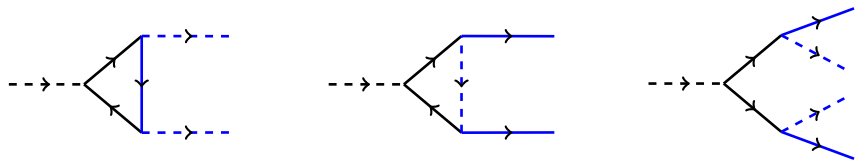

As a result of the imposed symmetry, the masses of the right-handed neutrinos $N_1$ and $N_2$ become equal and their coupling with the inflaton gets an opposite sign. Furthermore, in the couplings with the SM, the two fermions will have equal Yukawa couplings at each lepton flavour generation. During inflation the $N$ mass Lagrangian becomes
\begin{equation}
\mathsf{\mathcal{L}}_{N\,Mass}=-\frac{1}{2}(M_{1}+h\phi)N_{1}N_{1}^{c}-\frac{1}{2}(M_{1}-h\phi)N_{2}N_{2}^{c}-\frac{1}{2}M_{3}N_{3}N_{3}^{c}~.
\end{equation}
The symmetry forbids inflaton linear couplings with other fields, such that if we take $\phi=0$ to be the minimum of the potential, for $M_1>M_\phi/2$  the inflaton is stable at the minimum, $M_\phi$ being the inflaton mass in the vacuum. In particular, as illustrated in Fig. \ref{fig:FDecaysIDM}, the contributions of virtual $N_1$ and $N_2$ inflaton decay modes into other lighter particles cancel each other, due to the opposite sign coupling to $\phi$. However, while the inflaton is oscillating about the minimum of its potential, after the slow-roll dynamics, the $\mathbb{Z}_{2}$ symmetry is broken and the right-handed neutrinos obtain an effective mass $M_\pm=\left|M_1\pm h\phi\right|$, which can become smaller than $M_\phi$. Thus, inflaton decay will be possible for a certain range of $\phi$ values. When the field value drops below a certain threshold the decay becomes blocked. This mechanism allows for an efficient reheating after inflation while keeping a stable remnant to account as DM.

As discussed in Ref. \cite{Manso:2018cba}, the $\nu$IDM model may require a coupling between the inflaton and the two right-handed neutrinos, $N_1$ and $N_2$, to be $\cal{O}$(1). This may lead to significant radiative corrections to the inflaton self-coupling, that may spoil the compatibility between small self-coupling values and the amplitude of the primordial spectrum of curvature perturbations.
A possible solution to this problem may come from a supersymmetric extension of the $\nu$IDM model, where the radiative corrections coming from the fermion coupling can be partially cancelled with the contribution of its bosonic superpartner, the right-handed sneutrino \cite{effpot}. Here, we develop such a scenario.

\section{Inflation within Supergravity }
\label{sec3}

We extend the symmetry of the fields into a symmetry of the correspondent superfields. Under $C_{2}\subset\mathbb{Z}_{2}\times S_{2}$, 
\begin{equation}
\Phi  \leftrightarrow-\Phi~, \qquad 
\mathrm{N}_{1} \leftrightarrow \mathrm{N}_{2}~.
\end{equation}
As a consequence, besides the inflaton and the right-handed neutrinos also their superpartners are restricted by the symmetry. The inflaton and the inflatino will only be allowed to interact with the neutrinos and sneutrinos. 

As mentioned, the simple chaotic inflation model with a quartic potential cannot provide an observational consistent description of inflation.  In order to develop an analogous evolution as the non-minimal coupling to gravity used in the $\nu$IDM model within a supersymmetric description, we are naturally induced to consider an inflation model in the framework of supergravity. We closely follow the analysis in \cite{Kallosh:2013yoa}, working with the family of Superconformal Inflationary $\alpha$-Attractors models, and summarize here the main results. This results in a non-canonical kinetic term for the inflaton, mimicking an evolution akin to Starobinsky inflation \cite{Starobinsky:1980te,Pallis:2016mvm}, and leading to a spectral index and tensor-to-scalar ratio consistent with Planck data \cite{planck}. The symmetries of the model will also allow a quadratic mass term for the inflaton in the superpotential. This mass parameter, $M_\phi$,  will be constrained later in section VI in order to have the right abundance of inflaton DM, such that $M_\phi < O(1-10)$ TeV.

Take the two right-handed neutrinos superfields as auxiliary fields to stabilize the inflaton superfield during the accelerated expansion period. The K\"ahler potential is given by 
\begin{equation}
K=-3\alpha\log\left[1-\ \frac{\Phi\bar{\Phi}+\mathrm{N}_1\bar{\mathrm{N}}_1 +\mathrm{N}_2\bar{\mathrm{N}}_2}{3}\right]\ . 
\end{equation}
Keeping the symmetries of the $\nu$IDM model we write the relevant superpotential for inflation as
\begin{equation}
W=\frac{1}{2\sqrt{2}}\kappa\Phi^2\left(\mathrm{N}_1+ \mathrm{N}_2\right)\left(3-\Phi^2\right)^{\frac{3\alpha-1}{2}}+ \frac{h}{4}\Phi(\mathrm{N}_2^2-\mathrm{N}_1^2)+\frac{M_1}{2}(\mathrm{N}_1^2+\mathrm{N}_2^2)\ .\label{Eq:SuperPot}
\end{equation} 
 In \cite{Kallosh:2013yoa} $\alpha=1/3$ was excluded, since $\alpha>1/3$ is required for a stable inflation behavior. For the sake of simplicity we then take $\alpha=1$, and follow their analyses on inflation.

The effective Lagrangian for inflation, for real fields $\Phi=\bar \Phi= \varphi$ and at $\mathrm{N}_1=\mathrm{N}_2=0$ yields 
\begin{equation}
{\cal{L}}=\sqrt{g}\left[ \frac{1}{2}R-\frac{1}{\left( 1-\varphi^2/3\right) ^2}(\partial\varphi)^2-\frac{1}{4}\kappa^2\varphi^4\right]\ .
\end{equation}
To better analyze our dynamics we ought to canonically normalize the kinetic components. The relation between the geometric field $\varphi$ and the canonical one $\phi$ is
\begin{equation}
	\frac{\varphi}{\sqrt{3}}=\tanh\frac{\phi}{\sqrt{6}}\label{Eq:field_trans}\,.
\end{equation}The expression is analogous to the rapidity ($\phi$) and velocity ($\varphi$) in special relativity.
As a result, the effective Lagrangian in the Einstein frame is
\begin{equation}
	{\cal{L}}=\sqrt{g}\left[ \frac{1}{2}R-\frac{1}{2}(\partial\phi)^2-\frac{9}{4}\kappa^2 \tanh^4\left[ \frac{\phi}{\sqrt{6}}\right] \right] \ . 
\end{equation}
We may now perform the standard analysis of inflation in the slow-roll regime, where the slow-roll parameters and the number of e-folds of inflation are given by:
\begin{equation}
\epsilon=\frac{1}{2}m_P^{2}\left(\frac{V'(\phi)}{V(\phi)}\right)^{2}~,\qquad\eta=m_P^{2}\frac{V''(\phi)}{V(\phi)}~,\qquad N_e=\frac{1}{m_P^{2}}\int_{\phi_{e}}^{\phi_{*}}\frac{V(\phi)}{V'(\phi)}d\phi~,\label{Eq:slowroll IDM}
\end{equation}
where $m_P$ is the reduced Planck mass and and $V(\phi)= 9/4\,\kappa^2 \tanh^4\left[ \phi/\sqrt{6}\right]$; $\phi_\star$ and $\phi_e=\sqrt{\frac{3}{2}}\,m_P\mathrm{Arcsinh}\left[\frac{4}{\sqrt{3}}\right]$ represent the field values when the CMB scales exit the horizon during inflation and at the end of inflation, respectively.  While inflating we want to generate the observed amplitude for the spectrum of the scalar curvature perturbations, $\Delta^2_{\cal{R}}=2.2\times10^{-9}$.  This results on the condition $V/\epsilon=(0.0269m_P)^4$ \cite{planck}, that restricts the value of the inflaton self-interaction coupling to $\kappa\simeq3.5\times10^{-6}$, at 60 e-folds before the end.
The other observables that connect the theory to the observations, the scalar to tensor ratio $r$ and the scalar spectral index $n_s$, are given by:
\begin{flalign}
r & = 16\epsilon = \frac{48}{3+N_{e}(\sqrt{57}+4N_{e})}\simeq {\frac{12}{N_e^2}} \,,\\
n_s& =1-6\epsilon+2\eta =1 - \frac{24}{-3+\sqrt{57}+8N_{e}}+\frac{8}{3+\sqrt{57}+8N_{e}}
\simeq 1-{\frac{2}{N_e}}\,,
\end{flalign}
where the last expressions have the leading results for large $N_e$, which provides the expected results for the Superconformal Inflationary $\alpha$-Attractors models \cite{Kallosh:2013yoa}. In particular for $N_e =60$ one gets $r \simeq 3 \times 10^{-3}$ and $n_s \simeq 0.967$, well within the parameter space allowed by the Planck data. 

For the above analysis, as an effective one field model, we have assumed the stability of the truncation to real fields and $\mathrm{N}_1=\mathrm{N}_2=0$. We now address such considerations. 

The scalar fields potential is obtained from $K$ and $W$ using
\begin{equation}
	V=e^K\left[K^{ij^\star} D_iWD_j^\star W^\star -3 |W|^2\right],
\end{equation}
where $K_{ij^\star}$, the K\"ahler metric, is defined as 
\begin{equation}
	K_{ij^\star}=\frac{\partial^2K}{\partial\phi_i\partial\phi^\star_j}
\end{equation}
and $K^{ij^\star}$ is its inverse. The covariant derivative is defined as
\begin{equation}
D_{i}=\frac{\partial}{\partial\phi_i}+\frac{\partial K }{\partial\phi_i}\ .
\end{equation}
The scalar fields masses may be obtained trough the second derivative with respect to the fields. For our truncation conditions
\begin{equation}
	\small
		M_{S}^2[\varphi]=\left(
	\begin{array}{ccc}
	\frac{\kappa ^2 \varphi ^2 \left(2 \varphi^4-9 \varphi ^2+18\right)}{2 (\varphi^2-3)^2} & \frac{3\kappa  \varphi  (h \varphi -2 M_1)}{2\sqrt{2}(\varphi ^2-3)} & -\frac{3\kappa  \varphi  (h \varphi +2 M_1)}{2\sqrt{2}(\varphi ^2-3) } \\
        \frac{3\kappa  \varphi  (h \varphi -2 M_1)}{2\sqrt{2}(\varphi ^2-3) }
        & \frac{108 M_1 (M_1 - h \varphi) + \varphi^2 (27 h^2 - 2 \kappa^2 (\varphi^2 -3)^3}{12 \left(\varphi ^2-3\right)^2}
        & -\frac{\kappa ^2 \varphi ^2 \left(\varphi ^4-9 \varphi ^2+9\right)}{6 (\varphi^2 -3)} \\
        -\frac{3\kappa  \varphi  (h \varphi +2 M_1)}{2\sqrt{2}(\varphi ^2-3) }
	&
-\frac{\kappa ^2 \varphi ^2 \left(\varphi ^4-9 \varphi ^2+9\right)}{6 (\varphi^2 -3)}
&  \frac{108 M_1 (M_1 + h \varphi) + \varphi^2 (27 h^2 - 2 \kappa^2 (\varphi^2 -3)^3}{12 \left(\varphi ^2-3\right)^2}
\\
	\end{array} 
	\right) \,,
\end{equation}
where $M^2_S[\varphi]=\partial^2 V/\partial \Phi_I \partial \bar \Phi_J$ at $\Phi=\bar \Phi=\varphi$ and $N_1=N_2=0$, and for easy of notation here $\varphi$ and $M_1$ are given in $m_P$ units. We can then verify that during inflation off-diagonal terms for the sneutrinos only present a very small correction to the diagonal masses as far as $h \gg \kappa$, and they will have large, positive squared masses. Once they are set to zero during inflation,  the inflaton $\phi$ mass is much smaller than the Hubble rate for large $\phi$ values, which ensures the assumed stable inflationary trajectory. We have also checked that the squared mass of the inflaton imaginary component is large and positive, without the need of adding any stabilizing extra term in the K\"ahler potential \cite{Kallosh:2013yoa}.

At the end of the slow-roll regime, $\phi_e\simeq m_P$, the inflaton field value will rapidly decrease. When it becomes subplackian it will turn the K\"ahler potential into a minimal construction. In this scenario the quartic term will become dominant, and the inflaton energy density will redshift as radiation. As discussed in \cite{Manso:2018cba}, its decay hopefully will lead to the formation of the relativistic thermal bath in the so called reheating period. In this period, the inflaton field value will continue to decrease, such that the non-canonical kinetic terms can safely be neglected in the post-inflationary evolution of the inflaton dynamics. 

\section{Reheating}
\label{sec4}


As discussed at the end of last section, when the slow-roll conditions are violated, at $\phi\sim m_P$, the inflaton field values will be allowed to decrease, and will start to oscillate about the origin of its potential. When field values $\ll m_P$ the potential will, essentially, be quartic and redshift as radiation. We will recover the inflaton final behavior as a dark matter fluid later, when the field independent mass term $M_\phi$ in the potential starts dominating over the self-interactions. 

Neglecting the sub-leading mass term in the inflaton potential, we now proceed with a an analysis of the reheating period, first discussing the fields dynamics and interactions, to then introduce the Boltzmann equations and their numerical solutions. 

At the relevant scales for the reheating interactions the superpotential is
\begin{equation}
W \simeq \frac{1}{2\sqrt{2}}\kappa\Phi^2\left(\mathrm{N}_1+ \mathrm{N}_2\right)+ \frac{h}{2}\Phi(\mathrm{N}_2^2-\mathrm{N}_1^2)+\frac{M_1}{2}(\mathrm{N}_1^2+\mathrm{N}_2^2)+y\mathrm{H_uL}\left(\mathrm{N}_1+ \mathrm{N}_2\right)\,.\label{Eq:SPotential}
\end{equation}
The first difference with the non-SUSY case comes with the possible inflaton interactions with other particles. Here, the inflaton will be allowed to decay into both bosonic and fermionic degrees of freedom. The relevant Lagrangian is obtained from the superpotential in equation \eqref{Eq:SPotential}:
 \begin{multline}
{\cal{L}} ={\cal{L}}_{kin}+{\cal{L}}_{\mathrm{N}_i\rightarrow SM} +V(\phi,\tilde{N}_1,\tilde{N}_2)+
M_1 (N_1 N_1^c+N_2 N_2^c) +
h\phi (N_2 N_2^c-N_1 N_1^c)\\+
(\tilde{N}_2\, \psi_\phi N_2^c-\tilde{N}_1\, \psi_\phi N_1^c)+
\frac{\kappa}{\sqrt{2}}\phi(\psi_\phi N_1^c+\psi_\phi N_2^c)+
\frac{\kappa}{\sqrt{2}}(\tilde{N}_1+\tilde{N}_2)\psi_\phi\psi_\phi^c + h.c.
\label{Lagrangian SUSY}
\end{multline}
where the potential is given by
\begin{multline}
V(\phi,\tilde{N}_1,\tilde{N}_2)=
M_1^2(\tilde{N}_1^2+\tilde{N}_2^2)+
\frac{h^2}{4}(\tilde{N}_1^4+\tilde{N}_2^4)-
\frac{h^2}{2}\tilde{N}_1^2\tilde{N}_2^2+ 
\frac{\kappa^2}{4}\phi^4+2hM_1\phi(\tilde{N}_2^2-\tilde{N}_1^2)\\+ h \frac{\kappa}{\sqrt{2}}\phi(\tilde{N}_2^2\tilde{N}_1+\tilde{N}_1^2\tilde{N}_2)+
h^2\phi^2(\tilde{N}_1^2+\tilde{N}_2^2)+
M_1 \frac{\kappa}{\sqrt{2}}\phi^2(\tilde{N}_1+\tilde{N}_2)+
\frac{\kappa^2}{2}\phi^2(\tilde{N}_1^2+\tilde{N}_2^2)+
\kappa^2\phi^2\tilde{N}_1\tilde{N}_2\,.
\end{multline}

\subsection*{Particle Interactions}

We now study all relevant particle interactions to then proceed to the evolution of the system with the Boltzmann equations. 
With the imposition of the interchange symmetry, the inflaton superfield can only interact with the right-handed neutrino superfields. Thus, only decays into right-handed neutrinos, sneutrinos and inflatinos are allowed. 
	 
\begin{figure}[H]
	\centering
	\begin{tikzpicture}[line width=1 pt, scale=1.3]
	\draw[fermion] (0:0)--(-40:1);
	\draw[fermionbar] (0:0)--(40:1);
	\draw[scalarnoarrow] (180:1)--(0:0);
	\node at (-40:1.2) {$N_i$};
	\node at (40:1.2) {$N_i^c$};
	\node at (180:1.2) {$\phi$};
	\node at (150:0.3) {$\pm h$};
	\end{tikzpicture}	
	\begin{tikzpicture}[line width=1 pt, scale=1.3]
	\draw[fermion] (0:0)--(-40:1);
	\draw[fermionbar] (0:0)--(40:1);
	\draw[scalarnoarrow] (180:1)--(0:0);
	\node at (-40:1.2) {$N_i$};
	\node at (40:1.2) {$\psi_{\phi}^c$};
	\node at (180:1.2) {$\phi$};
	\node at (130:0.25) {$\kappa$};
	\end{tikzpicture}
	\begin{tikzpicture}[line width=1 pt, scale=1.3]
	\draw[scalarnoarrow] (0:0)--(-40:1);
	\draw[scalarnoarrow] (0:0)--(40:1);
	\draw[scalarnoarrow] (180:1)--(0:0);
	\node at (-40:1.2) {$\tilde{N}_i$};
	\node at (40:1.2) {$\tilde{N}_i$};
	\node at (180:1.2) {$\phi$};
	\node at (155:0.4) {$\pm hM_1$};
	\end{tikzpicture}
	\begin{tikzpicture}[line width=1 pt, scale=1.3]
	\draw[scalarnoarrow] (0:0)--(-40:1);
	\draw[scalarnoarrow] (0:0)--(40:1);
	\draw[scalarnoarrow] (180:1)--(0:0);
	\node at (-40:1.2) {$\tilde{N}_i$};
	\node at (40:1.2) {$\tilde{N}_j$};
	\node at (180:1.2) {$\phi$};
	\node at (145:0.4) {$\left\langle \phi \right\rangle h^2$};
	\node at (215:0.4) {$\left\langle \phi \right\rangle \kappa^2$};
	\end{tikzpicture}
	\begin{tikzpicture}[line width=1 pt, scale=1.3]
	\draw[scalarnoarrow] (0:0)--(-45:1);
	\draw[scalarnoarrow] (0:0)--(0:1);
	\draw[scalarnoarrow] (0:0)--(45:1);
	\draw[scalarnoarrow] (180:1)--(0:0);
	\node at (-45:1.2) {$\tilde{N}_i$};
	\node at (0:1.2) {$\tilde{N}_i$};
	\node at (45:1.2) {$\tilde{N}_i$};
	\node at (180:1.2) {$\phi$};
	\node at (155:0.40) {$\pm h\kappa$};
	\end{tikzpicture}

	\caption{Inflaton decay channels} \label{fig:Feyn1}
\end{figure}
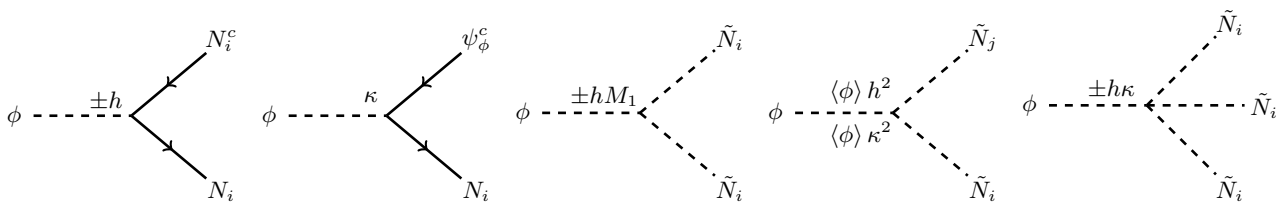

To further proceed with our analysis we must now study the boson and fermion mass matrices. Recall the truncation conditions stated in the previous section, $ \Phi-\bar{\Phi}=\mathrm{N_1}=\mathrm{N_2}=0$. Taking in consideration that $\kappa\simeq 3.5 \times 10^{-6}$, from the amplitude of the scalar curvature perturbations, and $h \gg \kappa$  we can diagonalize the mass matrix and simplify the results. From the second derivative of the potential with respect to the fields we obtain the bosonic squared masses 
\begin{equation}
	M_{B}^2 =\left(
	\begin{array}{ccc}
	\kappa^2  \phi ^2 & \frac{ \phi \kappa(M_1-h \phi )}{\sqrt{2}} & \frac{\phi \kappa  (M_1+h \phi )}{\sqrt{2}} \\
	\frac{\phi \kappa (M_1-h \phi )}{\sqrt{2}} & \frac{\kappa^2  \phi ^2}{2}+(M_1-h \phi )^2 & \frac{\kappa^2  \phi ^2}{2} \\
	\frac{\phi \kappa  (M_1+h \phi )}{\sqrt{2}} & \frac{\kappa^2 \phi ^2}{2} & \frac{\kappa\phi ^2}{2}+(M_1+h \phi )^2 \\
	\end{array}
	\right)\simeq
	 \left(
	\begin{array}{ccc}
\frac{	\kappa^2  \phi ^2}{2} &0 & 0
	\\
0 & (M_1-h \phi )^2 & 0 \\
	0 & 0 &(M_1+h \phi )^2 \\
	\end{array}
	\right).
\end{equation}

\begin{figure}[H]
	\centering
	\includegraphics[totalheight=6cm]{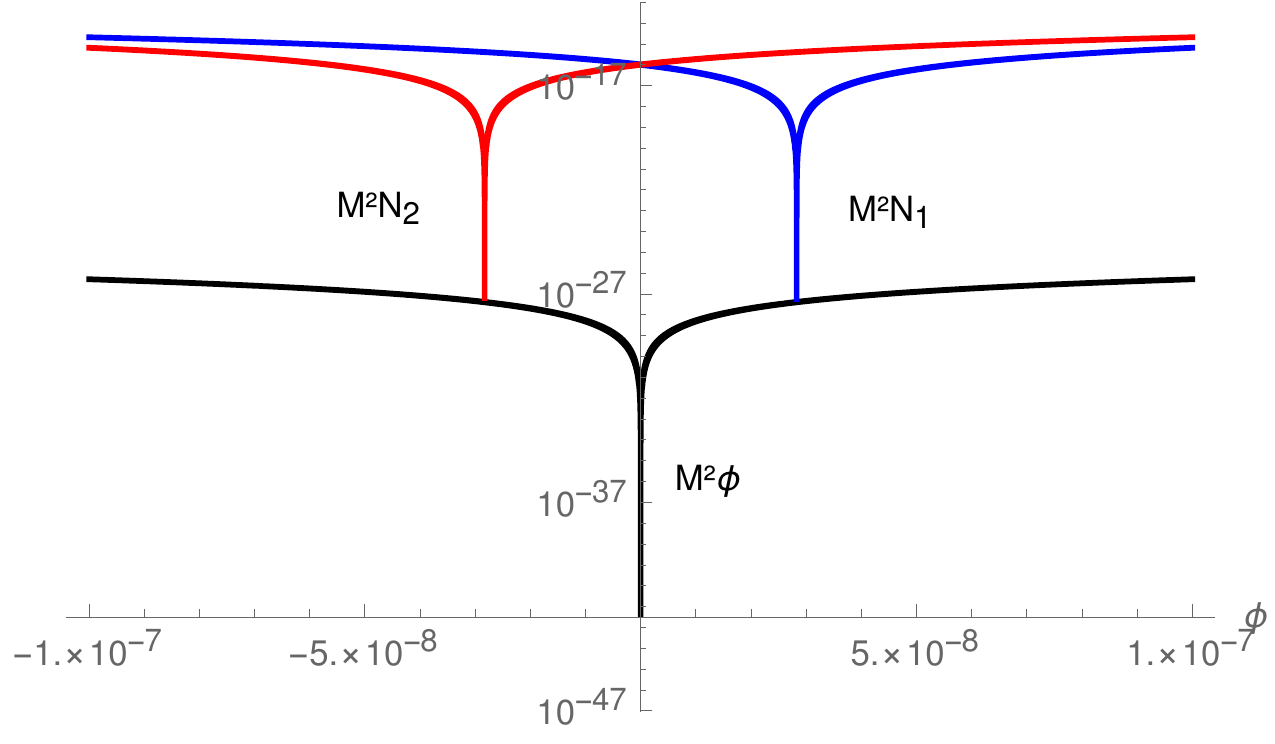}
		\caption{Evolution of the inflaton and right-handed superfield masses as the inflaton field value decreases. There is a (small) region where the inflaton mass is lower than each of the right-handed (s)neutrinos masses and where the decay will be allowed. Results are given in Planck units. We have taken $h=1$ and $M_1 = 10^{-8}m_P$.} \label{fig:Inflaton_decay}
		
\end{figure}

For the fermions, we can collect the terms for the mass matrix from the Lagrangian in equation \eqref{Lagrangian SUSY} and we may then simplify it under the assumption that $h \gg \kappa$,
\begin{equation}
M_{F} =\left(
\begin{array}{ccc}
0 & \frac{\kappa}{\sqrt{2}}\phi &  \frac{\kappa}{\sqrt{2}}\phi \\
 \frac{\kappa}{\sqrt{2}}\phi & M_1-h\phi  & 0 \\
 \frac{\kappa}{\sqrt{2}}\phi & 0 & M_1+h \phi \\
\end{array}
\right)\simeq
 \left(
 \begin{array}{ccc}
 \frac{\kappa}{\sqrt{2}}\phi &0 & 0
 \\
 0 & M_1-h \phi  & 0 \\
 0 & 0 &M_1+h \phi  \\
 \end{array}
 \right).
\end{equation}	
To a very good degree of approximation, supersymmetry is preserved, and bosonic and fermionic masses are the same. 
For the right-handed neutrino superfields, far from the origin of the potential, we have a mass splitting between the two species
\begin{equation}
	M_\pm = \left|M_1 \pm h\phi \right|. 
	\label{Eq:M+-}
\end{equation}
The inflaton can only decay when its mass is greater than two of the right-handed species. This may only occur for very small regions, at $\phi\simeq \pm M_1/h$, as verified in Fig.  \ref{fig:Inflaton_decay}.

With the introduction of the inflaton and right-handed neutrino superpartners we have allowed new decay channels to produce back inflaton particles. Therefore a more rigorous analysis on the reheating period is required. 
As we can see in Fig.  \ref{fig:Inflaton_decay}, on the scalar field first passage at $\phi\sim M_1/h$ the $N_1$ sparticles are produced. Since, at this point, their masses are approximately zero, these particles are produced as relativistic degrees of freedom. However, when the $\phi$ values change, the (s)neutrino masses rapidly increase. They can, for instance, immediately decay into SM particles, if the relevant decay rate is larger than $H$, stay as relativistic particles or become non-relativistic, depending on the thermal bath temperature.
When the $N_1$ sparticle masses become again larger than the inflaton and inflatino masses, the latter may be produced. Once again, the generated sparticles may be either non-relativistic or relativistic. The $N_2$ (s)neutrinos production is entirely analogous, for negative field values.  
Once the inflaton oscillation amplitude becomes lower than $M_1/h$ it can no longer decay into right-handed (s)neutrinos. Nonetheless, the latter can still decay into inflatons and inflatinos. However, all the interaction rates related to the inflatinos are $\kappa$ suppressed. We will thus ignore their production and neglect their interactions in the rest of our analysis.


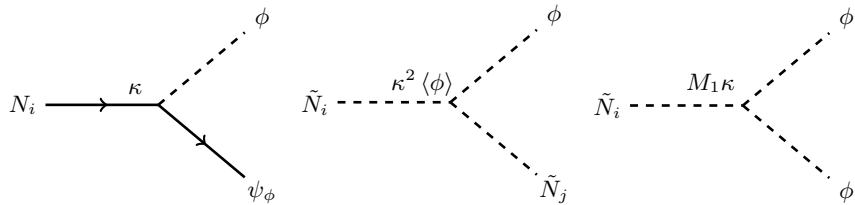
\begin{figure}[H]
	\centering
	\begin{tikzpicture}[line width=1 pt, scale=1.5]
	\draw[fermion] (0:0)--(-40:1);
	\draw[scalarnoarrow] (0:0)--(40:1);
	\draw[fermion] (180:1)--(0:0);
	\node at (-40:1.2) {$\psi_{\phi}$};
	\node at (40:1.2) {$\phi$};
	\node at (180:1.2) {$N_i$};
	\node at (145:0.25) {$\kappa$};
	\end{tikzpicture}
	\begin{tikzpicture}[line width=1 pt, scale=1.5]
	\draw[scalarnoarrow] (0:0)--(-40:1);
	\draw[scalarnoarrow] (0:0)--(40:1);
	\draw[scalarnoarrow] (180:1)--(0:0);
	\node at (-40:1.2) {$\tilde{N}_j$};
	\node at (40:1.2) {$\phi$};
	\node at (180:1.2) {$\tilde{N}_i$};
	\node at (145:0.3) {$\kappa^2\left\langle \phi \right\rangle $};
	\end{tikzpicture}
	\begin{tikzpicture}[line width=1 pt, scale=1.5]
	\draw[scalarnoarrow] (0:0)--(-40:1);
	\draw[scalarnoarrow] (0:0)--(40:1);
	\draw[scalarnoarrow] (180:1)--(0:0);
	\node at (-40:1.2) {$\phi$};
	\node at (40:1.2) {$\phi$};
	\node at (180:1.2) {$\tilde{N}_i$};
	\node at (145:0.35) {$M_1\kappa$};
	\end{tikzpicture}
	\caption{Inflaton production channels} 
	\label{fig:Feyn2}
\end{figure}
As already discussed, the inflaton will decay out of equilibrium into the right-handed (s)neutrinos for specific $M_\pm$ values. We have different decay rates for decays into fermions $\Gamma_\phi^{N_i}$, and bosons $\Gamma_\phi^{\tilde{N}_i}$:
\begin{flalign}
\Gamma^{N_i}_\phi=&\frac{h^2}{16\pi}m_\phi\left(1- \frac{4M_\pm^2 }{m_\phi^2} \right)^{3/2},\\
\Gamma^{\tilde{N}_i}_\phi=&	\left( \frac{h^2M_1^2}{2\pi m_\phi}+\frac{h^4\phi^2}{8\pi m_\phi}+\frac{\kappa^4\phi^2}{32\pi m_\phi} \right )
\left(1- \frac{4M_\pm ^2 }{m_\phi^2} \right)^{1/2}+
\frac{\kappa^4\phi^2}{16\pi m_\phi}  \left(1+ \frac{\left( M_+ ^2-M_-^2\right)^2 }{m_\phi^4} - 2\frac{M_+ ^2+M_-^2 }{m_\phi^2}\right)^{1/2},
\end{flalign}

In Fig.  \ref{fig:Feyn2} we have represented the possible inflaton particles production channels from (s)neutrino decays. These new scalar particles will not rejoin the homogeneous inflaton condensate. Instead, they will establish a "new" particle species, $\delta\phi$, that depending on the background temperature may join or decouple from the thermal bath. The relevant decay rates are represented by $\Gamma_{\tilde{N}_i}^{\delta\phi}$. By neglecting $\psi_\phi$ production we have $\Gamma_{N_i}^{\delta\phi}=0$ and
\begin{equation}
\Gamma^{\delta\phi}_{\tilde{N}_i}=\frac{M_1^2\kappa^2}{16\pi} \frac{1}{M_\pm}\left(1-4\frac{m_\phi^2}{M_\pm^2}\right) ^{1/2}
+\frac{\kappa^4\left<\phi\right>^2  }{16\pi}\frac{1}{M_\pm}\left( 1+\frac{\left(m_\phi^2-M_\mp^2 \right)^2}{M_\pm^4}-2\frac{m_\phi^2+M_\mp^2}{M_\pm^2}\right)^{1/2}\,.
\end{equation}

During reheating we must ensure that quarks, leptons and gauge bosons are produced before they are required for the generation of light elements at BBN. Therefore, a transition through a Yukawa coupling between the right-handed superfields and the Minimal Supersymmetric Standard Model (MSSM) particles is required. This is achieved through the interaction terms derived from the superpotential \eqref{Eq:SuperPot},
\begin{flalign}
{\cal{L}}_{SM\leftrightarrow N}=y(\tilde{N}_1+\tilde{N}_2)\tilde{H}_uL^c+y\tilde{L}\tilde{H}_u(N_1^c+N_2^c)+yH_uL(N_1^c+N_2^c)+h.c.\,.
\label{Lagrangian SUSY_rhn}	
\end{flalign}
From this Yukawa terms we will generate Dirac mass terms, and the right-handed fields may now decay into the Higgs and lepton super-doublets. These decays will be introduced in the Boltzmann equations through $\Gamma^{SM}_{N_i}$ ($\Gamma^{SM}_{\tilde N_i}$),
\begin{equation}
  \Gamma^{SM}_{N_i}=\Gamma^{SM}_{\tilde N_i}= \frac{y_{eff}^2 }{8\pi}M_\pm
  \label{gammaSMN}
\end{equation}
where and we have implicitly assumed that (s)neutrinos are much heavier than (s)leptons and (s)higges, and  we have hid all the (s)flavour details inside the effective parameter $y_{eff}$. For instance, due to the interchange symmetry $N_1\leftrightarrow N_2$, we have 
\begin{equation}
	y_{eff}^2=3 \sum_{l=1}^{3}y^2_{1l}. 
\end{equation}
 
Other processes that may lead to reheating and contribute to the production of inflaton particles, are the scatterings between the inflaton condensate and particles in the medium, which may excite the former into higher momentum states \cite{Bastero-Gil:2015lga,Manso:2018cba}. These are called evaporation processes and will convert the $\phi$ into $\delta\phi$ states,  dissipating the inflation energy density. Depending on the particle content at the time, different interactions may come into play. One could have scatterings with the right-handed superfields or even interactions at one loop with the Standard Model, see Fig. \ref{fig:Feyn4}.
\begin{figure}[H]
	\centering
	
	\begin{tikzpicture} [line width=1 pt, scale=1.5]
\draw[fermion] (-140:1)--(0,0);
\draw[scalarnoarrow] (140:1)--(0,0);
\draw[fermion] (0:0)--(1,0);
\node at (-138:1.2) {$N_{i}$};
\node at (138:1.2) {$\left\langle \phi\right\rangle $};
\node at (.5,.2) {$N_{i}$};	
\node at (180:0.25) {$h$};
\begin{scope}[shift={(1,0)}]
\draw[fermion] (0,0)--(-40:1);
\draw[scalarnoarrow] (0,0)--(40:1);
\node at (-42:1.2) {$N_{i}$};
\node at (42:1.2) {$\phi$};	
\node at (0:0.25) {$h$};
\end{scope}
\end{tikzpicture}\quad
\begin{tikzpicture} [line width=1 pt, scale=1.5]
\draw[scalarnoarrow] (-140:1)--(0,0);
\draw[scalarnoarrow] (140:1)--(0,0);
\draw[scalarnoarrow] (-40:1)--(0,0);
\node at (-138:1.2) {$\tilde{N}_i$};
\node at (138:1.2) {$\left\langle \phi\right\rangle $};
\draw[scalarnoarrow] (0,0)--(40:1);
\node at (-42:1.2) {$\tilde{N}_i$};
\node at (42:1.2) {$\phi$};	
\node at (90:0.25) {$h^2$};
\node at (-90:0.25) {$\kappa^2$};
\end{tikzpicture}

	\begin{tikzpicture}[line width=1 pt, scale=1]
		\draw[scalarnoarrow] (-1,2.3) -- (0,2);
		\draw[fermion] (0,2) -- (2,2);
		\draw[scalarnoarrow] (2,2) -- (3,2.3);
		\node at (-0.9,2.5) {$\left\langle \phi\right\rangle $};
		\node at (2.9,2.5) {$\phi$};
		\node at (1,2.4) {$N_i$};
		
		\draw[scalarnoarrow] (-1,-.3) -- (0,0);
		\draw[fermion] (2,0) -- (0,0);
		\draw[scalarnoarrow] (2,0) -- (3,-.3);
		\node at (-0.9,0) {$H_u$};
		\node at (2.9,0) {$H_u$};
		\node at (1,0.3) {$\nu_{\ell}(e_{\ell})$};
		\draw[fermion] (0,0) -- (0,2);
		\draw[fermion] (2,2) -- (2,0);
		\node at (-0.5,1) {$N_i$};
		\node at (2.5,1) {$N_i$};
		\end{tikzpicture}
		\begin{tikzpicture}[line width=1 pt, scale=1]
		\draw[scalarnoarrow] (-1,2.3) -- (0,2);
		\draw[fermion] (0,2) -- (2,2);
		\draw[scalarnoarrow] (2,2) -- (3,2.3);
		\node at (-0.9,2.5) {$\left\langle \phi\right\rangle $};
		\node at (2.9,2.5) {$\phi$};
		\node at (1,2.4) {$N_i$};
		
		\draw[fermion] (-1,-.3) -- (0,0);
		\draw[scalarnoarrow] (2,0) -- (0,0);
		\draw[fermion] (2,0) -- (3,-.3);
		\node at (-0.9,0) {$\nu_{\ell}(e_{\ell})$};
		\node at (2.9,0) {$\nu_{\ell}(e_{\ell})$};
		\node at (1,0.2) {$H_u$};
		\draw[fermion] (0,0) -- (0,2);
		\draw[fermion] (2,2) -- (2,0);
		\node at (-0.5,1) {$N_i$};
		\node at (2.5,1) {$N_i$};	
		\end{tikzpicture}
		\begin{tikzpicture}[line width=1 pt, scale=1]
		\draw[scalarnoarrow] (-1,2.3) -- (0,2);
		\draw[scalarnoarrow] (0,2) -- (2,2);
		\draw[scalarnoarrow] (2,2) -- (3,2.3);
		\node at (-0.9,2.5) {$\left\langle \phi\right\rangle $};
		\node at (2.9,2.5) {$\phi$};
		\node at (1,2.4) {$\tilde{N}_i$};
		
		\draw[scalarnoarrow] (-1,-.3) -- (0,0);
		\draw[scalarnoarrow] (2,0) -- (0,0);
		\draw[scalarnoarrow] (2,0) -- (3,-.3);
		\node at (-0.9,0) {$H_u$};
		\node at (2.9,0) {$H_u$};
		\node at (1,0.2) {$\tilde{\nu}_{\ell}(\tilde{e}_{\ell})$};
		\draw[scalarnoarrow] (0,0) -- (0,2);
		\draw[scalarnoarrow] (2,2) -- (2,0);
		\node at (-0.5,1) {$\tilde{N}_i$};
		\node at (2.5,1) {$\tilde{N}_i$};
		\end{tikzpicture}
	\caption{Dominant Feynman diagrams for evaporation processes. 
	For the square diagrams a coupling  $h^2y_{eff}^2$ arises.} \label{fig:Feyn4}
\end{figure}
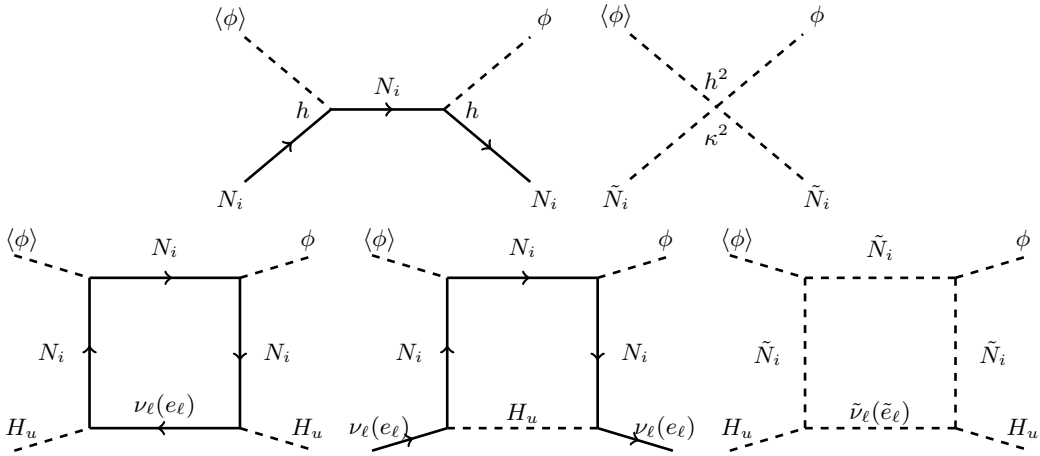 We can estimate these interaction rates as 
\begin{equation}
	\Gamma^i_{evap}=n_i\left<\sigma v\right>_{evap}^{i}\,,  \label{gamevp} 
\end{equation}
where $n_i$ represents the relevant number density and $\left<\sigma v\right>_{evap}$ the cross section. For the scatterings with the right-handed particles we have
\begin{flalign}
	\left<\sigma v\right>^{\tilde{N}_i}_{evap} \simeq &\frac{ \kappa^4+4 \ h^4}{64\pi E_{\tilde{N}_i}^2},\\
	\left<\sigma v\right>^{N_i}_{evap} \simeq &\frac{h^4}{16\pi E_{N_i}^2}. 
\end{flalign} 
The number densities will depend on their changing mass $M_\pm$, and on the temperature $T$ of the thermal bath.

The scatterings with the standard model particles, massless at this stage, and our thermal bath content, can be derived from the super-potential \eqref{Eq:SPotential} (for more details check Appendix \ref{Evapcalc}). These interactions with the inflaton particles are only induced at one loop level. 

Let us start by tackling the scattering with the Higgs boson. We can estimate, to a very good approximation, the loop diagram as yielding an effective four-point interaction. Making $g'\sim\mathcal{M}_{\phi H_u}/16\pi^2$ our effective coupling, we have that the resulting cross section, with a center of mass energy $\sqrt{s}\sim 3T$, is (see Appendix \ref{Evapcalc})
\begin{equation}
\sigma_{\phi H_u}=\frac{\left|\mathcal{M}_{\phi H_u}\right|^2}{256\pi^5 (3T)^2}\simeq \frac{h^4y_{eff}^4}{256\pi^5 (3T)^2}\left\{
\begin{array}{ll}
64 & M_1>3T\,, \\16
\left|\frac{ \left(M_1^2-9T^2\right)}{9T^2} \log \left(\frac{M_1^2-9T^2}{M_1^2}\right)\right|^2  & M_1<3T\,.
\end{array} 
\right.
\label{sigphiH}
\end{equation}
Since the Higgs spontaneous symmetry breaking is still to happen, the scattering with the sleptons, either charged or neutral, is the same. 

Moving to the left-handed neutrino scattering, considering again the effective operator, we see that $\phi^2\nu_\ell^2$ is a dimension-5 operator. This means that the effective coupling has a dimension-1 mass suppression, going like $M_1^{-1}$. This leads to a $T^2/M_1^2$ suppression when comparing with the Higgs scattering cross section. This naturally applies whenever we have fermionic external legs. 

Finally, in the scalar loops, besides the $h$ and $y_{eff}$ factors in the effective coupling, we will have $M_1^2\left\langle\phi\right\rangle^2$ and  $M_1^4$ proportionality, coming from the scalar potential, that when $M_1<3T$ makes these amplitudes drop very fast as we increase $T$. When $M_1>3T$ by solving the loop integrals one can see numerically that scalar contribution is constant, as a function of $T$, but some orders of magnitude smaller than the fermion counterpart. We therefore neglect the contributions from these interactions.   

Since this is a scattering with particles in the thermal bath, i.e. relativistic species, using $n_i\simeq T^3/\pi^2$ and Eq. (\ref{sigphiH}) in Eq. (\ref{gamevp}) we have that for the Higgs-inflaton interaction rate 
\begin{equation}
\Gamma_{\phi H_u}\simeq \frac{\left|\mathcal{M}_{\phi H_u}\right|^2T}{2304\pi^7}
\end{equation}
We can count all the possible interactions with this dominant amplitude to determine our degrees of freedom. Having a symmetry between the first two diagrams, an electromagnetic charge invariance, and the $N_1\leftrightarrow N_2$ symmetry, it results in $g_{evap}=2\times2\times2=8$. Thus
\begin{equation}
\Gamma^{SM}_{evap}\simeq \frac{g_{evap}\left|\mathcal{M}_{\phi H}\right|^2T}{2304\pi^7}\simeq\frac{ h^4y_{eff}^4 T }{18\pi^7 }\left\{
\begin{array}{ll}
4 & M_1>3T\,, \\
\left|\frac{ \left(M_1^2-9T^2\right)}{9T^2} \log \left(\frac{M_1^2-9T^2}{M_1^2}\right)\right|^2  & M_1<3T\,.
\end{array}
\right.
\label{gammaevapSM}
\end{equation}

Finally, we have the scatterings, $\left<\sigma v\right>^{eff}_a,  $  that result from annihilations and inverse annihilations of our particle content (see Appendix \ref{appendixA}). These, if efficient enough, may lead the system into equilibrium.

\subsection*{\textbf{Boltzmann Equations}}

We now wish to describe the above dynamics in a set of Boltzmann equations. For the sake of simplicity and clarity let us consider first a simplified model, where we only include either the coupling to neutrinos or sneutrinos, that we will denote generically by $N$. At the end of this section we will present the full set of equations we have used for the numerical analysis.
 
The inflaton starts oscillating when inflation ends, and decays into $N$ with decay rate $\Gamma_\phi^{N}$. The inflaton condensate (oscillating field) will also loose energy through evaporation processes when $N$ or radiation particles scatter an inflaton particle $\delta \phi$ from the condensate. The evolution Eq. for the inflaton field is then given by 
\begin{equation}
  \ddot{\phi}+3H\dot{\phi}+\Gamma_\phi\dot{\phi}+\kappa^2\phi^3=0.
  \label{infeom}
\end{equation} 
with $\Gamma_\phi=\Gamma_\phi^N+\Gamma_{evap}$,  
and therefore we have for the inflaton energy density:
\be
\dot \rho_\phi + 3 H (\rho_\phi + p_\phi) = - \Gamma_\phi (\rho_\phi + p_\phi) \,,
\ee
where $\rho_\phi + p_\phi = \dot \phi^2$, and $p_\phi$ is the pressure. The energy lost by the inflaton is converted into $N$ and $\delta\phi$ particles, and radiation is produced through the decay of $N$ into SM particles. Denoting by $\rho=\rho_{N}+ \rho_{\delta \phi} + \rho_R$ the energy density of all of them, energy conservation gives: 
\be
\dot \rho + 3 H (\rho + p) =  \Gamma_\phi (\rho_\phi + p_\phi) \,,  \label{dotrho}
\ee
with $p= p_{N} + p_{\delta \phi} + p_R$, and  
the Hubble parameter  
\begin{equation}
  H^2=\frac{1}{3 m_P^2}\left(\rho_\phi+ \rho \right)
  \,. 
\end{equation}


Based on the detailed balance principle and the total energy density conservation we wish to derive the evolution equation for the radiation energy density \cite{Arcadi:2011ev,Drees:2018dsj}. In other words, we want to determine the temperature $T$ of the thermal bath. We then need the evolution Eqs of $\rho_{N}$ and $\rho_{\delta \phi}$: 
\begin{flalign}
&\dot{\rho}_{N}+3H\left(\rho_N+p_N\right)  = \Gamma_\phi^N\left(\rho_\phi+p_\phi\right)+\Gamma_\phi^N\left(\rho_{\delta\phi}-\rho_{\delta\phi}^{eq}\right)-\left[\Gamma_N^{\delta\phi}+\Gamma^{SM}_N\right]\left(\rho_N-\rho^{eq}_N\right)-\frac{\left\langle \sigma v \right\rangle_{N}^{SM}}{E_N}\left[\rho_{N}^2-(\rho_{N}^{eq}) ^2 \right]\,,\label{Eq:neutrinoenergydensityeq}\\
  &\dot{\rho}_{\delta\phi}+3H\left(\rho_{\delta\phi}+p_{\delta\phi}\right)  = \Gamma^\phi_{evap}\left(\rho_\phi+p_\phi\right)+\Gamma_N^{\delta\phi}\left(\rho_N-\rho^{eq}_N\right)-\Gamma_\phi^N\left(\rho_{\delta\phi}-\rho_{\delta\phi}^{eq}\right)-\frac{\left\langle \sigma v \right\rangle_{\delta\phi}^{SM}}{E_{\delta\phi}}\left[\rho_{\delta\phi}^2-(\rho_{{\delta\phi}}^{eq}) ^2 \right]\,.\label{Eq:inflatonparticlenergydensityeq}		
\end{flalign}
The energy densities for $N$ and $\delta \phi$ particles are given by the relation:
\be
\rho_a \simeq E_a n_a \,,
\ee
where $E_a$ is the mean thermal energy of each component\footnote{The factor of 3 in front of the temperature comes when comparing the number density and the energy density for a dof in equilibrium:
	\be
	\langle E \rangle_{B} = \rho^{eq}/n^{eq} \simeq 2.701 T \,,\;\;\;
	\langle E \rangle_{F} = \rho^{eq}/n^{eq} \simeq 3.151 T \,,
	\ee
	where B/F are bosons/fermions respectively.
}
\begin{equation}
E_a\simeq 3 T+m_a\frac{K_1\left( \frac{m_a}{T}\right) }{K_2\left( \frac{m_a}{T}\right), } \label{Ea}
\end{equation} 
$K_n(x) $ being the modified Bessel functions. For the equilibrium number densities we have   
\begin{equation}
n_a^{eq}=\frac{g_am_a^2T}{2\pi^2}K_2\left(\frac{m_a}{T}\right),
\end{equation}
where $g_a$ is the no. of dof.

Notice that, in equations \eqref{Eq:neutrinoenergydensityeq}  and \eqref{Eq:inflatonparticlenergydensityeq} we have introduced a term describing the inflaton particles decay into right-handed neutrinos and sneutrinos. This is exactly the same interaction that we have between our inflationary fluid and the (s)neutrinos. However, in this case it can be an equilibrium decay. Note that this decay can only happen if the inflaton field amplitude is larger that $\sim M_1/h$. Thus, if reheating, meaning the relevant inflationary field dissipation, is successfully driven by evaporation we will not observe this decay channel.

The last step is therefore to set the equation for the radiation and temperature. For that we use Eq. (\ref{dotrho}) and the total energy conservation
\be
\dot{\rho}_T + 3 H( \rho_T + p_T) =0 \,,\label{Eq:totalenergyconservation}
\ee
where $\rho_T$ and $ p_T$ are the sum of all contributions to the energy densities and pressures, respectively. This leads us into:
\begin{flalign}
\dot{\rho}_{R}+3H\left(\rho_R+p_R\right)&=3H\left(\rho_R+p_R\right)+\dot{\rho}_T-\dot{\rho}_N-\dot{\rho}_{\delta\phi}-\dot{\rho}_\phi \\
&=\dot{\rho}_T+3H\left(\rho_T+p_T\right)+\Gamma^{SM}_N\left(\rho_N-\rho^{eq}_N\right)+\frac{\left\langle \sigma v \right\rangle_{N}^{SM} }{E_N}\left[\rho_{N}^2-(\rho_{N}^{eq}) ^2 \right]
+\frac{\left\langle \sigma v \right\rangle_{\delta\phi}^{SM}} {E_{\delta\phi}}\left[\rho_{\delta\phi}^2-(\rho_{\delta\phi}^{eq}) ^2 \right]\,,\\
&=\Gamma^{SM}_N\left(\rho_N-\rho^{eq}_N\right)+\frac{\left\langle \sigma v \right\rangle_{N}^{SM} }{E_N}\left[\rho_{N}^2-(\rho_{N}^{eq}) ^2 \right]
+\frac{\left\langle \sigma v \right\rangle_{\delta\phi}^{SM}} {E_{\delta\phi}}\left[\rho_{\delta\phi}^2-(\rho_{\delta\phi}^{eq}) ^2 \right]\,,	
\end{flalign}
Finally, the pressure of each particle component is given by
\be
p_a = -\frac{\dot E_a}{E_a} \frac{\rho_a}{3H}, \label{pressurei}
\ee
which using the definition of the mean thermal energy $E_a$ in equation \eqref{Ea} has the right relativistic and non-relativistic limits: when $m_a \gg T$ and $E_a \simeq m_a$ then $p_a\simeq 0$; and when $m_a \ll T$, we have $p_a \simeq - (\rho_a/3H) (\dot T/T)$. 
To find the evolution of $T$ we will use the radiation energy density, 
\be
\rho_R = \frac{\pi^2}{30} g_{\star R}(T) T^4 \,, \label{rhoRT}
\ee
where $g_{\star R}(T)$ is the effective number of relativistic degrees of freedom (dof), such that the equation of motion for $T$ is given by: 
\be
\frac{\dot T}{T} = \frac{1}{4}\left ( 1 + \frac{\dot g_{\star R}}{g_{\star R}} \right)^{-1} \frac{\dot \rho_R}{\rho_R} \,,
\ee
which we do not write explicitly. 

We can now write the full set of equations
\begin{flalign}\label{Eq:fullbotzmannn1}
&\ddot{\phi}+3H\dot{\phi}+\kappa^2\,\phi^3  = -\left[\Gamma^{N_i}_\phi+\Gamma^{\tilde{N}_i}_\phi+\Gamma_{evap} \right]\dot{\phi}\,,  \\
&\dot \rho_{N_i} + 3 H (\rho_{N_i} + p_{N_i}) = \Gamma_\phi^{N_i} \dot{\phi}^2+\Gamma^{N_i}_\phi(\rho_{\delta\phi}-\rho_{\delta\phi}^{eq})
-\Gamma_{N_i}^{\delta \phi} (\rho_{N_i} - \rho_{N_i}^{eq})
-\Gamma_{N_i}^{SM} (\rho_{N_i} - \rho_{N_i}^{eq}) 
-\frac{\left\langle \sigma v \right\rangle_{N_i}^{SM}}{E_{N_i}} \left[\rho_{N_i}^2-(\rho_{N_i}^{eq})^2 \right]
\,, \label{drhoNi}\\
&\dot \rho_{\tilde{N}_i} + 3 H (\rho_{\tilde{N}_i} + p_{\tilde{N}_i}) = \Gamma_\phi^{\tilde{N}_i} \dot{\phi}^2 +\Gamma^{\tilde{N}_i}_\phi(\rho_{\delta\phi}-\rho_{\delta\phi}^{eq})
-\Gamma_{\tilde{N}_i}^{\delta \phi} (\rho_{\tilde{N}_i} - \rho_{\tilde{N}_i}^{eq}) 
-\Gamma_{\tilde{N}_i}^{SM} (\rho_{\tilde{N}_i} - \rho_{\tilde{N}_i}^{eq}) 
-\frac{\left\langle \sigma v \right\rangle_{\tilde{N}_i}^{SM}}{E_{\tilde{N}_i}}\left[\rho_{\tilde{N}_i}^2-(\rho_{\tilde{N}_i}^{eq})^2 \right]
\,, \label{drhophi}\\
&\dot \rho_{\delta \phi} + 3 H (\rho_{\delta \phi} + p_{\delta \phi}) = \Gamma_{evap} \dot{\phi}^2-\left[\Gamma^{N_i}_\phi+\Gamma^{\tilde{N}_i}_\phi\right](\rho_{\delta\phi}-\rho_{\delta\phi}^{eq})
+\Gamma_{\tilde{N}_i}^{\delta \phi} (\rho_{\tilde{N}_i} - \rho_{\tilde{N}_i}^{eq}) 
-\frac{\left\langle \sigma v \right\rangle_{\delta \phi}^{SM}}{E_{\delta \phi}}\left[\rho_{\delta \phi}^2-(\rho_{\delta \phi}^{eq})^2\right]\,, \\
&\dot \rho_R + 3H (\rho_R + p_R) = \Gamma_{N_i}^{SM} ( \rho_{N_i}- \rho_{N_i}^{eq}+\rho_{\tilde{N}_i}- \rho_{\tilde{N}_i}^{eq})
+\frac{\left\langle \sigma v \right\rangle_{a}^{SM}}{E_{a}}\left[\rho_{a}^2-(\rho_{a}^{eq})^2 \right] 	\label{Eq:fullbotzmannn2}
\end{flalign}
where $\Gamma_{evap}=\Gamma^{N_i} _{evap}+ \Gamma^{\tilde{N}_i} _{evap}+\Gamma^{SM}_{evap}$, $a=\left\{ N_i,\tilde{N}_i,\delta\phi\right\}$, $i=\left\{ 1,2\right\}$ represent the two (s)neutrinos species with different masses. All decay/evaporation rates and scattering cross-sections are summarized in Appendix \ref{appendixA}. 

The complete set of equations can be solved numerically. However, a semi-analytical approach can be useful to get a handle off the entire dynamics for different parameters. As we have briefly discussed in the introduction, the chosen set of free parameters can lead to different cosmological outcomes with repercussions on Dark Matter physics. We will then first describe in the next section the possible reheating scenarios into a radiation dominated Universe, and in section VI we will combine those with the constraints to have a viable Dark Matter candidate (WIMP, FIMP or OSF). 

\section{ S-$\nu$IDM paths to Standard Cosmology}
\label{sec5}
We will start with an introduction of the semi-analytical approach, generalizing what was done in \cite{Bastero-Gil:2015lga,Manso:2018cba}, and then move to the relevant scenarios introduced in Fig.  \ref{fig:Evolution}.

While oscillating about the minimum of the potential, here a quartic function, and before it decays significantly, the inflaton evolution is approximately described by a damped harmonic oscillator with varying frequency,
\begin{equation}
\ddot{\phi}+3H\dot{\phi}+\omega^{2}\phi=0\,.\label{Eq:phi^4 dp osc}
\end{equation}
The time dependent frequency $\omega$ depends on the amplitude of field oscillations, $\Phi$, similarly to what happened in the non-supersymmetric case. Numerically $\omega\simeq \sqrt{3}\kappa \Phi/2$ is in good agreement with the exact solution for the Klein-Gordon equation for a homogeneous field under a quartic potential \cite{Ichikawa:2008ne}. 
We then obtain
\begin{equation}
\phi(t)\simeq\Phi(t)\sin(\omega t+\alpha),\qquad\Phi(t)=\sqrt{\frac{\sqrt{3}m_P}{2 \kappa t}}\,.\label{Eq:phi^4 solution}
\end{equation}

\begin{figure}[H]
	\centering
	
	\begin{tikzpicture}[line width=1 pt, scale=1]
	\node at (-2.2,1) {$\phi $};
	\draw[fermion] (-2,1) -- (1.5,2);
	\node at (2,2) {$N_i,\tilde{N}_i$};
	\draw[fermion] (2.5,2) -- (3.7,2);
	\node at (4.2,2) {$SM$};
	\draw[fermion] (-2,1) -- (-0.3,0);
	\node at (0,0) {$\delta\phi$};
	\draw[fermion] (0.3,0.1) -- (1.95,1.75);
	\draw[fermion] (2.05,1.75) -- (3.7,0.1);
	\draw[fermion] (0.3,0.1) -- (3.7,2);
	\draw[scalar] (0.3,0) -- (3.7,0);
	\node at (4.1,0) {$\delta\phi$};

	\node at (-0.7,.6) {$\Gamma_{evap}$};
	\node at (-0.6,1.7) {$\Gamma^{N_i,\tilde{N}_i}_{\phi}$};
	\node at (0.9,1.2) {$\Gamma^{N_i,\tilde{N}_i}_{\phi}$};
	\node at (3.3,1) {$\Gamma^{\tilde{N}_i}_{\phi}$};
	
	\node at (3.1,2.3) {$\Gamma^{SM}$};
	\node at (2.1,0.6) {$\left\langle\sigma v\right\rangle^{SM}  $};
	\draw[scalarnoarrow][blue] (3.8,2.5) -- (3.8,-0.5);
	\node at (4.8,2.7) {$Stable\ particles$};
	\end{tikzpicture}
	
	\caption{Interactions scheme} \label{fig:Interacscheme}
\end{figure}
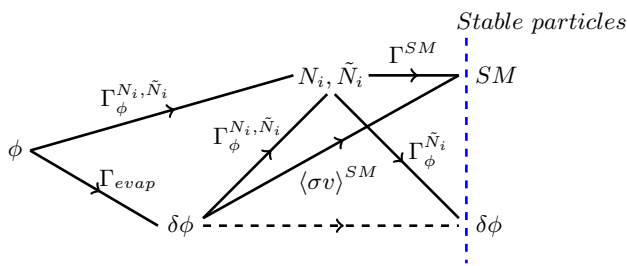

We now want to estimate the radiation energy density at the point of the reheating transition. 
Recall the radiation energy density equation,
\begin{equation}
\dot \rho_R + 3H (\rho_R + p_R) = \Gamma_{N_i}^{SM} ( \rho_{N_i} - \rho_{N_i}^{eq} +\rho_{\tilde{N}_i}- \rho_{\tilde{N}_i}^{eq})
+\frac{\left\langle \sigma v \right\rangle_{a}^{SM}}{E_{a}}\left[\rho_{a}^2-(\rho_{a}^{eq})^2 \right]\,, 
\end{equation}
where $a=\left\{ N_i,\tilde{N}_i,\delta\phi\right\}$ and $i=\left\{1,2\right\}$.
Differently from the reduced analysis done for the simpler $\nu$IDM model, there is no direct decay from the inflaton energy density into the radiation degrees of freedom. Moreover, in this more intricate scenario we have also introduced the evaporation interactions in our Boltzmann equations. We can get a feeling of the evolution in our particle content in Fig.  \ref{fig:Interacscheme}.

To a very good degree, almost all the energy lost by the inflationary fluid, either from out of equilibrium decays into right-handed (s)neutrinos or from the evaporation processes that promote the condensate into excited particle states, ends up as radiation. In the decay scenario, right-handed particles will decay into the SM states, whereas in a evaporation dominant case, although not as straightforwardly, through the inflaton particles decay into $N_i$, $\tilde{N}_i$ and through thermal scatterings, these degrees of freedom will join the SM thermal bath.

The full equation for the radiation energy density has no analytical solution. However, by studying the full numerical system, \eqref{Eq:fullbotzmannn1}-\eqref{Eq:fullbotzmannn2}, we realize that evaporation is only relevant either to start reheating for a very small parameter range ($h,y_{eff}\sim 1)$,  or much later after the inflaton decay is blocked ($h,y_{eff}\lesssim 1)$,  i.e. we have a clear separation of scales. As reported in Fig.  \ref{fig:Evolution}, if the inflaton decay is the dominant interaction after inflation this may lead directly into a radiation dominated Universe or into an early matter, controlled by the (s)neutrino. After the right-handed (s)neutrino dominant stage there can be a transition into a Universe dominated by inflaton particles that will need to thermalize, or move directly into a SM Universe. It is the hierarchy between the different decay rates, $\Gamma_\phi^N$, $\Gamma_N^{SM}$ and $\Gamma_\dphi^{SM}$ which will determine in which scenario we end. We will now describe each of them until the reheating transition. 

\subsection*{Early Matter Universe}

In order to have an Universe temporarily dominated by sneutrinos and neutrinos we will require that during the inflaton oscillating phase we have $\Gamma^N_\phi>H$ before the decay is blocked, but $\Gamma_{evap},\,\Gamma^{SM}_N<H$. These will result in the bounds
\begin{flalign}
\frac{3h\, y_{eff}^3}{8\sqrt{32\pi}\kappa} \cdot \sqrt{\frac{\left|4h^2-3\kappa^2\right|}{10h^2+3\kappa^2}} &<\frac{M_1}{m_P} <\frac{h^4}{2\sqrt{3}  \pi^{2}\kappa }\cdot \frac{\left(10h^2+3\kappa^2\right)}{\left|4h^2-3\kappa^2\right|}\label{Eq:Early_matter constrains}\,.
\end{flalign}
\\
Since the initial temperature and N number densities are small, the low evaporation rate condition is easily verified if $h,y_{eff}\lesssim {\cal{O}} (0.1)$. The RHS condition in Eq. \eqref{Eq:Early_matter constrains} comes from demanding $\Gamma^N_\phi >H$ in order to have enough (s)neutrinos, while the LHS lower limit comes from imposing $\Gamma^{SM}_N <H$ until (s)neutrinos domination.  

Let us now discuss how the Universe evolves for these parameters and then the way-out of this early-matter period, until it reaches the Standard Model radiation Universe. We will trace the system evolution with the inflaton equation \eqref{Eq:phi^4 dp osc} and the simplified equations for radiation and $N$, which in this analysis will represent both sneutrinos and neutrinos.
\begin{flalign}
	&\dot{\rho}_{N}+3H\rho_{N}=\Gamma_\phi^N\dot{\phi}^{2}-\Gamma^{SM}_N\rho_{N}~,\\
	&\dot{\rho}_{R}+4H\rho_{R}=\Gamma^{SM}_N\rho_{N}~.
\end{flalign}
Considering first the decays, as in \cite{Manso:2018cba}, in each oscillation period, $\tau_\phi= 2 \pi/\omega$, the decays into the (s)neutrinos are allowed in two occasions for each particle during a time $\delta t$ with an average decay, estimated with the maximum rate,
	\begin{equation}
	\langle \Gamma^N_\phi \rangle = 2\times2\, \Gamma_{max}\frac{\delta t}{\tau_\Phi}.
	\end{equation}
	The maximum decay width is obtained for field values $\phi=M_1/h$ and summing both neutrinos and sneutrinos contributions. 
	\begin{equation}
	\Gamma_{max}=\frac{h\left(10 h^2+3 \kappa^2 \right)}{16 \sqrt{3} \pi \kappa }M_1\,.\\
	\end{equation}
	The $\delta t$ factor is obtained when we compute the field values for which the kinematic condition fails $M_\pm=m_\phi/2$, 
	Keeping with $M_+$ for the sake of simplicity, we have
        \be
        \phi=-\frac{M_1}{h \pm\frac{\sqrt{3}}{2}\kappa}\,.
	  \ee
	If we now expand $\phi(t)$, to first order, with equation \eqref{Eq:phi^4 solution} we obtain
	\begin{equation}
	t_\pm=\frac{-M_1}{\omega \Phi (h \pm\frac{\sqrt{3}}{2}\kappa)}\,.
	\end{equation}
	Then, defining $\delta t = \left|t_+-t_-\right|$, we have for the average decay rate into (s)neutrinos:
	\begin{flalign}
	\langle \Gamma_\phi^N \rangle=\frac{h}{4 \pi^2}\cdot \frac{10 h^2+3 \kappa^2 }{\left| 4h^2-3 \kappa^2 \right|}\cdot \frac{M_1^2}{\Phi}\,.\label{Eq:Inflatondecay}
	\end{flalign}
        The decay width into SM particles is given by Eq. (\ref{gammaSMN}). 
Before the inflaton decays significantly with $h \Phi\gg M_1$, we may consider in our equations an average decay rate
	\begin{equation}
	\langle \Gamma^{SM}_N \rangle=\frac{y_{eff}^2}{8\pi}h \Phi.
\label{Eq:NSMdecay}
	\end{equation}
		
Finally, since the inflaton energy density behaves like radiation, we have for the average kinetic energy $\langle {\dot \phi}^2/2 \rangle \simeq \kappa^2 \Phi^4/2$ \cite{Ichikawa:2008ne}. It is useful to evaluate the equations in terms of the scale factor $a$. Since the inflaton, dominant at this stage, behaves as radiation under a quartic potential we have  $H= H_ex^{-2}$ and $\Phi=\Phi_ex^{-1}$ where we have defined $x=a/a_e$ with $a_e$ the moment when the inflaton left the slow roll behavior. Then the evolution equations can be written as
\begin{flalign}
&x{\rho}'_{N}+3\rho_{N}\simeq \bar{\Gamma}^{N}\frac{\Phi_e}{H_e}\Phi_e^{2}m_P^2~ x^{-1}-\bar{\Gamma}^{SM}\frac{\Phi_e}{H_e}\rho_{N}~ x\label{Eq:rhnenergydensityscalefactor}\,,\\
&x{\rho}'_{R}+4\rho_{R}\simeq \bar{\Gamma}^{SM}\frac{\Phi_e }{H_e}\rho_{N}x \,. \label{Eq:radenergydensityscalefactor}
\end{flalign}
where we have introduced the dimensionless constants:
\begin{flalign}
  &\bar{\Gamma}^{N}=\frac{ \kappa^2\ h}{3\pi ^2} \cdot \frac{10 h^2+3 \kappa^2}{\left|4h^2-3\kappa^2\right|} \cdot \frac{M_1^2}{m_P^2}\,,\label{Eq:Gamma^N}\\
  &\bar{\Gamma}^{SM}=\frac{y_{eff}^2}{8 \pi}h\,.
\end{flalign} 	
For the right-handed particles it yields
\begin{equation}
  \rho_N(x)= x^{-3} \frac{\bar{\Gamma}^{N}\,\Phi_e\,H_e m_P^2}{(\bar{\Gamma}^{SM})^2}\left(e^{\bar{\Gamma}^{SM}\frac{\Phi_e}{H_e}(1-x)}\left(1-\bar{\Gamma}^{SM}\frac{\Phi_e}{H_e}\right)-1+\bar{\Gamma}^{SM}\frac{\Phi_e}{H_e}x\right)\,,
\end{equation}
with the initial condition $\rho_N(1)=0$. We may then plug it in Eq. (\ref{Eq:radenergydensityscalefactor}) and get the energy density in radiation, 
\be 
\rho_R= \frac{\bar{\Gamma}^{N}\Phi_e^3 m_P^2}{H_e} ~ x^{-4} \left[\left(\frac{
    1}{3}\left(x^3-1\right)	- \frac{H_e\left(1+x^2\right)}{2\,\bar{\Gamma}^{SM}\Phi_e}\right)+\frac{e^{\bar{\Gamma}^{SM}\frac{\Phi_e}{H_e}(1-x)}\left(\bar{\Gamma}^{SM}\frac{\Phi_e}{H_e}-1\right)\left(1+\bar{\Gamma}^{SM}\frac{\Phi_e}{H_e}x\right)+1}{ \left(\bar{\Gamma}^{SM}	\frac{\Phi_e}{H_e}\right)^3}\right]\,.
\label{rhoRsol}
\ee
where again we have taken as initial condition $\rho_R(1)=0$. 

By observing the differential equations we see that if we have an efficient decay into (s)neutrinos, i.e. a large $\bar{\Gamma}^{N}$,  the Universe can move either into an early matter era or into a radiation phase. In other words, we could have a right-handed (s)neutrino dominant stage, that will latter decay into the SM degrees of freedom, in time for BBN. 

Let us now look for the condition for an efficient inflaton decay. We focus first in the limit $\Gamma^{SM}_N \ll \Gamma_\phi^N$, such that 
	\bea
	\rho_N &\simeq &\frac{\bar{\Gamma}^{N}\Phi_e^3 m_P^2}{2 H_e}\left(x^2-1\right)x^{-3}\,,\label{Eq:rhnenergydensitysol} \\
        \rho_R &\simeq &\bar \Gamma^{SM}\frac{\bar{\Gamma}^{N}\Phi_e^4 m_P^2}{8 H_e^2}\left(x^2-1\right)^2 x^{-4}\,.\label{Eq:rhrenergydensitysol}
	\eea
To move from an inflaton dominated Universe we need $\rho_N \geq\rho_\phi$ and the equality will happen at 
	\begin{flalign}
		&\Phi= \left(\frac{\sqrt{3}}{2}\bar{\Gamma}^{N}\right)^{1/3} \frac{m_P}{\kappa} \,,\label{Eq:Phireheating}
	\end{flalign}
	where we have used the relation on the initial parameters obtained through the Friedmann equation at the end of inflation
\be
\frac{\Phi_e^2}{H_e}=\frac{\sqrt{3}m _P}{\kappa}
\,.\label{Eq:InflatonHubble_ratio}
\end{equation}	
Recall that  $\bar{\Gamma}^{N}$ is only non-zero for $\Phi\gtrsim M_1/h$. Therefore, imposing this condition on Eq. \eqref{Eq:Phireheating} we obtain
\begin{flalign}
  & \frac{M_1}{m_P}< \frac{h^4}{2\sqrt{3} \pi^{2}\kappa} \cdot \frac{10h^2+3\kappa^2}{\left|4h^2-3\kappa^2\right|} \,.\label{Eq:boundM1}
\end{flalign}
To have the (s)neutrino dominance before recovering a radiation dominated Universe, we have to avoid the decay into SM particles becoming too efficient while the inflaton was still dominant, i.e. $\Gamma_N^{SM} < H$. From Eq. \eqref{Eq:NSMdecay}, and $H \simeq  \kappa \Phi^2/(\sqrt{3} m_P)$, this will happen for field amplitudes: 
\begin{flalign}
	&\Phi > \frac{\sqrt{3}y_{eff}^2 h}{8\pi \kappa} m_P\,.\label{Eq:PhinoSMdecay}
\end{flalign}
We can now finally get the bounds for an early matter era\footnote{Although such temporary RH sneutrino domination could  be interesting for non-thermal leptogenesis, it will be not consistent with light neutrino masses, see section VI and Fig. \ref{fig:parameters seesaw}.}. First we make sure we fulfill Eq. \eqref{Eq:boundM1}, and then impose that the decay into SM model particles is still negligible when $\rho_N = \rho_\phi$. By combining Eqs. \eqref{Eq:Phireheating} and \eqref{Eq:PhinoSMdecay} we obtain the lower limit on $M_1$:
\be
\frac{3h\, y_{eff}^3}{4\sqrt{2\pi}\kappa} \cdot \frac{\left|4h^2-3\kappa^2\right|^{1/2}}{(10h^2+3\kappa^2)^{1/2}} <\frac{M_1}{m_P} \,,
\ee
and therefore the condition Eq. \eqref{Eq:Early_matter constrains}.

To exit this (s)neutrino dominated stage we may either have a direct transition into the radiation era or an intermediate incursion into a $\delta\phi$ dominant period, which at this stage should be redshifting as radiation due to the inflaton potential shape. 
By comparing $\Gamma^{SM}_N$ and $\Gamma^{\delta\phi}_N$ we find the correct parameter range for each trajectory.
Since at this point the inflaton field amplitude has dropped and $h\Phi\ll M_1$ the decay rates become
\begin{flalign}
	\Gamma^{SM}_N=&\frac{y_{eff}^2}{8\pi}M_1\,,\\
	\Gamma^{\delta\phi}_N=&\frac{\kappa^2}{16\pi}M_1\,,
\end{flalign}
and to have the direct radiation production we require that
\begin{equation}
\Gamma^{SM}_N>\Gamma^{\delta\phi}_N\Leftrightarrow y_{eff}>\frac{\kappa}{\sqrt{2}}\,.\label{Eq:Earlymatter_yeff}
\end{equation}
Since we have now a constant decay width, this system behaves as in standard reheating. We can then estimate the reheating temperature as usual by equating the Hubble parameter at this stage with $\Gamma_N^{SM}$, giving 
\begin{equation}
	T_R\simeq 1.6\times 10^8\, y_{eff}\left(\frac{100}{g_{\star R}}\right)^{1/4}\left(\frac{M_1}{\mathrm{GeV}}\right)^{1/2}\,\mathrm{GeV}\,,\label{Eq:ReheatingT_early_matter}
\end{equation}
where $g_{\star R}$ is the number of relativistic degrees of freedom at reheating. This temperature must larger than $T_{BBN}\sim 100\, \mathrm{MeV} $.

\subsection*{Transition into Standard Cosmology}
With an efficient decay the other possible scenario is a direct transition into a radiation dominated Universe when both $\Gamma_\phi^N$, $\Gamma_N^{SM} > H$. This happens when the SM particle production occurs while the inflaton field is still dominant,
\begin{equation} 
  \frac{M_1}{m_P} < {\rm min}(
  \frac{3h\, y_{eff}^3}{4\sqrt{2\pi}\kappa} \cdot \sqrt{\frac{\left|4h^2-3\kappa^2\right|}{10h^2+3\kappa^2}},\,\frac{h^4}{2\sqrt{3} \pi^2\kappa }\cdot \frac{\left(10h^2+3\kappa^2\right)}{\left|4h^2-3\kappa^2\right|}
)
   \label{Eq:Direct radbound}
\end{equation}

Again with $y_{eff} > \kappa/\sqrt{2}$, (s)neutrinos decay dominantly into radiation and Eq. (\ref{rhoRsol}) gives 
\begin{equation}
\rho_R\simeq \bar{\Gamma}^{N}\frac{\Phi_e^2 m_P^2}{3H_e}\Phi\,.\label{Eq:radenergydensitysol}
\end{equation}
Using Eqs. \eqref{Eq:Phireheating}, \eqref{Eq:InflatonHubble_ratio} and $\rho_R=(\pi^2 g_{\star R} /30)T^4$, the reheating temperature is given by the condition $\rho_\phi \simeq \rho_R$:
\be
T_R\simeq \left(\frac{30}{\pi^2 g_{\star R}} \right)^{1/4}
\left(
\frac{\sqrt{\kappa} h}{3 \pi^2} \cdot
  \frac{(10 h^2+3\kappa^2)}{\left|4h^2-3\kappa^2\right|}
  \cdot \, M_1^2 m_P
  \right)^{1/3} 
\simeq \,4.8 \times 10^4\, h^{1/3} \left(\frac{M_1}{\mathrm{GeV}}\right)^{2/3}\left(\frac{100}{g_{\star R}}\right)^{1/4} \,\mathrm{GeV}\,.
	\label{Eq:ReheatingT_Directradiation}
\ee
Notice that Eq. \eqref{Eq:Direct radbound} only impose an upper bound on the mass parameter $M_1$, but to have an efficient (s)neutrino decay into EW states we must require $M_\pm = |M_1 \pm h \Phi| \gtrsim T_{EW} \sim O(100)$ GeV until reheating ends. For mass parameters $M_1$ below the EW scale, we need $h \Phi \gtrsim T_{EW}$. At reheating we have
\be
\kappa^2 \Phi^4 \simeq \frac{\pi^2}{30} g_{*R} T_R^4 \,,
\ee
and therefore the condition on $h \Phi$  when $M_1 < T_{EW}$ gives a lower bound on $T_R$,
\be
T_R \gtrsim 7.8\times 10^{-4} \left(\frac{100}{g_{*R}}\right)^{1/4}\cdot \frac{T_{EW}}{h} \,,
\ee
which can be translated into a lower bound on the coupling $h$:
\be
h \gtrsim 1.44\times 10^{-6} \left(\frac{T_{EW}}{100\,\gev}\right)^{1/4}\,.
\ee

\begin{figure}[th]
  \centering
  	\includegraphics[scale=0.3]{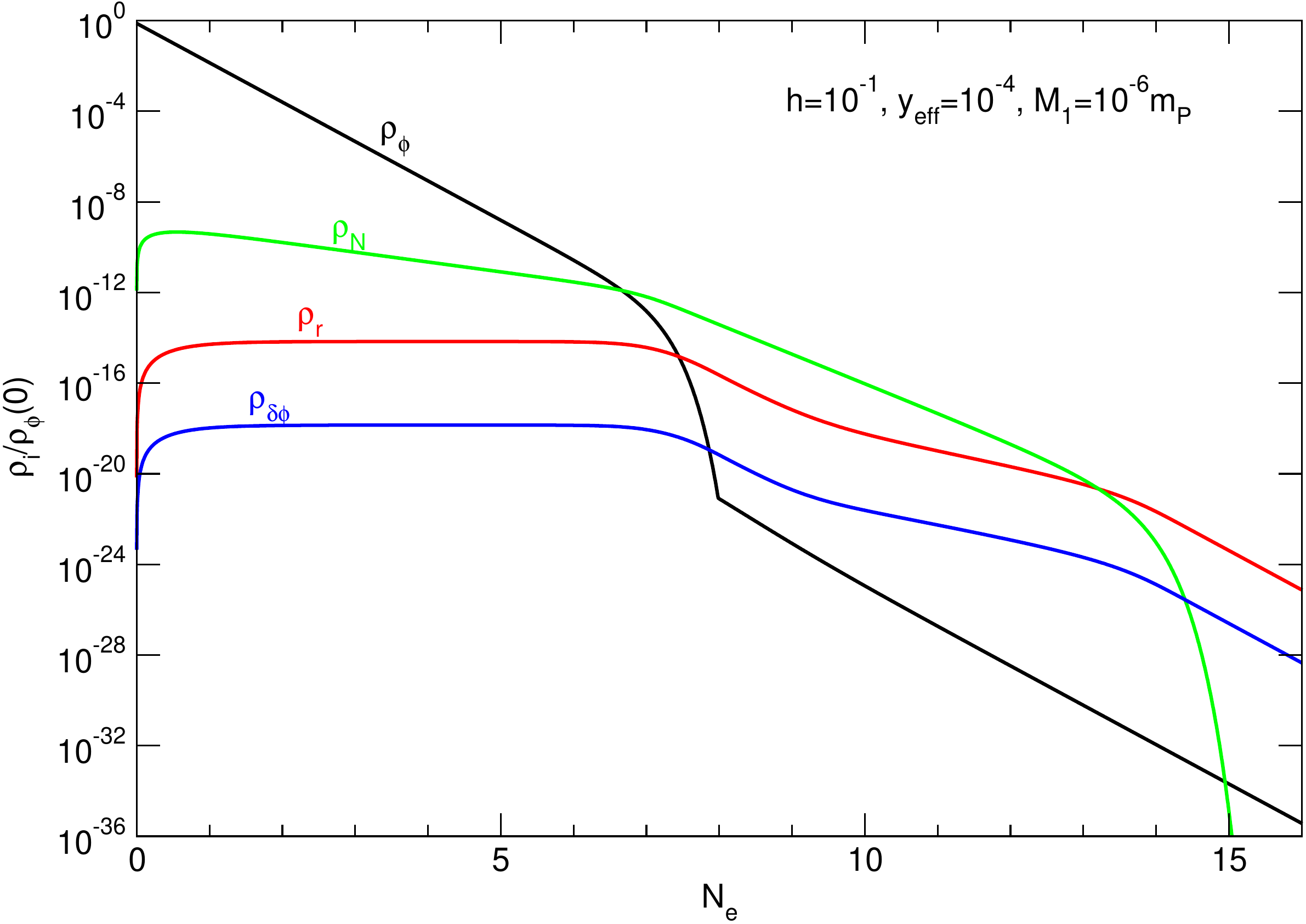}
	\includegraphics[scale=0.3]{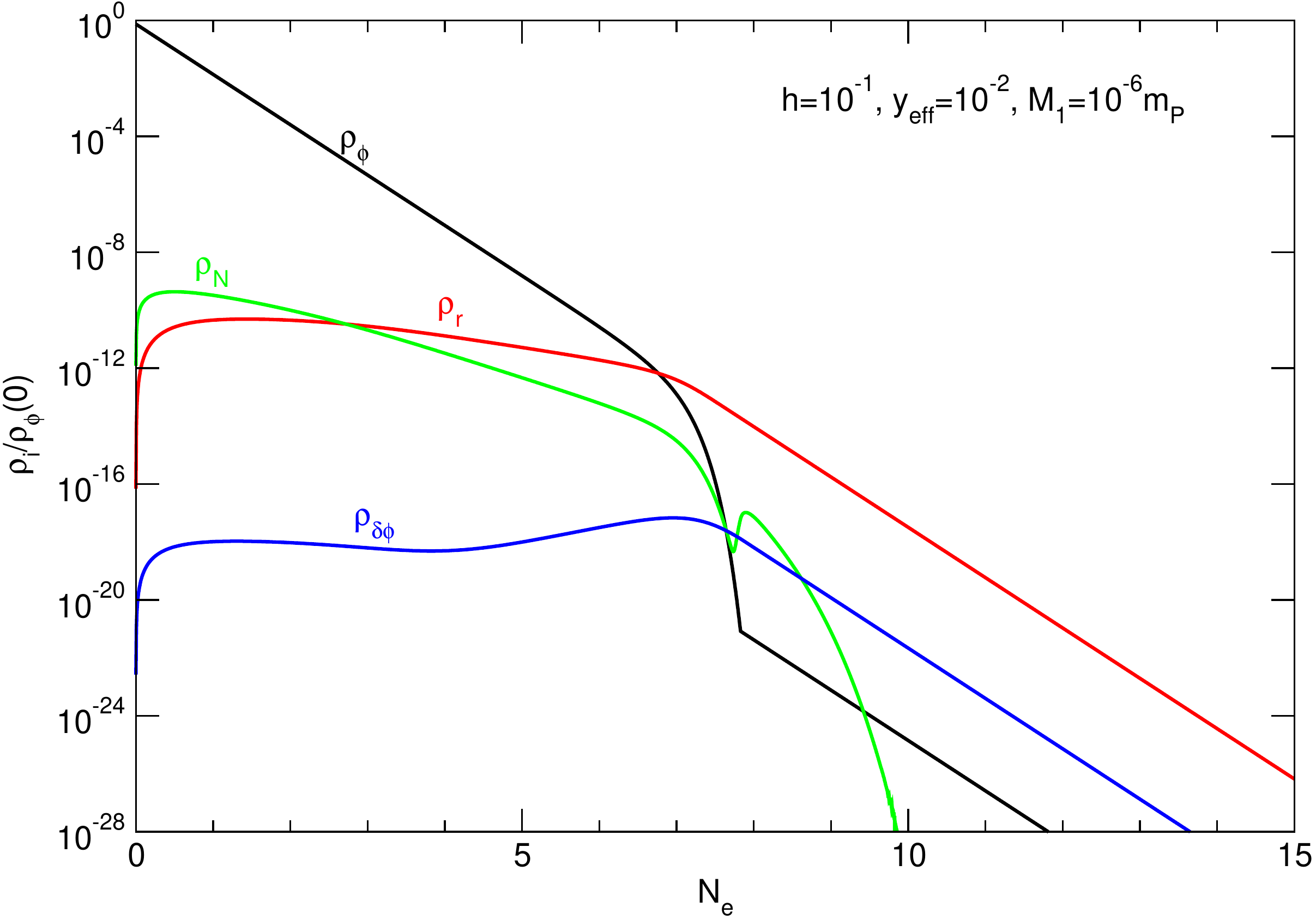}
	\caption{Evolution of the energy densities after inflation ends versus the no. of efolds $N_e=\ln a/a_e$, where $a_e=1$ at the end of inflation. $\rho_\phi$ (black) is the inflaton condensate energy density, $\rho_R$ (red) that of radiation, $\rho_N$ (green) denotes both sneutrinos and neutrinos, and $\rho_\dphi$ (blue) refers to inflaton particles. All energy densities normalized by the initial total one at $a_e$. We have taken $\kappa =3.5\times 10^{-6}$, $M_1 = 10^{-6}m_P$, $h=0.1$ and $y_{eff}=10^{-2}$ on the RHS while $y_{eff}= 10^{-4}$ on the LHS.
        } \label{plotrhoM16}
\end{figure}

In Fig. (\ref{plotrhoM16}) we have plotted an example of an early matter dominated Universe on the LHS, and a direct transition to radiation on the RHS. We have integrated the full system of equations, including decay and scattering rates, but with averaged inflaton oscillations. For the numerical examples we have used the thermal averaged decay rates:
\be
\langle \Gamma_D \rangle = \Gamma_D \frac{K_1(m_D/T)}{K_2(m_D/T)} \,.
\ee
Nevertheless, we have checked that the evolution for the inflaton condensate, radiation and (s)neutrinos until either (s)neutrino or inflaton decay (whatever happens later) is well reproduced only keeping $T=0$ decay rates, and the preceding analyses holds. 

\subsubsection*{\textbf{Incursion on a $\delta\phi$ dominated Universe ?}}

\begin{figure}[th]
  \centering
  	\includegraphics[scale=0.3]{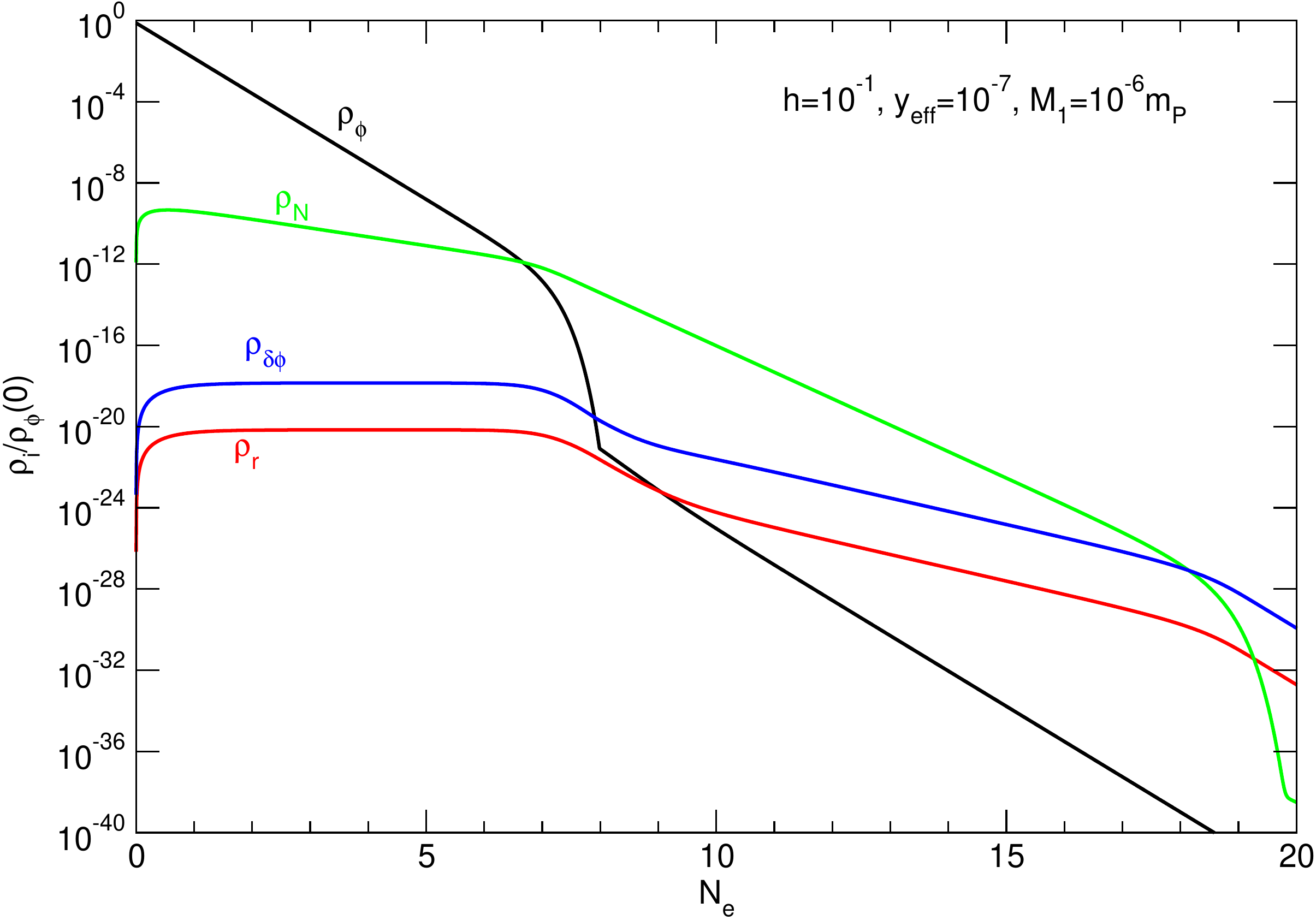}  	\includegraphics[scale=0.3]{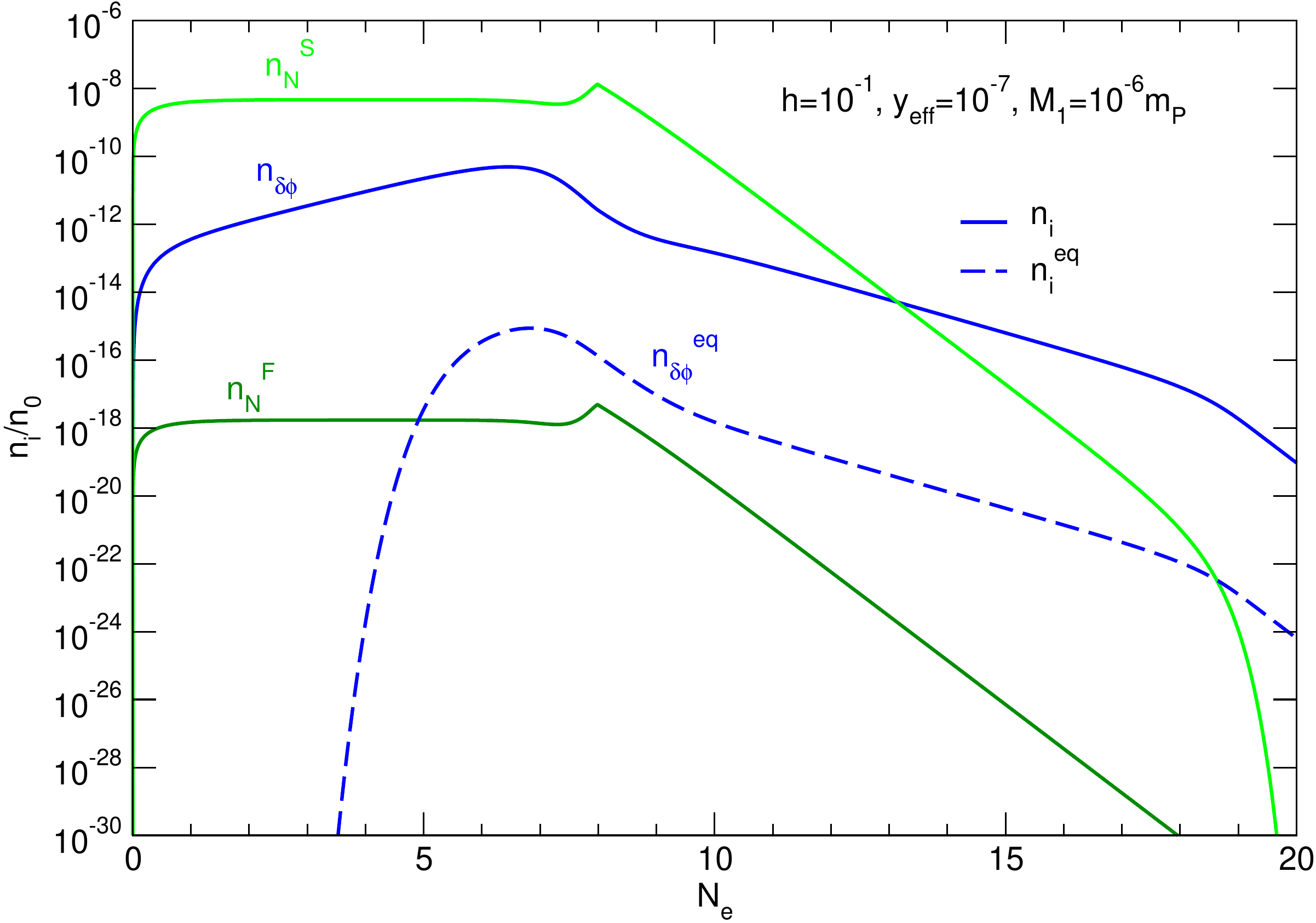}
	\caption{Evolution of the energy densities after inflation ends when $\Gamma_N^{SM} < \Gamma_N^{\delta \phi}$ on the LHS, and the number densities on the RHS, versus the no. of efolds. $\rho_N$ denotes both the sneutrinos and neutrinos contribution, while for the no. densities $n_N^S$ is that of the sneutrinos and $n_N^F$ that of the neutrinos; solid lines are the no. densities $n_i = \rho_i/E_i$, while the dashed line is the equilibrium one for inflaton particles (neutrinos and sneutrinos are very heavy particles and their equilibrium no. densities are Boltzmann suppressed and negligible). Energy densities are normalized by the initial total one at $a_e$, while no. densities are normalized by $n_0=\rho_\phi(0)/m_\phi$. We have taken: $\kappa =3.5\times 10^{-6}$, $M_1 = 10^{-6}m_P$, $h=0.1$ and $y_{eff}=10^{-7}$.
        } \label{plotrhohsm1e7M16}
\end{figure}

When $\Gamma_N^{SM} < \Gamma_N^{\delta \phi}$, i.e., for small values $y_{eff} < \kappa/\sqrt{2}$, we will have more inflaton particles than radiation both during inflaton decay and sneutrino decay.  This can be seen in the example in Fig. \eqref{plotrhohsm1e7M16}, where on the RHS we have plotted the energy densities, while on the LHS we show the number densities and their equilibrium values. Once the sneutrinos decay, inflaton particles are out-of-equilibrium and their energy density redshifts as radiation. In order to recover the standard cosmological evolution, we will need them first to thermalize before $T_{EW}$, with their thermalization rate given by:
\be
n_\dphi^{eq}\langle \sigma v \rangle_\dphi \simeq \frac{\kappa^2 y_{eff}^2}{8 \pi^4} T \,,
\ee
which will only become larger than $H$ when
\be
T \lesssim 4.4 \times 10^{-7} \left(\frac{100}{g_{\star R}}\right)^{1/2}\left(\frac{y_{eff}}{\kappa}\right)^2 \left(\frac{\rho_R}{\rho_\dphi} \right)^{1/2} \,\gev\,.
\ee
This is clearly much smaller than $T_{EW}$ for $y_{eff} < \kappa$, and therefore this scenario is ruled-out. 

\subsection*{Inefficient inflaton decay: Reheating with Evaporation} 

 We explore now parameter values for which the inflaton decay is not large enough to deplete its energy density, i.e,
\be
\frac{M_1}{m_P} >  \frac{\ h^4} {2\sqrt{3} \pi^2\kappa} \cdot\frac{10h^2+3\kappa^2}{\left|4h^2-3\kappa^2\right|} \,,
\ee
which requires $h \lesssim 0.1$ if we want to keep $M_1 \lesssim m_P$. When the field amplitude reaches its threshold value, the inflaton becomes stable and will start behaving as dark radiation as it still oscillates about the minimum of a quartic potential. However, the remaining particles produced in its decay, either right-handed (s)neutrinos or radiation, depending on $\Gamma^{SM}_N$, may scatter off the low-momentum states.
These scatterings (evaporation processes)  may lead to the excitation of the inflaton particles, thus reducing the energy density of the dominant inflationary homogeneous fluid. Moreover, thermal scatterings and annihilations will bring into equilibrium the inflaton particles transferring its energy to the thermal bath, whenever their thermalization rate $n_\dphi^{eq}\langle \sigma v \rangle_{\dphi}^{SM} $ equals the Hubble expansion rate. Taking the inflaton particles to be still relativistic at this point and $\langle \sigma v \rangle_{\dphi}^{SM} $ in Eq. \eqref{sigmadphi} in Appendix \ref{appendixA}, we have  :
\be
T_{R} \simeq 3.6 \times 10^4 y_{eff}^2 \left( 1 + \left(\frac{h}{0.004}\right)^4 y_{eff}^2 \right)\left(\frac{100}{g_{\star R}}\right)^{1/2} \, \gev \,,\label{Eq:ReheatwevapTR}
\ee
Therefore, in order to be able to evaporate the condensate and thermalize the inflaton particles through scatterings with the thermal bath before $T_{EW} \simeq 100$ GeV, we need couplings $y_{eff}$ larger than $O(0.1)$.   

\subsection*{Inefficient inflaton decay: Early Matter Universe} 

\begin{figure}[th]
  \centering
  	\includegraphics[scale=0.3]{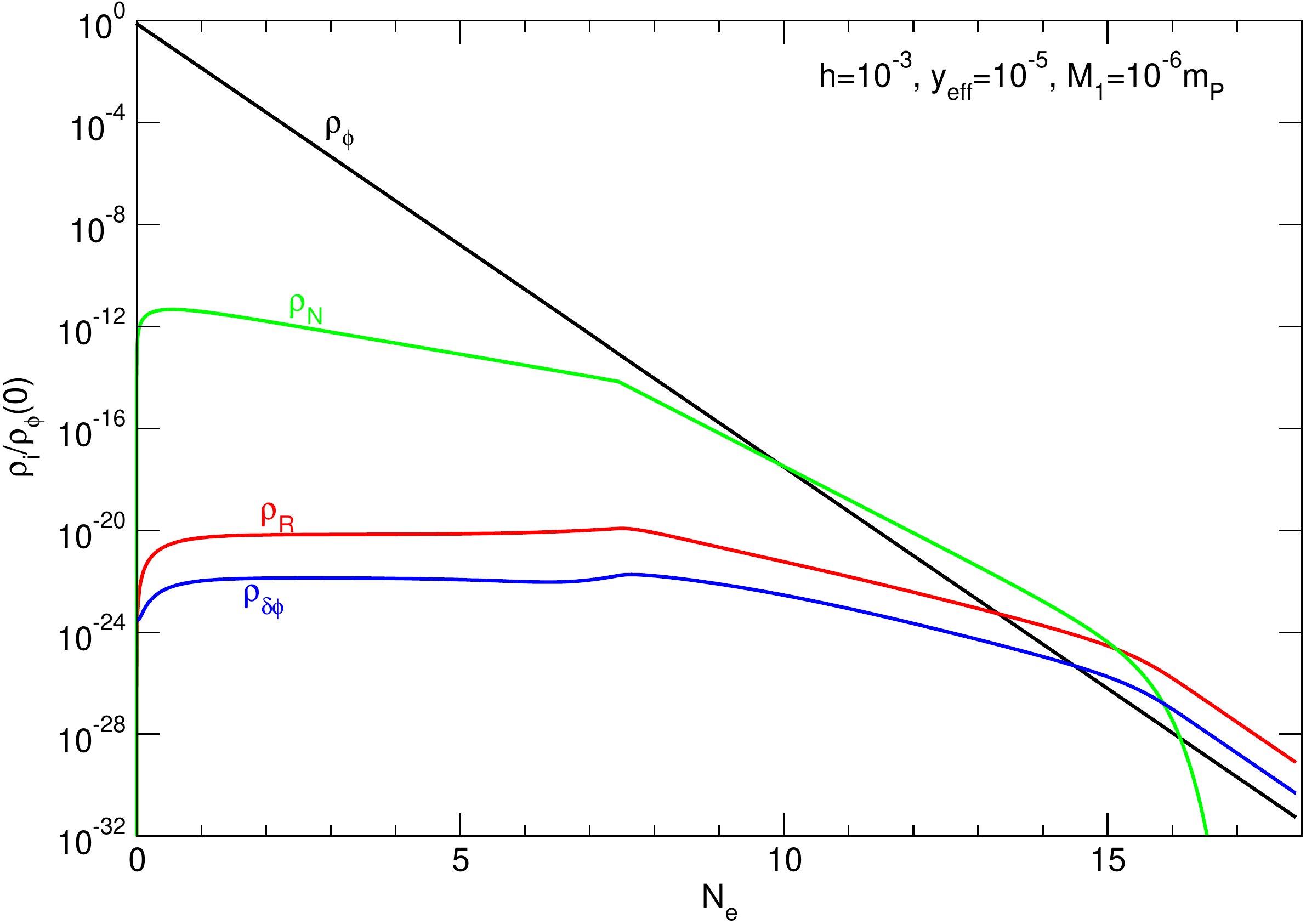}  	\includegraphics[scale=0.3]{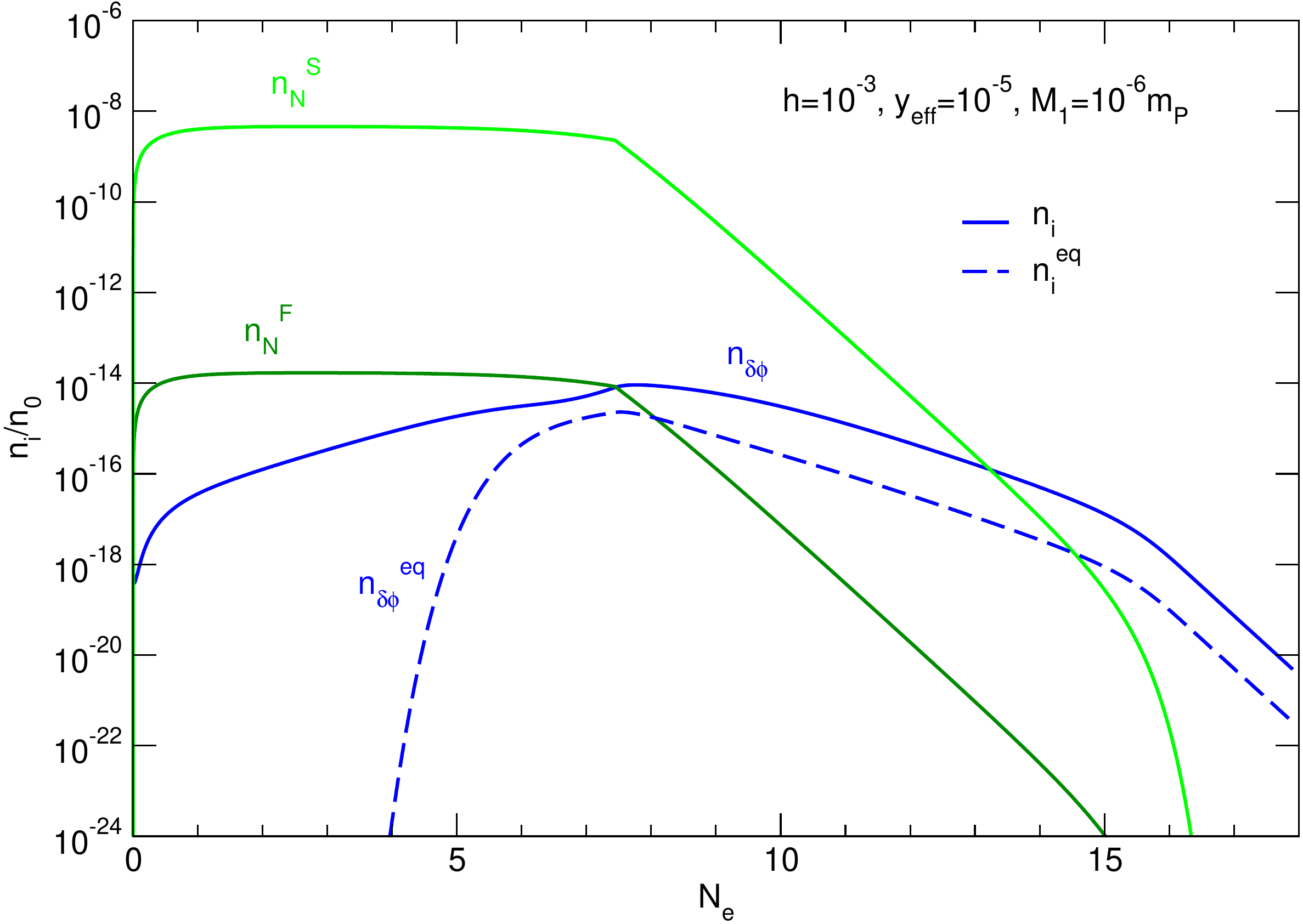}
	\caption{Evolution of the energy densities after inflation ends on the LHS, and the number densities on the RHS, versus the no. of efolds, for inefficient inflaton decay but a late (s)neutrino decay. Same convention than in Fig. (\ref{plotrhohsm1e7M16}). We have taken: $\kappa =3.5\times 10^{-6}$, $M_1 = 10^{-6}m_P$, $h=10^{-3}$ and $y_{eff}=10^{-5}$.
        } \label{plotrhohsm1e5M16}
\end{figure}

However, smaller values $y_{eff} \ll 0.1$ could still give rise to a viable reheating scenario, through the decay of the (s)neutrinos instead of evaporation. For a late decay, the (s)neutrinos may dominate the total energy density before decaying, in which case we have again an early matter dominated period and the transition to a radiation one, with $T_{R}$ given by \eqref{Eq:ReheatingT_early_matter}. We require $y_{eff} > \kappa/\sqrt{2}$ as before to ensure that the decay produces mainly radiation instead of inflaton particles. The condition to ensure (s)neutrino domination before decay is given by  $\Gamma_N^{SM} < H_{N\phi}$, where the subindex ``$N\phi$'' denotes the time at which $\rho_\phi=\rho_N$, 
\be
H_{N\phi} = H_e \left(\frac{a_e}{a_{N\phi}}\right)^2 = H_e \left(\frac{a_e}{a_{D\phi}}\right)^2\left(\frac{\rho_N}{\rho_\phi}\right)^2_{D\phi}\,,
\ee
and the subindex ``$D\phi$'' is the time when inflaton decay ends. Using $a_{D\phi}/a_e \simeq \Phi_e h/M_1$, $\rho_\phi(a_{D\phi})= \rho_\phi(a_e) (a_e/a_{D\phi})^4$ and Eq. \eqref{Eq:rhnenergydensitysol} for $\rho_N$ we have:
\be
H_{N\phi} \simeq 1.4 \times 10^2~h^6 m_P \,,
\ee
and from $H_{N_\phi} > \Gamma_N^{SM}$:
\be
y_{eff} \lesssim 59.5  ~h^3  \left(\frac{m_p}{M_1}\right)^{1/2} \,. 
\ee
And example is given in Fig. (\ref{plotrhohsm1e5M16}), for the parameter values $h=10^{-3}$, $y_{eff}=10^{-5}$ and $M_1= 10^{-6} m_P$. In this example we have $T_{R}\simeq 2.5\times 10^9$ GeV, and $\rho_\dphi > \rho_\phi$ after the (s)neutrino decay; both of them behaving like dark radiation while they are still relativistic.

\subsubsection*{\textbf{Early Evaporation}}

A final and special case comes when $h,y_{eff}\sim 1$. In this scenario, during the first oscillations about the minimum of the potential,  some (s)neutrinos will be produced when passing through  $\phi=\pm M_1/h$. These new particles will enable the evaporation processes, and due to the large couplings, the scatterings may become very efficient, larger than $H$, thus exciting and thermalizing the condensate states right after inflation. Moreover, as in the other evaporation scenarios, the excited inflaton particles, through annihilation processes, will move into the equilibrium, generating the SM degrees of freedom that lead to reheating. In any case, we go directly from a inflaton dominated Universe, to a radiation dominated Universe, and  inflaton particles thermalize soon after. The reheating temperature can be estimated just with the condition $\Gamma_{evp}=H$,  with the 1-loop contribution (box diagram) dominating the scattering with the thermal bath at large couplings, so that 
\be
  T_{R} \simeq 4.8\times 10^{14} h^4 \yeff^4\left(\frac{100}{g_{\star R}}\right)^{1/2}\, \gev \,.
  \label{Eq:ReheatingT_h1Gamevp}
\ee

\section{Dark Matter Production}
\label{sec6}
In the previous section we have unraveled the possible paths to reheat the S-$\nu$IDM Universe. We will now take a look on the consequences from such possibilities on our Dark Matter candidate, the inflaton. Then, we will combine all the bounds and constrains for each scenario to obtain the viable parameter space.

Starting with the cases where the inflaton decay was the only driving force leading to reheating, moving into the cases where evaporation was crucial for this transition, we have seen that we have always kept a stable remnant of inflatons. Due to an incomplete decay, protected by our interchange symmetry and a kinematic condition on the masses, we were able to ensure a weakly interacting and neutral DM candidate. With a mass, $M_\phi$, below a certain field amplitude or under a certain temperature the scalar degrees of freedom will finally behave as non-relativistic matter. The condition $M_\phi<M_1/2$ must be imposed to ensure the required stability.

Depending on the evolution during reheating there are two possible end states for the inflaton: either we produce inflaton particles, through evaporation, thermal scatterings and sneutrino decay (where in the latter the $\delta\phi$ may not couple with the bath); or the remnant inflaton degrees of freedom remain as a condensate, in the coherent low-momentum oscillation states. 
We will start by restating the conditions required to achieve the so called WIMPlaton scenario \cite{Bastero-Gil:2015lga}, and seeing where they fit in our parameter space. Then, we shall discuss the uncoupled dark matter scenarios, either as feebly interacting massive particles (FIMP) or as an oscillating scalar field (OSF). Finally, we address the consequences of a key feature in the right-handed neutrino portal, the generation of the observed neutrino masses through the seesaw mechanism \cite{Manso:2018cba}.

\subsection*{The WIMPlaton DM scenario}

As already discussed,  through evaporation interactions the low-momentum inflaton states in the oscillating condensate are promoted to higher momentum states. If sufficiently fast in comparison with the Hubble rate, these processes lead to the depletion of the remnant inflaton condensate, transferring its energy density into the inflaton particles. At the same time, the thermalization scatterings, with smaller but similar scattering rates, come into play thermalizing $\delta\phi$ with the rest of the thermal plasma. However, once the elastic scatterings and annihilation processes become inefficient, i.e. slower than the Hubble rate, $\delta\phi$ particles will decouple from the thermal bath and their abundance will freeze out as a standard WIMP candidate,  the so called "WIMPlaton scenario" \cite{Bastero-Gil:2015lga}.

For the efficient inflaton hypothesis, we have obtained either an early-matter epoch, dominant (s)neutrinos, or directly a reheating after depleting the inflaton condensate without requiring  either evaporation or thermalization for the energy transfer. To ensure a WIMP candidate we must extend the thermalization constrain to these scenarios. We recover  
the condition in Eq. \eqref{Eq:ReheatwevapTR}, now for the thermalization temperature $T_{\delta\phi}$.


When the temperature of the relativistic bath becomes lower than $M_\phi$, the inflaton particles will start behaving as a non-relativistic fluid, and once they decouple from the thermal bath their abundances will freeze out.
At the decoupling moment both right-handed neutrinos and sneutrinos are non-relativistic, yielding an annihilation cross-section 
\begin{equation}
		\sigma_{\delta\phi\delta\phi}\simeq\frac{h^4}{8\pi M_\phi^2}\,.\ 
\end{equation}
With the standard computation of the thermal relic abundance of a decoupled species \cite{WIMPabundance} we get 
\begin{equation}
	 M_\phi \simeq 1.4h^2\left(\frac{\Omega_{\delta\phi}h_0^2}{0.1}\right)^{1/2}\left(\frac{g_{\star F}}{10}\right)^{1/4}\left(\frac{x_F}{25}\right)^{-3/4}\mathrm{TeV},\label{Eq:MassrelationWIMP}
\end{equation}
where $g_{\star F}$ represents the number of relativistic degrees of freedom at freeze-out, $x_F=M_\phi/T_F$ and $T_F$ the freeze out temperature.

So far we have discussed a total of four different reheating scenarios that may lead us to a WIMP candidate. The S-$\nu$IDM model is defined by five parameters, the couplings $\kappa$, $h$, $y_{eff}$ and in addition the masses $M_1$ and $M_\phi$. Through the constrains on the scalar curvature perturbations we have fixed $\kappa \simeq 3.5 \times10^{-6}$. Moreover, with the measured thermal relic abundance, Eq. \eqref{Eq:MassrelationWIMP}, we can relate $h$ with $M_\phi$ and, thus reduce our independent parameters into three, $M_1$, $M_\phi$, and $y_{eff}$. 
In each scenario we have obtained constrains to ensure compatibility with the DM scenario and for each specific reheating evolution, which we now recover:
\\

\paragraph{\textbf{Early Matter Universe: $y_{eff}>\kappa/\sqrt{2}$} }
\begin{flalign}
&\frac{3h\, y_{eff}^3}{4\sqrt{2\pi}\kappa} \cdot \sqrt{\frac{\left|4h^2-3\kappa^2\right|}{10h^2+3\kappa^2}} <\frac{M_1}{m_P} <\frac{h^4}{2 \sqrt{3} \pi^{2}\kappa }\cdot\frac{\left(10h^2+3\kappa^2\right)}{\left|4h^2-3\kappa^2\right|} \label{M1EarlyMatt}\,, 
\\
&T_R\simeq 1.6\times 10^8\, y_{eff}\left(\frac{100}{g_{\star R}}\right)^{1/4}\left(\frac{M_1}{\mathrm{GeV}}\right)^{1/2}\,\mathrm{GeV}>T_{BBN}\,,
\label{TREarlyMatt}
\end{flalign}

\paragraph{\textbf{Direct Standard Model Universe}} 
\begin{flalign}
&\frac{M_1}{m_P} < {\rm min}(
\frac{3h\, y_{eff}^3}{4\sqrt{2\pi}\kappa} \cdot \sqrt{\frac{\left|4h^2-3\kappa^2\right|}{10h^2+3\kappa^2}} , \frac{h^4}{2 \sqrt{3} \pi^{2}\kappa }\cdot \frac{\left(10h^2+3\kappa^2\right)}{\left|4h^2-3\kappa^2\right|} ) \label{M1mpmin}\\
&T_R\simeq \,4.8 \times 10^4\, h^{1/3} \left(\frac{M_1}{\mathrm{GeV}}\right)^{2/3}\left(\frac{100}{g_{\star R}}\right)^{1/4} \,\mathrm{GeV}>T_{BBN}\,, \label{TRM1min}
\end{flalign}

\paragraph{\textbf{Inefficient inflaton decay}}
\begin{flalign}
	&\frac{M_1}{m_P}>\frac{h^4}{2 \sqrt{3} \pi^{2}\kappa }\cdot \frac{\left(10h^2+3\kappa^2\right)}{\left|4h^2-3\kappa^2\right|}  \label{M1Ineff}
\end{flalign}
\paragraph*{\quad \textit{c.2.}\qquad \textbf{\textit{Early Matter Universe}}:  $\kappa/\sqrt{2}<y_{eff} <59.5 ~h^3  \left(\frac{m_p}{M_1}\right)^{1/2}$}
\begin{flalign}
&T_R\simeq 1.6\times 10^8\, y_{eff}\left(\frac{100}{g_{\star R}}\right)^{1/4}\left(\frac{M_1}{\mathrm{GeV}}\right)^{1/2}\,\mathrm{GeV}>T_{BBN}\,,  \label{TRIneffEarlyMatt}
\end{flalign}
\paragraph*{\quad \textit{c.1.}\qquad \textbf{\textit{Reheating with Evaporation}}: $y_{eff} >59.5 ~h^3  \left(\frac{m_p}{M_1}\right)^{1/2}$}
\begin{flalign}
&T_R=T_{\delta\phi} >T_{EW}\,.
\end{flalign}

\begin{figure}[th]
	\centering
	\includegraphics[totalheight=8.1cm]{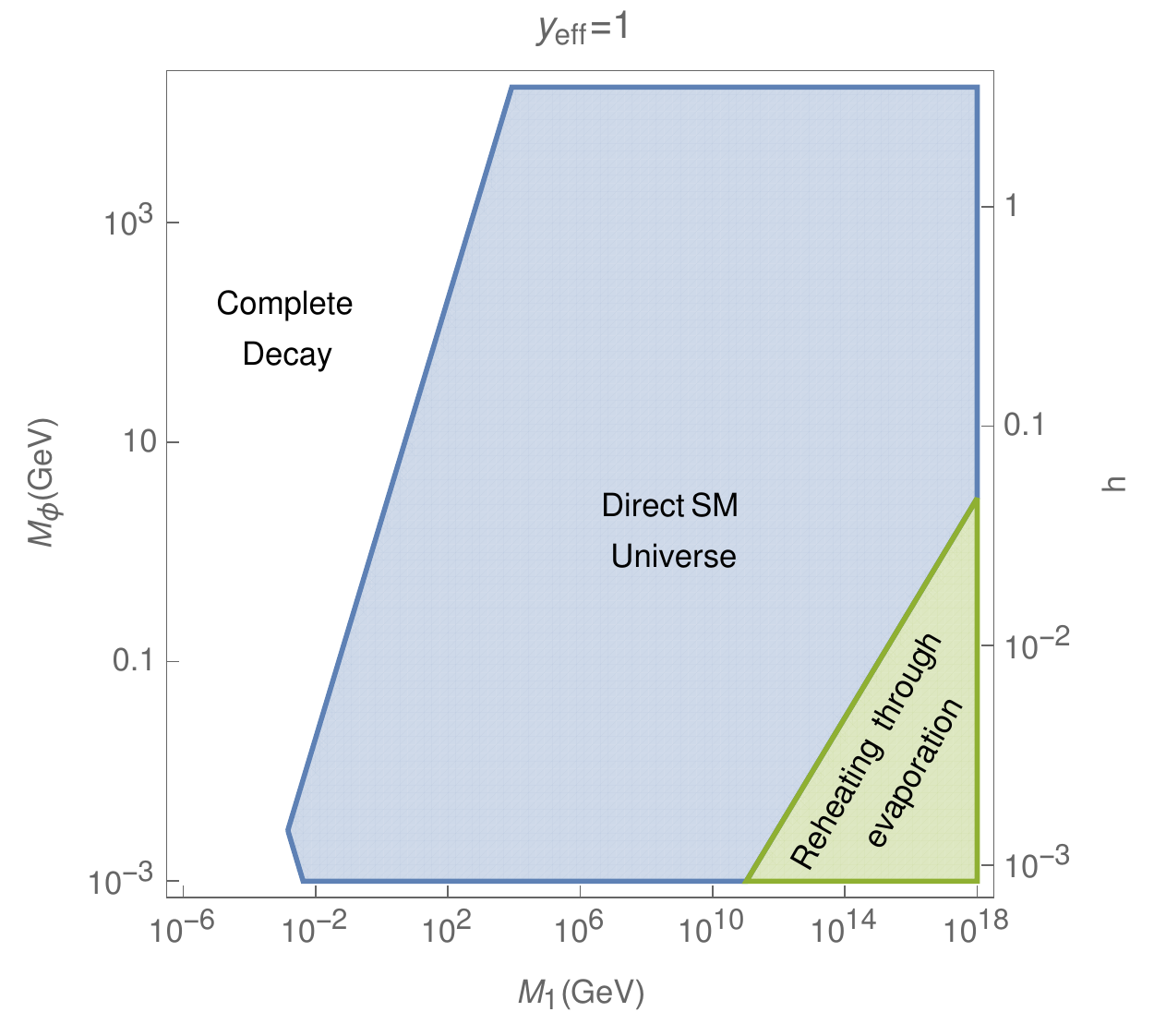}
	\includegraphics[totalheight=8.1cm]{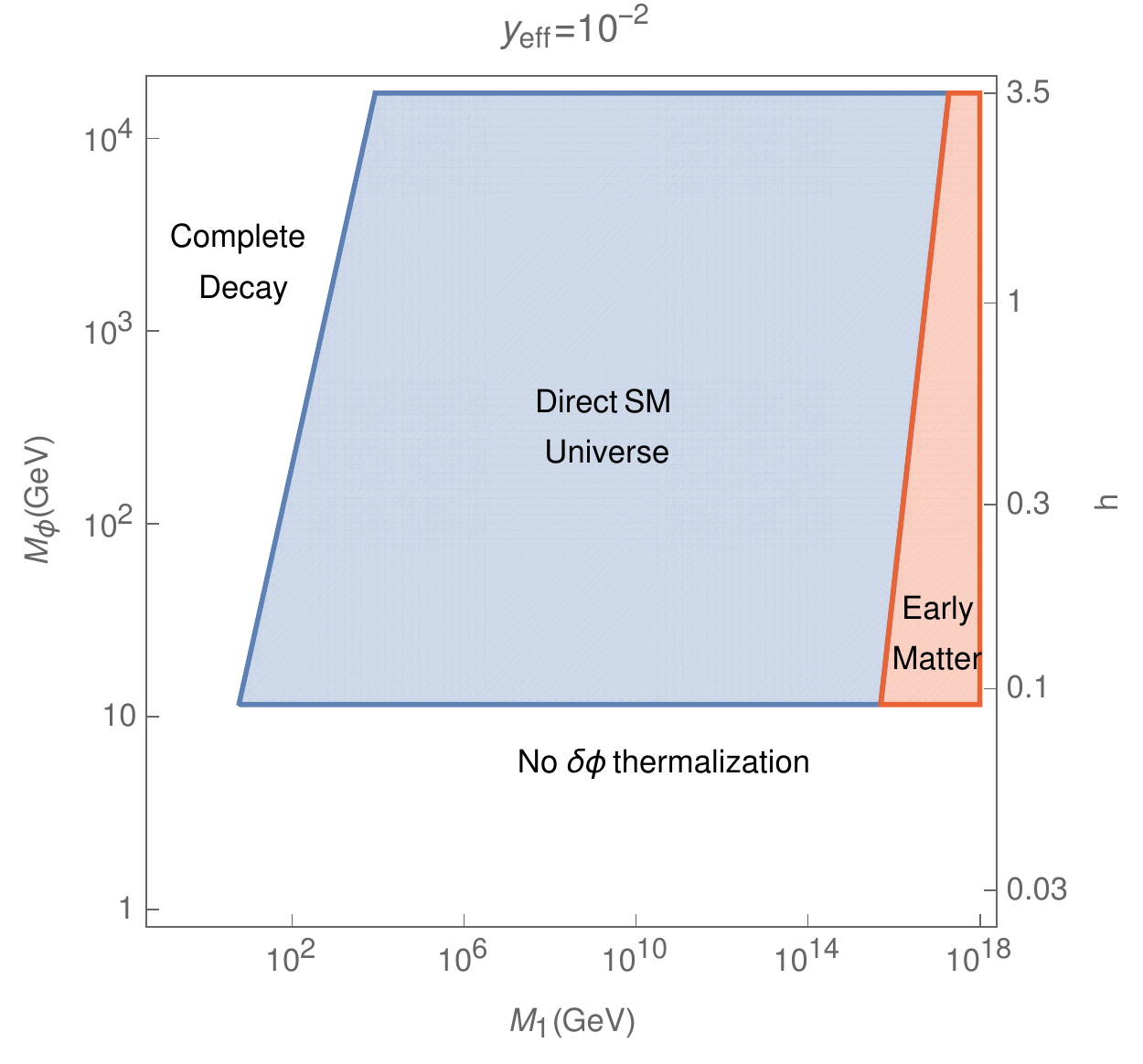}
	\includegraphics[totalheight=8.1cm]{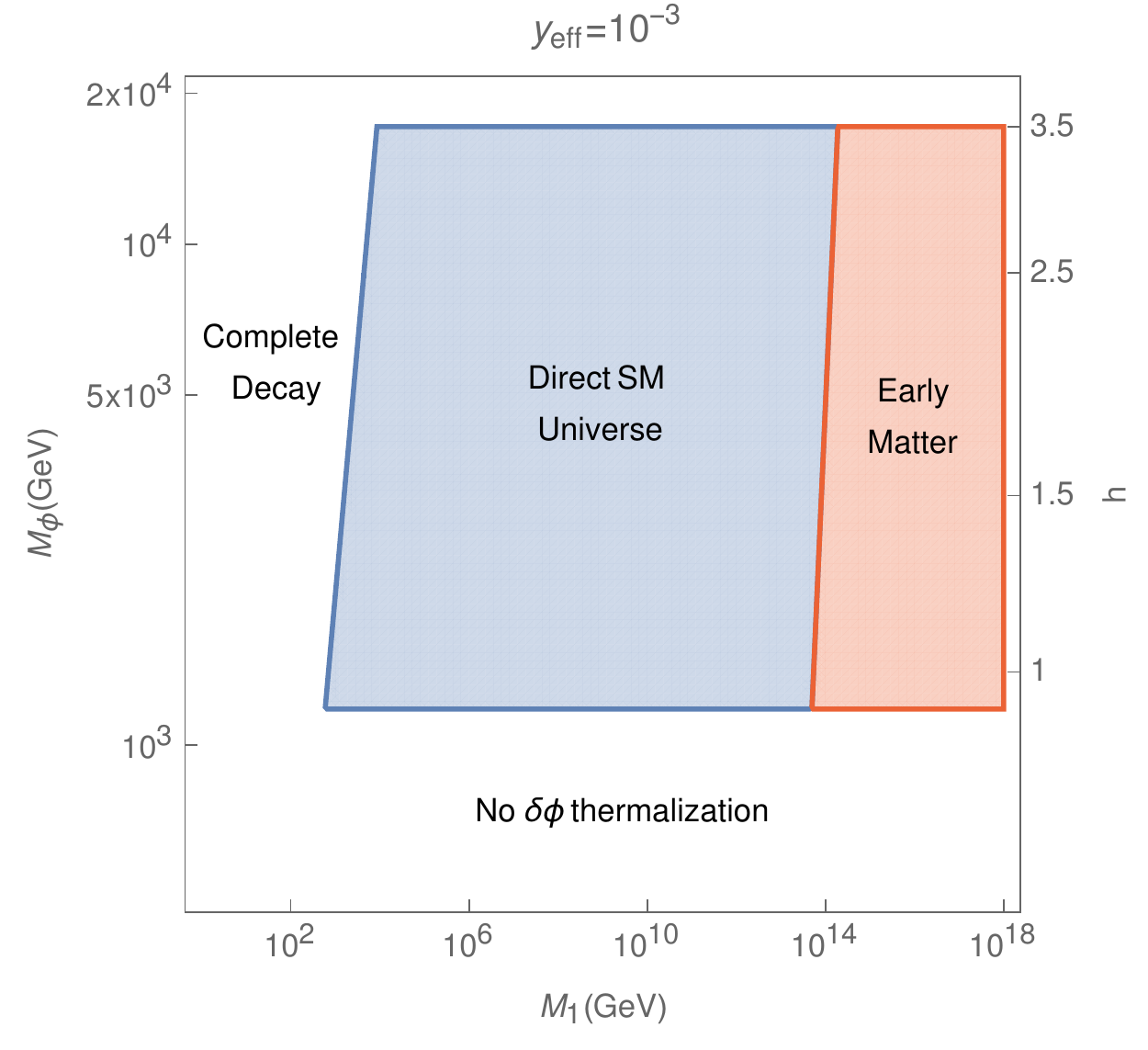}

	\caption{Collection of the allowed parameters all the reheating scenarios in the WIMPlaton scenario, taking $g_{\star R}=100$. The shaded blue region corresponds to a direct transition from inflation to a SM dominated Universe. The green region represents the inefficient inflaton decay where we have a late reheating through evaporation interactions. The red area indicates the parameters for an early matter Universe before the common radiation era. The conditions $M_\phi < 2 M_1$ and \eqref{Eq:thermalization} define "complete decay" and "no $\delta\phi$ thermalization" regions respectively.} \label{fig:parameters wimps}
	
\end{figure}

Finally, we have the common conditions to ensure a WIMP DM candidate: a bound to provide kinematical stability and the incomplete decay, $M_\phi < 2 M_1$  , and an efficient thermalization   
\begin{flalign}
	&T_{\delta\phi} \simeq 3.6 \times 10^4 y_{eff}^2 \left( 1 + \left(\frac{h}{0.004}\right)^4 y_{eff}^2 \right)\left(\frac{100}{g_{\star R}}\right)^{1/2} \, \gev>T_{EW}\,.\label{Eq:thermalization} 
\end{flalign}
Taking into account that all the couplings were taken within a perturbation theory we present the parameter space in Fig.  \ref{fig:parameters wimps} for fixed values of $y_{eff}$.

As one can see in the figure, the possible right-handed (s)neutrino masses go from GUT ($10^{16}\,\mathrm{GeV}$) to  $\mathrm{MeV}$ scales where as the inflaton mass goes from TeV to MeV scales. For high Yukawa couplings between (s)neutrinos and the standard model sector a SM production is preferred. For large $M_1$ and small $h$ the inflaton decay becomes insufficient to reheat requiring a hand from evaporation.  As we lower $y_{eff}$, we notice the appearance of evolution with a Early Matter period.
For low $M_\phi$ we observe a region where the thermalization of the inflaton particles is not achieved. If we are in the efficient inflaton decay region, we may have some hopes on a compatible Universe, with a different DM solution. However, if at some point the inflaton particles dominated the Universe after their decay was blocked, in the absence of thermalization there won't be a consistent transition into the Standard Cosmological model. We will further explore this regions without evaporation in the next subsections. We finally notice that in the case of an efficient inflaton decay we cannot reproduce an Early Matter Universe.

\subsection*{FIMPlaton Scenario}

Another possible scenario may come if we do have a  large production of inflaton particles, but the thermalization scatterings are inefficient. Meaning, the inflaton particles never couple with the cosmic plasma, thus forming a feebly interacting massive particle (FIMP) candidate \cite{Hall:2009bx,Bernal:2017kxu}.  
The inflaton particles, necessarily subdominant at BBN, can either be produced from the right-handed sneutrino decays or excited through evaporation. If these do not couple with the thermal plasma, the observed DM density can be produced by a freeze-in mechanism. 
In this scheme, the comoving inflaton particles number density freezes to a constant value when the number densities of particles producing the $\delta\phi$ become negligible due to Boltzmann suppression. Opposed to the freeze-out scenario, the contributing coupling will generally be very small.

For a FIMP candidate we must rule out the reheating with evaporation hypothesis, since this requires a thermalization with the bath to ensure a transition into standard cosmology. Thus, we may work with the scenarios of an early-matter Universe and a direct transition into SM Universe, as in Fig. \ref{plotrhoM16} and Fig. \ref{plotrhohsm1e5M16}.  As it is required to have a radiation dominated Universe at BBN, the evaporation with the SM particles will be the last available interaction for the $\delta\phi$ production. Following the analysis in \cite{Hall:2009bx} we write the relic density as
\begin{flalign}
\Omega_{\phi0}h_0^2\simeq1.01\times 10^{24} \frac{\zeta^2}{g_{\star S}\sqrt{g_{\star R}}}\,, \label{omegafimp}
\end{flalign}
where $\zeta$ represents the coupling of the interaction available before the freeze-in, and $g_{* S}$ the effective no. of dof in the entropy density. In the case of the scattering with the SM particles $\zeta^2=\frac{8}{\pi^4}h^4y_{eff}^4 + 2 \kappa^2y_{eff}^2$, and considering $g_{\star S}=g_{\star R}$, Eq. \eqref{omegafimp} results in the relation
\begin{equation}
\left(\kappa^2+\frac{4}{\pi^4}h^4y_{eff}^2\right)y_{eff}^2 \simeq5\times10^{-23}\left(\frac{\Omega_{\phi0}h_0^2}{0.12}\right)\left(\frac{g_{\star R}}{100}\right)^{3/2}\,.\label{Eq:FIMPrelation}
\end{equation}
Since $\kappa$ is fixed by scalar curvature perturbations, this holds a relation between $y_{eff}$ and $h$. For any $h$ we will have $y_{eff}\leq2.02\times 10^{-6}$. Recalling Eq. \eqref{Eq:Earlymatter_yeff}, this excludes an early matter era for a FIMP DM candidate.

As we impose that the inflaton particles do not thermalize with the rest of the bath, evaporation scatterings, having similar interaction rate, differing just by some symmetry factors, are likely not to be efficient as well.  Consequently, the inflaton condensate will not evaporate and will remain contributing for the DM abundance. Therefore, to realize a Universe with a dominant FIMP candidate one must impose the additional requirement that $\rho_{\delta\phi}>\rho_\phi$.

In a direct transition into a SM Universe, the moment inflaton decay is blocked, due to the already efficient decay into SM particles, the (s)neutrinos abundance will be depleted leaving as decay products the SM and inflaton particles, see Fig. \ref{plotrhoM16}. We can therefore compare the remnant inflaton condensate and $\delta\phi$ energy densities at reheating/ decay time, since with inefficient evaporation their relative ratio will not change. When the inflaton decay is blocked it yields,
\begin{flalign}
&\rho_\phi\simeq\frac{\kappa^2}{4}\left(\frac{M_1}{h}\right)^4 e^{\frac{4}{9}\frac{\bar{\Gamma}^N}{\kappa^2}\frac{m_P^2}{H_e\Phi_e}\left(1-\left(h\frac{\Phi_e}{M_1}\right)^3\right)},\label{Eq:FIMPrhophi}\\
&\rho_{\delta\phi}\simeq\frac{\Gamma_N^{\delta\phi}}{\Gamma_N^{SM}}\rho_R\simeq\frac{\pi^2}{60}\frac{\kappa^4}{h^2 y_{eff}^2 }\left(\frac{\sqrt{3}}{4\,\bar{\Gamma}^N}\right)^{2/3}\left(\frac{M_1}{m_P}\right)^2 g_\star^R\, T_R^4\,, \label{Eq:FIMPrhodeltaphi}
\end{flalign}
where $g_{\star R}$ are the relativistic degrees of freedom at reheating and $\bar{\Gamma}^N$, $T_R$ refer to equations \eqref{Eq:Gamma^N} and \eqref{Eq:ReheatingT_Directradiation}, respectively.

\begin{figure}[th]
	\centering
	\includegraphics[totalheight=8.85cm]{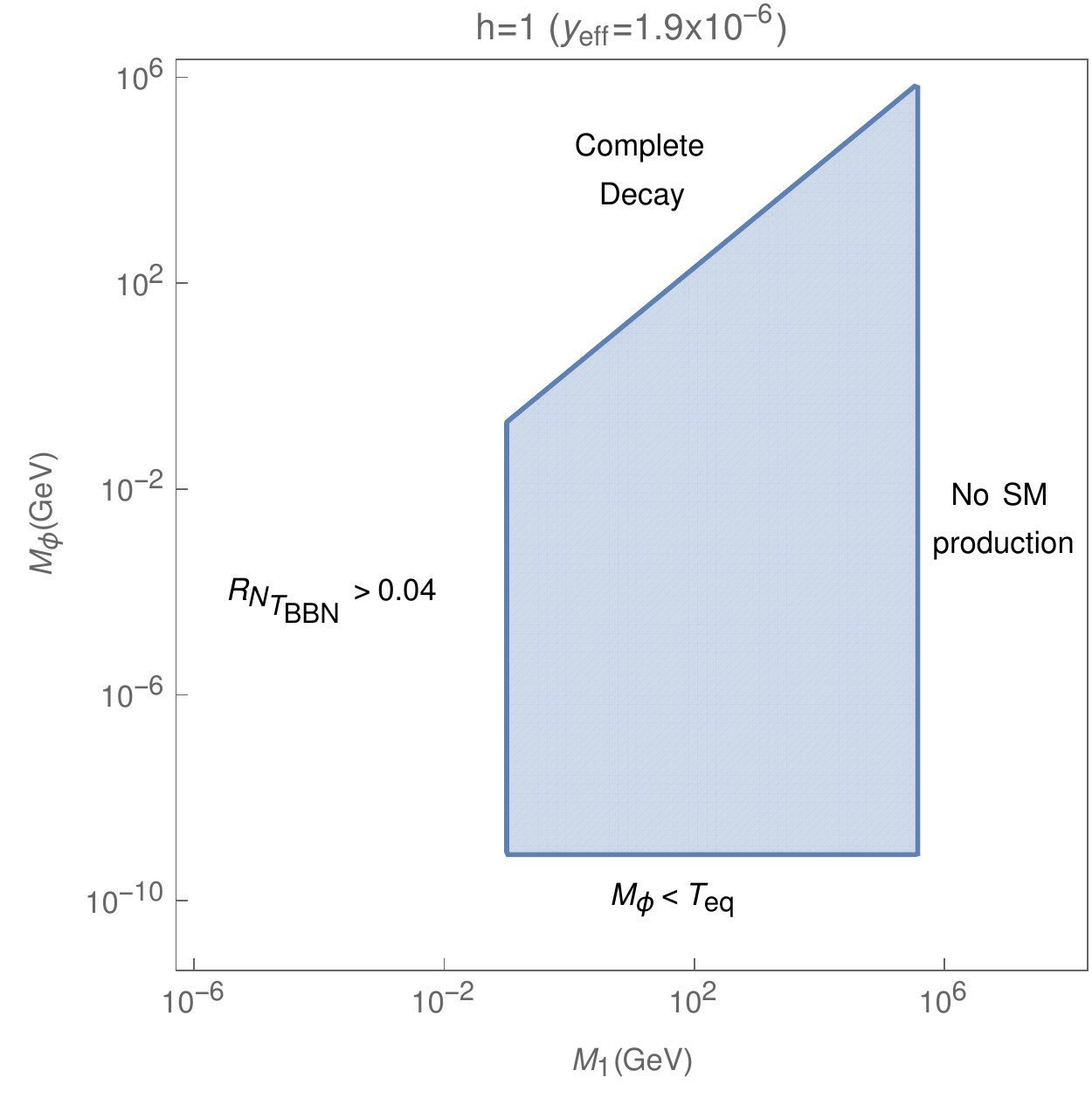}
	\includegraphics[totalheight=8.85cm]{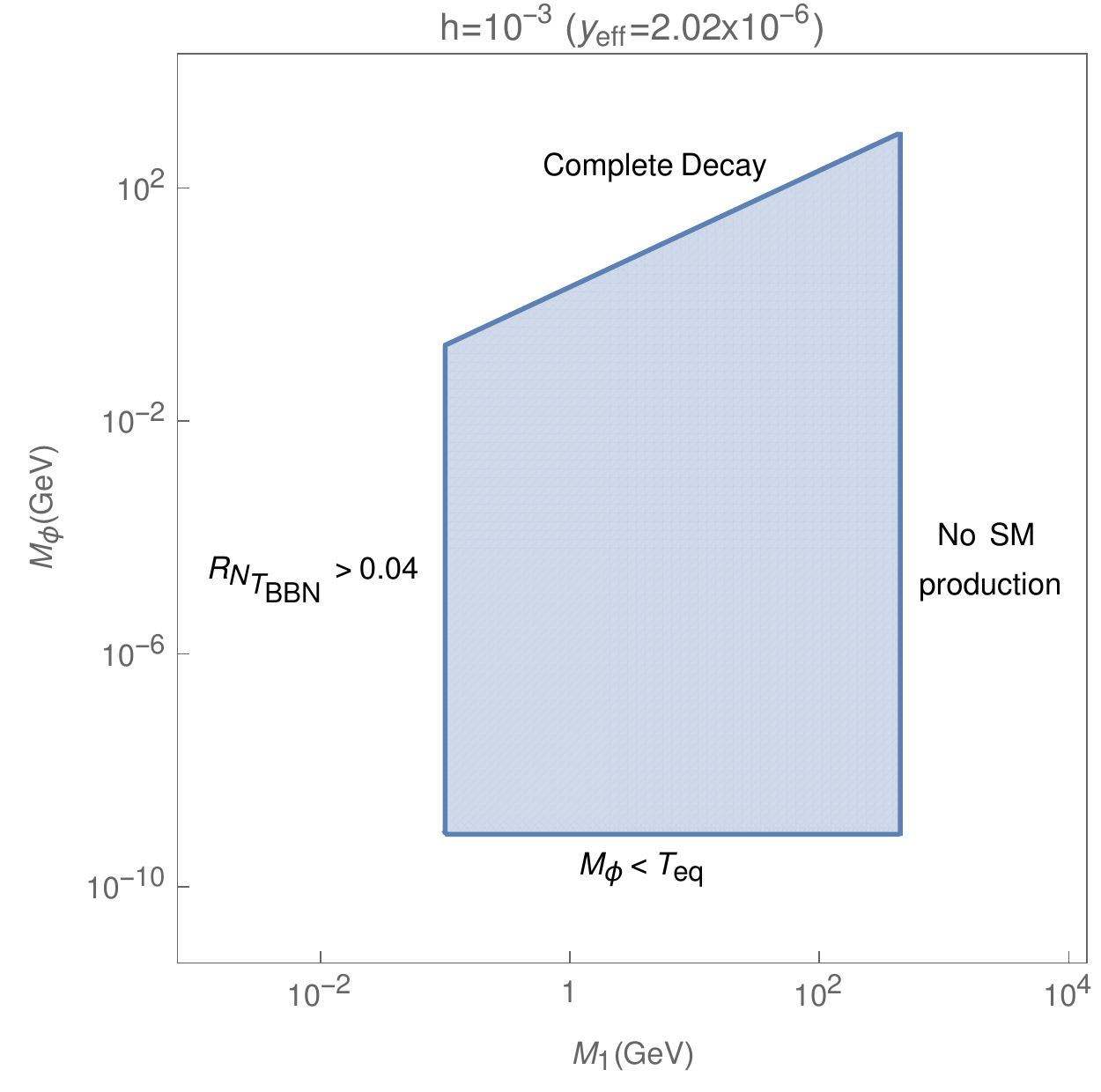}
	\caption{Resulting parameter space for FIMP DM scenario, taking $g_{\star R}=100$. Only with a direct SM reheating we find a fitting solution. The blue region corresponds to the compatible area. The two Yukawa parameters are related through Eq. \eqref{Eq:FIMPrelation}. Conditions $M_\phi > 2 M_1$ and \eqref{M1mpmin} define the ``complete decay'' and ``no SM production'' regions respectively. Constrains on $N_{eff}$ at BBN impose the limit on ratio $R_N$.} 
	\label{fig:parameters FIMP(D_SM)}
	
\end{figure}

Similarly to the previous section, we can study the parameter space by collecting all the bounds. However, in this scenario the relic density constrain relates our couplings $h$ and $y_{eff}$, reducing our free parameters to the masses $M_1$, $M_\phi$ and $h$. The parameter space is then constrained by the relations,
$\rho_{\delta\phi}>\rho_\phi$, $M_\phi/2 < M_1$ and Eqs. \eqref{M1mpmin}, \eqref{TRM1min}. 
%
The resulting parameter space is shown in Fig. \ref{fig:parameters FIMP(D_SM)}. 

As one can see in the figure we may find a suitable parameter space for $h\gtrsim10^{-3}$, compatible with both large $M_1$ and $M_\phi$ and sub-GeV masses. Due to the relic abundance condition we always have a small Yukawa $y_{eff}$ value, typical in non-thermal dark matter candidates. 

As we obtain small $M_1$ values, below $T_{BBN}$, we must ensure that the number of relativistic degrees of freedom respects the observational constrains at BBN. With such masses (s)neutrinos are relativistic at BBN and we find that their contribution to the radiation energy density  yields
\begin{equation}
	R_N\big|_{T_{BBN}}=\frac{\rho_N}{\rho_R}\bigg|_{T_{BBN}}\simeq 4.3\times10^{-5}\, y_{eff}^4\frac{m_P T_R^2}{E_N}\left(1-\left(\frac{T_{BBN}}{T_R}\right)^2\right),
\end{equation}
where $T_R$ is given by Eq. \eqref{TRM1min}. To avoid interference with $N_{eff}$ we must impose $R_N<0.04$, \cite{Aghanim:2018eyx}.

\subsection*{Oscillating Scalar field Dark Matter}
We have seen that the inflaton is allowed to decay until this is kinematically blocked at $\phi_{DR}=M_1/h$. It becomes stable with a dominant quartic term driving its potential, leading to a dark radiation behavior. Scatterings with the other particles may lead to excitations of the condensate states, as we have discussed. If neither sneutrino decay or evaporation are fast enough, i.e. the Universe is expanding even faster, the condensate will keep as a non-interacting radiation-like fluid. This lasts until the quadratic term becomes dominant, at
\begin{equation}
	\phi_{CDM}=\frac{\sqrt{2}}{\kappa}M_\phi \,. 
\end{equation}
From this moment on, the oscillating inflaton field will behave as cold dark matter (CDM), mimicking a non-relativistic fluid  and may dominate over the dark sector. Then, the number density to entropy ratio becomes constant until the present day, and can be related to the present and measured dark matter density. As in \cite{Manso:2018cba}, we have
\begin{equation}
\Omega_{\phi0}=\frac{\rho_{\phi0}}{\rho_{c0}}=\frac{M_{\phi}n_{\delta\phi_0}}{3H_{0}^{2}m_P^{2}}=\frac{M_{\phi}s_{0}}{3H_{0}^{2}m_P^{2}}\left(\frac{n_{\delta\phi}}{s}\right)_{CDM}.\label{Eq:energy ratio}
\end{equation}
Now, since between $\phi_{DR}$ and $\phi_{CDM}$ the inflaton redshifts as a radiation fluid
\begin{equation}
	\frac{\phi_{DR}}{\phi_{CDM}}=\frac{T_{R}}{T_{CDM}},\label{Eq:phi_DR/phi_CDM}
\end{equation}
where we used that $T_{DR}\simeq T_{R}$. The inflaton number density-to-entropy ratio is given by:
\begin{equation}
\left(\frac{n_{\delta\phi}}{s}\right)_{CDM}=\left(\frac{\rho_{\delta\phi}}{M_{\phi}s}\right)_{CDM}=\frac{\frac{1}{2}M_{\phi}\phi^2_{CDM}}{\frac{2\pi^{2}}{45}g_{\star CDM}T_{CDM}^{3}}~.\label{Eq:n e s ratio}
\end{equation}
Using Eqs. \eqref{Eq:phi_DR/phi_CDM}, \eqref{Eq:n e s ratio},  and $s_{0}=\frac{2\pi^{2}}{45}g_{\star0}T_{0}^{3}$ in Eq. \eqref{Eq:energy ratio}, we then obtain
\begin{equation}
\Omega_{\phi0}=\frac{M_{\phi}^2}{6H_{0}^{2}m_P^{2}}\frac{s_0}{s_{CDM}}\phi_{CDM}^2=\frac{\kappa\ M_{\phi}\ M_1^3}{\sqrt{2}\ 6\ h^3 H_{0}^{2}m_P^{2}}\frac{g_{\star\,0}}{g_{\star\,CDM}}\left(\frac{T_{0}}{T_R}\right)^3~.
\end{equation}
For $H_{0}=10^{-42}h_{0}\ \mathrm{GeV}$, $g_{\star 0}=3.91$ and $T_{0}=2.4\times10^{-13}\ \mathrm{GeV}$ it yields
\begin{equation}
\Omega_{\phi0}h_0^2=1.07\times 10^9 \frac{\kappa}{g_{\star CDM}h^3} \frac{M_1^3\ M_\phi}{T_R^3\ \mathrm{GeV}} ~.
\end{equation}
The inflaton mass in these never coupled dark matter scenarios is related to the other parameters through
\begin{equation}
M_\phi=1.32\times10^{-5}\, g_{\star CDM}\,h^3\,\frac{\Omega_{\phi 0}h_0^2}{0.1} \left(\frac{T_R}{M_1}\right)^3 \mathrm{GeV}\,,\label{Eq:MassrelationOSF}
\end{equation}
where the reheating temperature is given by \eqref{Eq:ReheatingT_early_matter} or \eqref{Eq:ReheatingT_Directradiation}, in the two compatible cases.
Moreover, since these particles are not coupled to the thermal bath they escape the stringent bounds on the effective number of degrees of freedom at the light nuclei production. In these last two scenarios the transition into the Dark Matter final behavior only has to be fulfilled before radiation matter-equality $T_{eq}=0,79$ eV. Using Eq. \eqref{Eq:phi_DR/phi_CDM} it yields
\begin{flalign}
&T_{R}>\left(\frac{\kappa^2}{2\,h^2}\right)^{1/2}\frac{M_1}{M_\phi}T_{eq}\,.\label{Eq:OSF_transition_Teq}
\end{flalign}

As inflaton particles, produced from the (s)neutrino decays, are also present in the particle content, one must require that the condensate dominates the dark matter component, $\rho_\phi>\rho_{{\delta\phi}}$. In a direct transition from inflation into a SM Universe we can use the expressions in Eqs. \eqref{Eq:FIMPrhophi} and \eqref{Eq:FIMPrhodeltaphi}. 
For the early-matter Universe hypothesis, the inflaton decay and the sneutrino decays happen at different times. Again, one may calculate the remnant condensate energy density after the decay. However, it then has to be redshifted, through an entire matter dominated epoch, until the reheating transition. At reheating it yields
\begin{flalign}\rho_\phi&\simeq\frac{\kappa^2}{4}\left(\frac{M_1}{h}\right)^4e^{\frac{4}{9}\frac{\bar{\Gamma}^N}{\kappa^2}\frac{m_P^2}{H_e\Phi_e}\left(1-\left(h\frac{\Phi_e}{M_1}\right)^3\right)}\left(\frac{a_R}{a_D}\right)^{-4}\nonumber\\&=\frac{\kappa^2}{4}\left(\frac{M_1}{h}\right)^4e^{\frac{4}{9}\frac{\bar{\Gamma}^N}{\kappa^2}\frac{m_P^2}{H_e\Phi_e}\left(1-\left(h\frac{\Phi_e}{M_1}\right)^3\right)}\left(\frac{ y_{eff}^4 \kappa^2}{48\pi^2 }\left(\frac{\sqrt{3}}{4 \bar{\Gamma}_N}\right)^{4/3}\left(\frac{M_1}{m_P}\right)^2\right)^{4/3} 	\\	\rho_{\delta\phi}&=\frac{\Gamma_N^{\delta\phi}}{\Gamma_N^{SM}}\rho_R\simeq\frac{\pi^2}{60}\frac{\kappa^2}{ y_{eff}^2 }g_{\star R} T_R^4\,,
\end{flalign}
where $T_R$ refers to \eqref{Eq:ReheatingT_early_matter}, $a_D$ and $a_R$ are defined at moment of the inflaton decay and reheating, respectively.

As before, we ought to combine all the constraints for each compatible scenario with the OSF hypothesis.
Once again, we have the tree independent parameters, $M_1$, $M_\phi$ and $y_{eff}$. However, no longer a simple relation between the DM inflaton mass and $h$ exists, $M_\phi$ is dependent on all the other parameters of the model, see Eq. \eqref{Eq:MassrelationOSF}. The relevant constraints are given by $M_\phi > 2 M_1$, $\rho_\phi > \rho_\dphi$,  Eqs. \eqref{M1EarlyMatt}-\eqref{TRIneffEarlyMatt}, and the common conditions 
\begin{flalign}
&T_{\delta\phi} \simeq 3.6 \times 10^4 y_{eff}^2 \left( 1 + \left(\frac{h}{0.004}\right)^4 y_{eff}^2 \right)\left(\frac{100}{g_{\star R}}\right)^{1/2} \, \gev<T_{EW}\,,\label{Eq:OSFnothermalization}\\
  &T_{R}>\left(\frac{\kappa^2}{2\,h^2}\right)^{1/2}\frac{M_1}{M_\phi}T_{eq}\,.
\end{flalign}

\begin{figure}[th]
	\centering
	\includegraphics[totalheight=8.85cm]{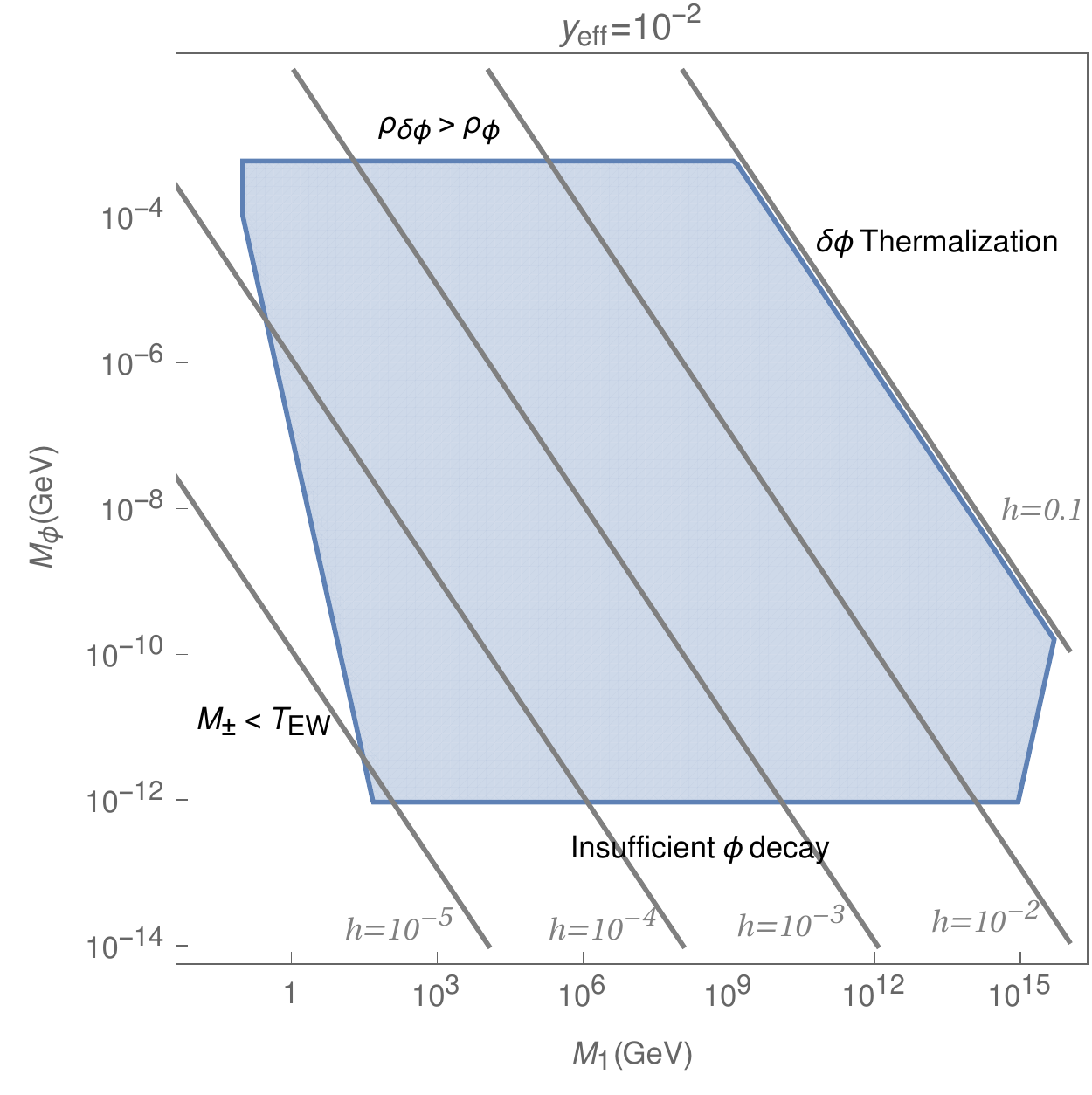}
	\includegraphics[totalheight=8.85cm]{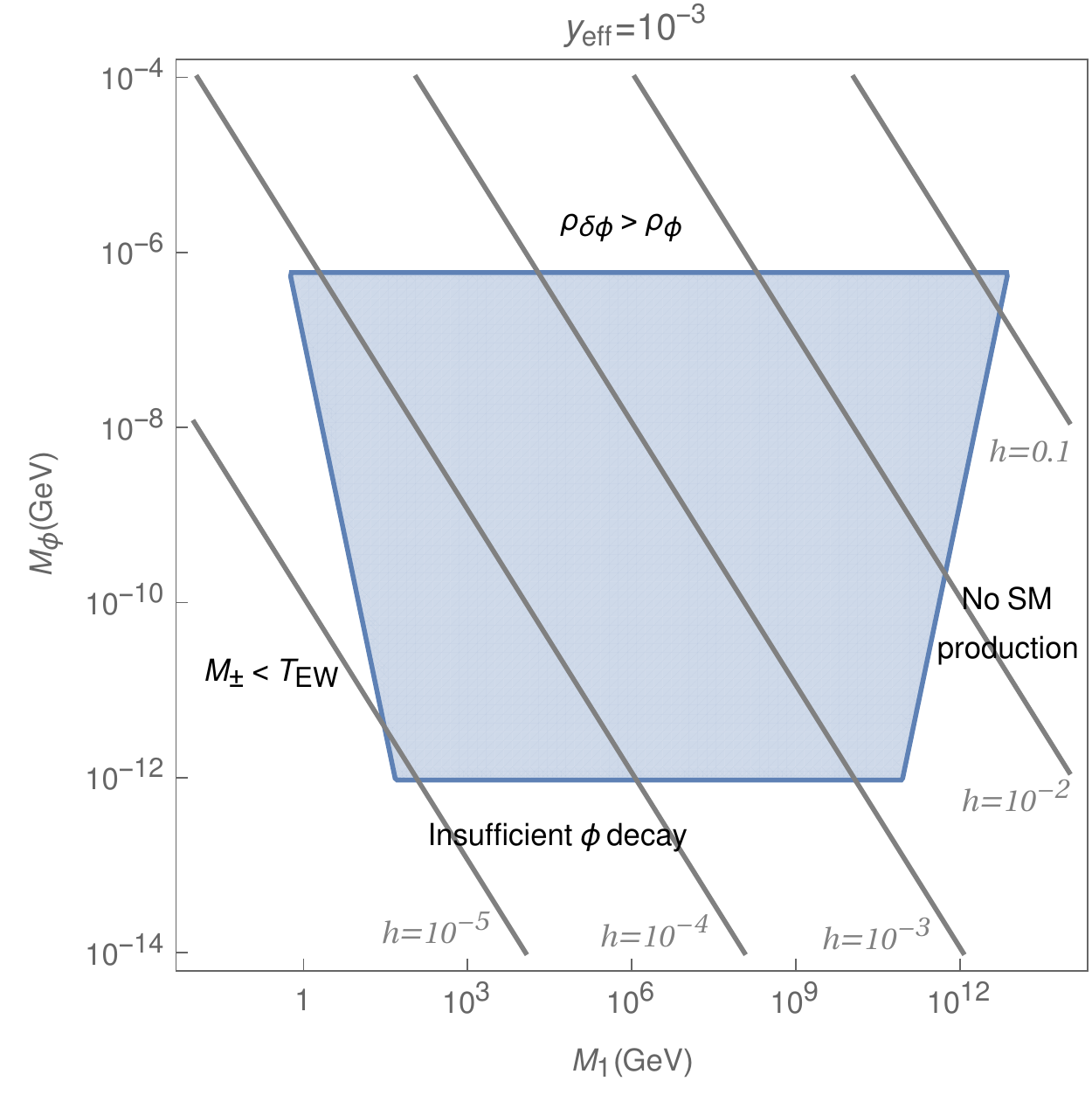}
	\caption{Resulting parameter space for OSF scenarios, taking $g_{\star R}=100$. Only with a direct SM reheating we find a fitting solution. The blue region corresponds to the compatible area. Gray lines represent the different $h$ values, obtained through equation \eqref{Eq:MassrelationOSF}. Conditions in Eqs. \eqref{Eq:OSF_transition_Teq} , \eqref{M1mpmin} define the "insufficient decay" and "no SM production", whereas Eq. \eqref{Eq:OSFnothermalization} describes the "$\delta\phi\ thermalization$" region. The (s)neutrino decay into SM particles requires that $M_\pm>T_{EW}$. For OSF dominant scenario we further impose $\rho_\phi>\rho_{\delta\phi}$.}
	\label{fig:parameters OSF(D_SM)}
	
\end{figure}

When combining all the constrains we only observe a compatibility for a direct transition into a SM Universe after inflation. By requiring that the condensate contribution dominates over the dark sector we exclude an early matter period in the early Universe.
In Fig. \ref{fig:parameters OSF(D_SM)} we can observe a region with  a broad range of masses, again lower than in the WIMP scenario. Although, with a larger freedom in the Yukawa couplings than for a FIMP candidate, we find these to be rather small. For lower masses we end up hitting the lower bound on the reheating temperature, imposed from a condition on the (s)neutrino masses, $M_\pm$.

\subsection*{Consequences of a seesaw mechanism to generate light neutrino masses}

Following what was done in \cite{Manso:2018cba} we further extend the applications of the model to generate the observed light-neutrino masses, through the seesaw mechanism. This provides a relation between the right-handed neutrino masses and the Yukawa coupling with the Higgs, 
\begin{equation}
M_{1}\simeq1.21\times10^{15}\,y_{eff}^{2}\mathrm{\ GeV},
\label{Eq:M1}
\end{equation}
which reduces the no. of free parameters. 
For both WIMP and OSF DM candidates we have the masses $M_1$ and $M_\phi$ as the free parameters, whereas for a FIMP candidate we have a coupling and a mass, $h$ and $M_\phi$.

\begin{figure}[H]
	\centering
	\includegraphics[totalheight=8.65cm]{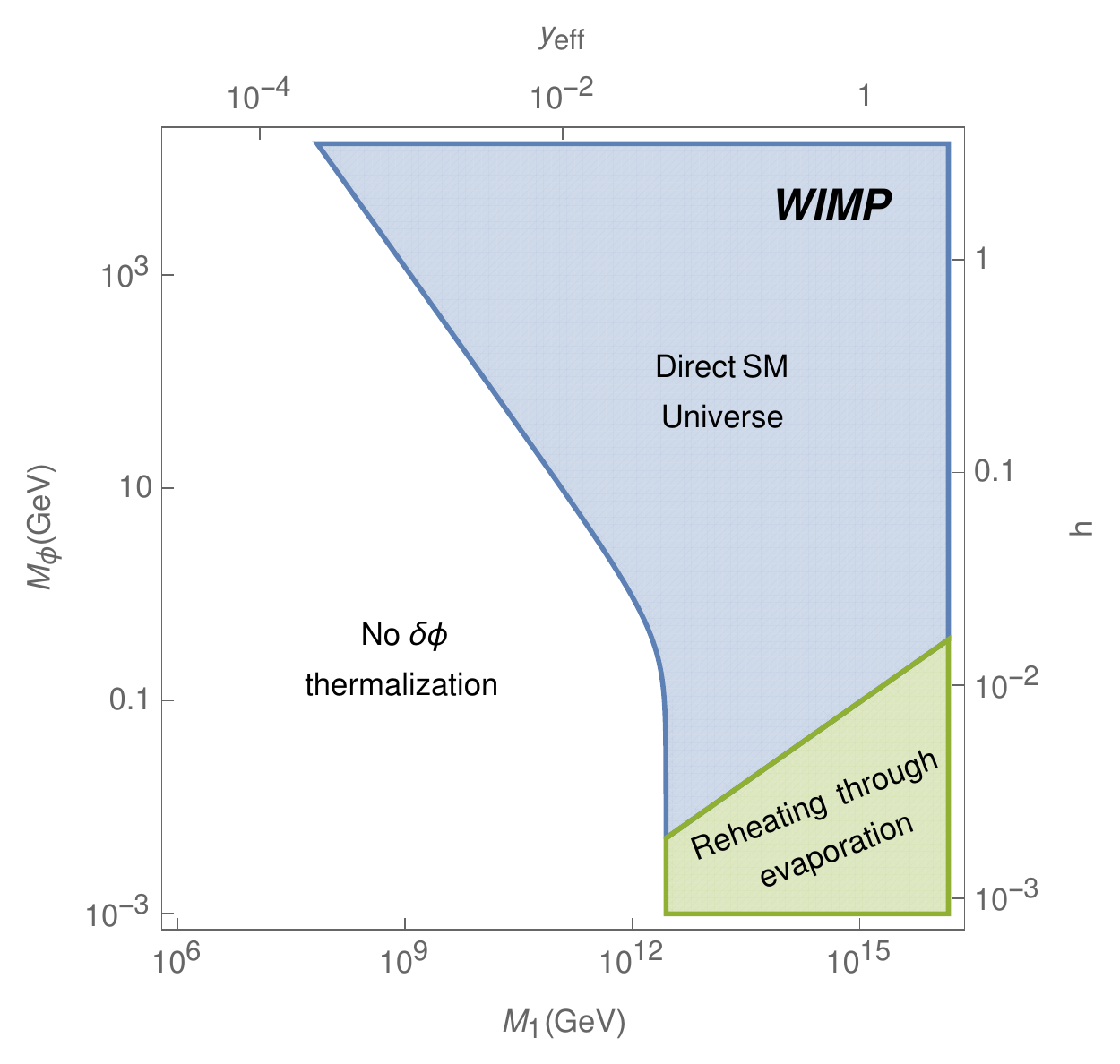}
	\includegraphics[totalheight=8.55cm]{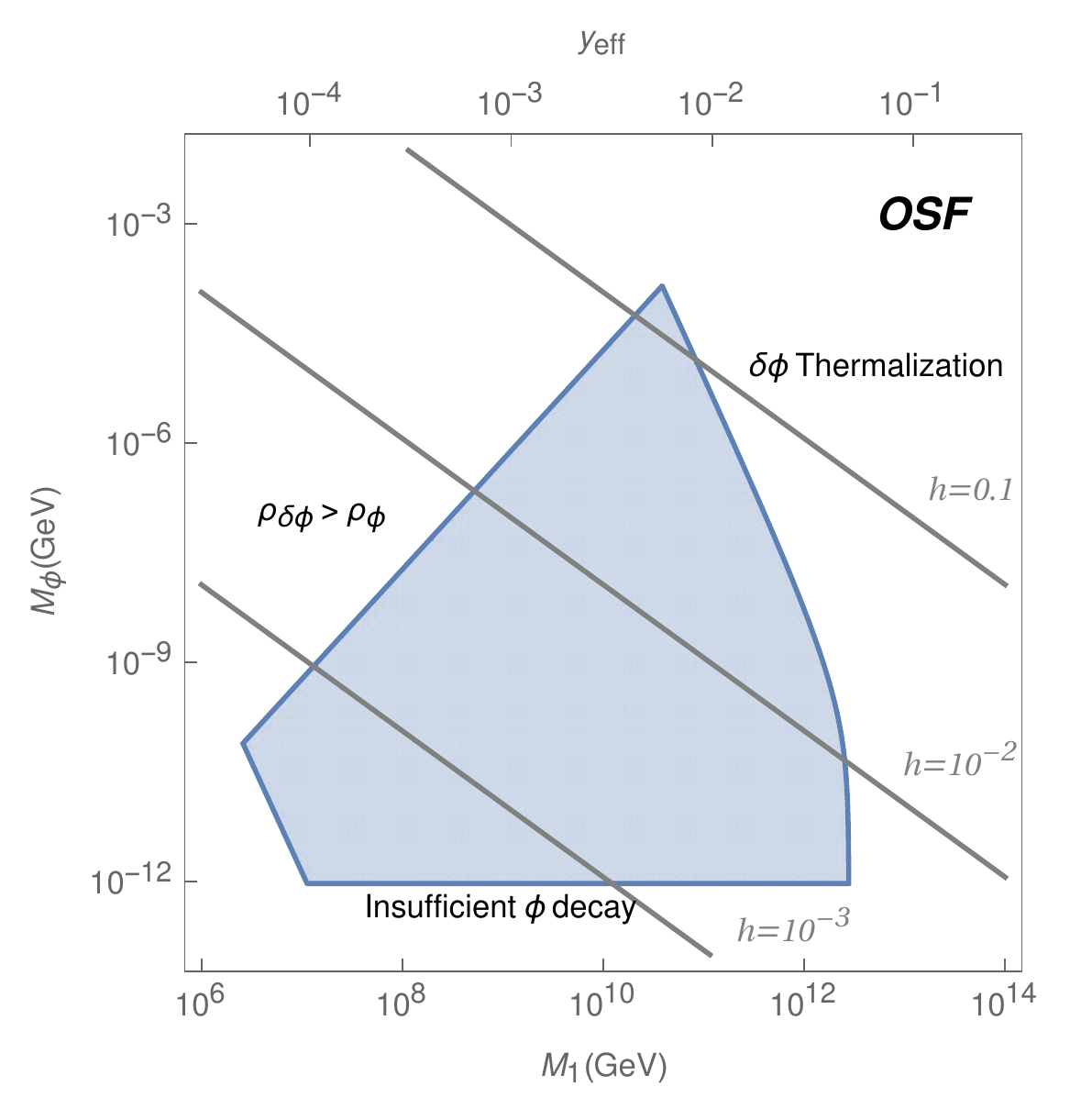}
	\includegraphics[totalheight=7.5cm]{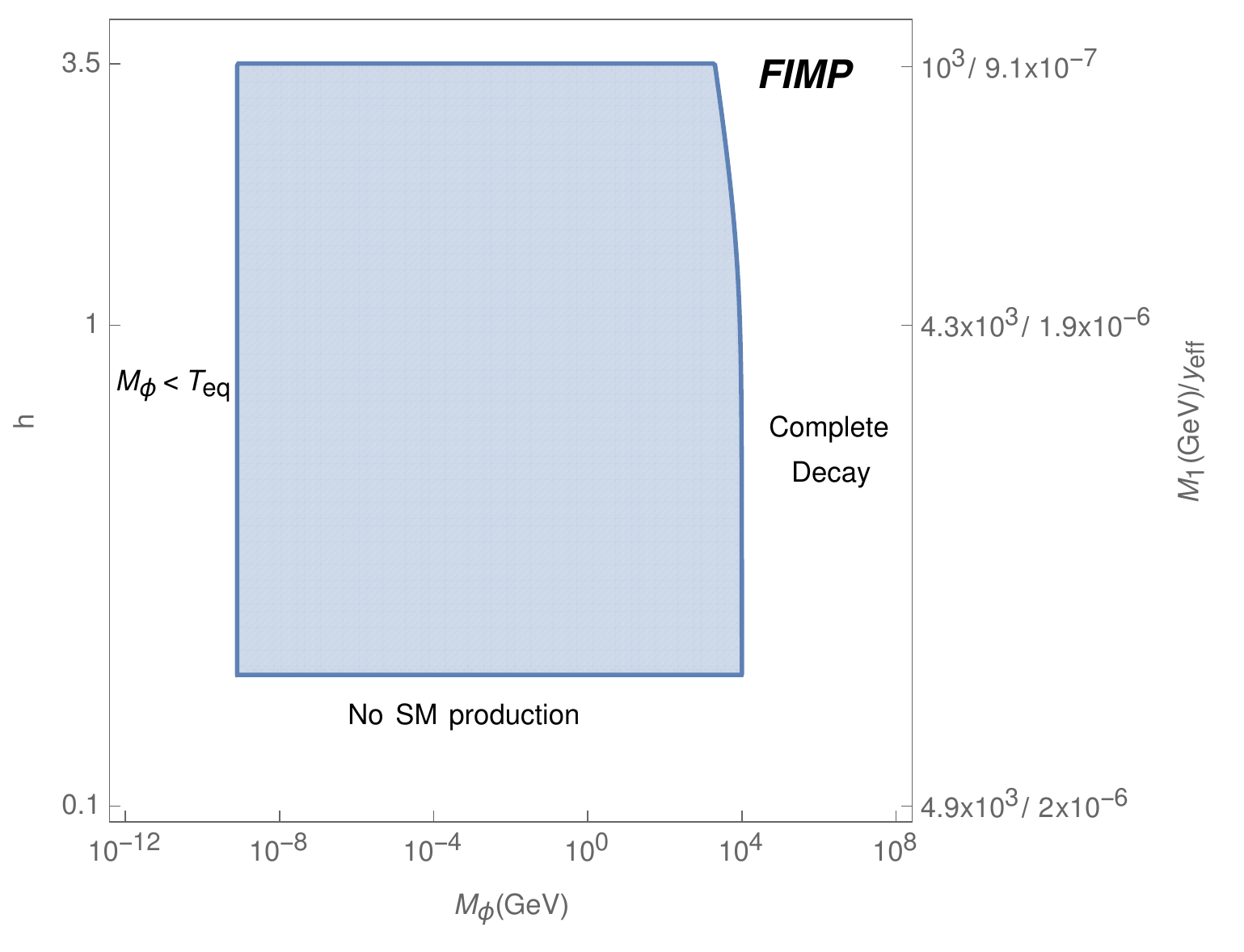}
	
	\caption{Collection of the allowed parameters for all DM scenarios when taking the seesaw formalism for the generation of light neutrino masses, $g_{\star R}=100$. The blue region corresponds to a direct transition from inflation into a SM Universe. In green we have a reheating through evaporation scenario only compatible with a WIMP DM candidate. Gray lines represent the different $h$ values for an OSF DM scenario.} \label{fig:parameters seesaw}
	
\end{figure}

Collecting again all the relevant bounds for each DM hypothesis we obtain the parameter space shown in Fig.  \ref{fig:parameters seesaw}. For a WIMP candidate one observes the direct transition into a SM Universe if the inflaton decay is efficient, and a possibility of reheating through evaporation, visible when lowering $h$. For such a DM candidate, with this seesaw relation, early-matter scenarios are ruled out. Large $M_1$ correspond to high $y_{eff}$ which will impose SM particle production and fast thermalization with the thermal bath. A direct transition is held at large reheating temperatures $T_R\gtrsim10^8-10^{10}$ GeV whereas if we reheat with evaporation, lower h, we find $T_R \gtrsim T_{EW}$. As we lower the parameters we have an appearance of a no-thermalization region that we shall explore for different DM proposals. 
At rather large (s)neutrino masses but lowering the inflaton mass we find that the model can hold an OSF DM scenario with a direct transition from inflation into a SM Universe, yielding $T_R\simeq10^9-10^{11}$ GeV. 
Finally, for a FIMP candidate we find that only very small Yukawas $y_{eff} \sim 10^{-6}$ are compatible with light neutrino masses, leading to RH neutrino masses of the order of $1-5$ TeV. On the other hand, it requires a  large $h$ coupling and we may observe a broad range of inflaton masses, from eV to TeV scales, resulting in $T_R \simeq10^6-10^7$ GeV.

\section{Conclusion}
\label{sec7}

We have extended the original $\nu$-IDM model, \cite{Manso:2018cba} to its Supersymmetric completion.
Using the inflaton incomplete decay into right-handed neutrinos and sneutrinos we have built an unified model for inflation and dark matter, where the same field describes both phases while avoiding the troublesome radiative corrections to the inflaton self coupling, strongly constrained by observations. With the introduction of (right-handed) sneutrinos we found a plethora of possible cosmological scenarios that may lead into a successful reheating. Moreover, we found that the inflaton can account for the DM content of the Universe as a common WIMP, a FIMP or as an oscillating scalar field. We focus here on the cosmology of the reheating period after inflation, taking place at $T \gtrsim T_{EW}$, without detailing any particular low scale SUSY model. We implicitly assume that the LSP, the natural DM candidate in SUSY models, is a negligible component of the total DM relic abundance. However other scenarios of multi-component DM, with a mixture of inflaton DM and LSP, could be possible. Nevertheless, the parameter space obtained in this work will not change substantially if we reduce the inflaton relic abundance by a factor $O(1)$.  

Recovering the idea of an inflaton incomplete decay, proposed in \cite{Bastero-Gil:2015lga}, the inflaton is allowed to decay into two of the three right-handed (s)neutrinos while keeping a discrete symmetry that forbids any additional decays. If we impose that the inflaton decay is kinetically blocked at the minimum of its potential, while still allowing a partial decay, we may realize $\phi$ as a stable DM candidate. Of crucial importance, distancing the dynamics from the non-SUSY version, are the sneutrino decays into the inflaton. These will allow a production of the latter after its decay becomes blocked. 

We have embedded the inflationary evolution within the superconformal $\alpha$-Attractors models \cite{Kallosh:2013yoa}, encompassed in a SUGRA framework. By considering a non-minimal K\"ahler potential and a superpotential compatible with the symmetries of the model, the inflaton acquires a non-canonical kinetic term, which leads to a flattening and stabilization of the potential at large field values. The inflaton follows the common slow-roll description  and we find a compatibility with the measurements in the CMB temperature and polarization spectra by fixing the quartic self coupling at $\kappa\simeq3.5\times10^{-6}$. At the end of inflation the scalar field will start oscillating under a quartic potential leading to the onset of reheating interactions.

After detailing all the possible interactions and developing the Boltzmann equations we were able to describe the reheating evolution for different parameter regimes. With a semi-analytical study, sustained by the numerical solutions, we found four different reheating scenarios compatible with BBN constrains. These were then matched with the three possible inflaton DM realizations.

For a direct transition from inflation into a standard model Universe, when the  (s)neutrinos produced immediately decay into SM particles, the remnant of inflaton  field can account for DM as a WIMP, a FIMP or as an OSF.  In the first two the DM relic density is accounted by inflaton particles $\delta\phi$, produced through evaporation or in the novel sneutrino decay. While as usual in the WIMP solution the scalar condensate evaporates and thermalizes with the cosmic plasma, leading to a freeze-out mechanism, in the FIMP scenario inflaton particles are mostly produced from sneutrino decays and never couple with the plasma. Thus, these may reproduce DM behavior after a freeze-in mechanism. Here, in the absence of evaporation, the produced inflaton particles must dominate over the remaining stable condensate, otherwise we have the OSF DM scenario.

An evolution with an early matter epoch, were we have a Universe dominated by heavy (s)neutrinos, can be achieved with or without an efficient inflaton decay. However, only in the former we find compatibility  with DM constrains, namely in the WIMPlaton scenario. Nonetheless, an inefficient inflaton decay can still lead into a successful transition into a radiation era and to connect with standard cosmology. In this peculiar scenario, after failing to reheat through the inflaton decay, the stable inflaton moves into a dark radiation phase. Even so, particles produced from the decay may be sufficient to sustain efficient evaporation and excitation of the inflationary homogeneous fluid. These could then thermalize with the subdominant radiation plasma, leveling the energy densities, to later freeze-out, leading again to a WIMP like candidate.

When combining all constrains we find a very broad range for (s)neutrino masses, from GUT to MeV scales. Typically, the WIMPlaton scenario is reproduced by MeV-GeV inflaton masses and large Yukawa couplings. Slightly lowering the coupling of the (s)neutrinos to Standard Model dof, $y_{eff}$, while keeping a high value of the inflaton coupling to the (s)neutrinos, $h$, we find the early-matter Universe regime. In the opposite case we find that reheating proceeds through evaporation after an inefficient inflaton decay. In a FIMPlaton DM scenario, $M_\phi$  may cover a broader range, from $O(10 \,\,{\rm eV}) $  to TeV scales, and small values $y_{eff}\sim10^{-6}$, characteristic of a freeze-in evolution. A case for very low inflaton masses comes with the oscillating scalar field. When both couplings are small, albeit larger than $10^{-5}$, we find this regime where $M_\phi$ ranges from sub-eV to keV masses.

We have developed a concrete model with a distinct prediction that dark matter is made of scalar particles only directly coupled with the right-handed neutrinos superfields. The early-matter and the intermediate phase where inflaton exited particles dominate the energy content provide distinct and interesting scenarios that may change how we see the evolution before BBN.
Finally, (s)neutrino interactions with the standard model particles may induce concise assertions on light neutrino masses and also on the generation of the cosmological baryon asymmetry, still left to explore in this supersymmetric extension. Nevertheless, a thermal lepton asymmetry production can be conceived within high reheating temperatures, as in WIMP or OSF DM scenarios. This would require a hierarchy on the right-handed (s)neutrino masses, $M_1=M_2\neq M_3$, with the lightest of these being larger than $O(10^{8})$ GeV, which implies $T_R \gtrsim 10^9\,\mathrm{GeV}$ \cite{leptoreheating}. On the other hand, since in the model right-handed (s)neutrinos are produced through inflaton decay, a non-thermal leptogenesis scenario may be available.  Thus, avoiding the need for such large temperatures at reheating, and possibly recovering the parameter space of a FIMP DM scenario. We thus hope that our work motivates further exploration of related effects to understand the unknown history of the early Universe. 

\appendix
\section{Decay and scattering rates}\label{appendixA}

We collect here all the decay rates and cross-sections used in the Boltzmann equations.
Starting with the inflaton decay
\begin{figure}[H]
	\centering
	\begin{tikzpicture}[line width=1 pt, scale=1.3]
	\draw[fermion] (0:0)--(-40:1);
	\draw[fermionbar] (0:0)--(40:1);
	\draw[scalarnoarrow] (180:1)--(0:0);
	\node at (-40:1.2) {$N_i$};
	\node at (40:1.2) {$N_i^c$};
	\node at (180:1.2) {$\phi$};
	\node at (150:0.3) {$\pm h$};
	\end{tikzpicture}	
	\begin{tikzpicture}[line width=1 pt, scale=1.3]
	\draw[fermion] (0:0)--(-40:1);
	\draw[fermionbar] (0:0)--(40:1);
	\draw[scalarnoarrow] (180:1)--(0:0);
	\node at (-40:1.2) {$N_i$};
	\node at (40:1.2) {$\psi_{\phi}^c$};
	\node at (180:1.2) {$\phi$};
	\node at (130:0.25) {$\kappa$};
	\end{tikzpicture}
	\begin{tikzpicture}[line width=1 pt, scale=1.3]
	\draw[scalarnoarrow] (0:0)--(-40:1);
	\draw[scalarnoarrow] (0:0)--(40:1);
	\draw[scalarnoarrow] (180:1)--(0:0);
	\node at (-40:1.2) {$\tilde{N}_i$};
	\node at (40:1.2) {$\tilde{N}_i$};
	\node at (180:1.2) {$\phi$};
	\node at (155:0.4) {$\pm hM_1$};
	\end{tikzpicture}
	\begin{tikzpicture}[line width=1 pt, scale=1.3]
	\draw[scalarnoarrow] (0:0)--(-40:1);
	\draw[scalarnoarrow] (0:0)--(40:1);
	\draw[scalarnoarrow] (180:1)--(0:0);
	\node at (-40:1.2) {$\tilde{N}_i$};
	\node at (40:1.2) {$\tilde{N}_j$};
	\node at (180:1.2) {$\phi$};
	\node at (145:0.4) {$\left\langle \phi \right\rangle h^2$};
	\node at (215:0.4) {$\left\langle \phi \right\rangle \kappa^2$};
	\end{tikzpicture}
	\begin{tikzpicture}[line width=1 pt, scale=1.3]
	\draw[scalarnoarrow] (0:0)--(-45:1);
	\draw[scalarnoarrow] (0:0)--(0:1);
	\draw[scalarnoarrow] (0:0)--(45:1);
	\draw[scalarnoarrow] (180:1)--(0:0);
	\node at (-45:1.2) {$\tilde{N}_i$};
	\node at (0:1.2) {$\tilde{N}_i$};
	\node at (45:1.2) {$\tilde{N}_i$};
	\node at (180:1.2) {$\phi$};
	\node at (155:0.40) {$\pm h\kappa$};
	\end{tikzpicture}
\end{figure}	 
\begin{flalign}
\Gamma^{N_i}_\phi=&\frac{h^2}{16\pi}m_\phi\left(1- \frac{4M_\pm^2 }{m_\phi^2} \right)^{3/2},\\
\Gamma^{\tilde{N}_i}_\phi=&	\left( \frac{h^2M_1^2}{2\pi m_\phi}+\frac{h^4\phi^2}{8\pi m_\phi}+\frac{\kappa^4\phi^2}{32\pi m_\phi} \right )
\left(1- \frac{4M_\pm ^2 }{m_\phi^2} \right)^{1/2}+
\frac{\kappa^4\phi^2}{16\pi m_\phi}  \left(1+ \frac{\left( M_+ ^2-M_-^2\right)^2 }{m_\phi^4} - 2\frac{M_+ ^2+M_-^2 }{m_\phi^2}\right)^{1/2}.
\end{flalign}
SM particle production comes through the right-handed (s)neutrinos decay into the Higg(ino)s and left-handed (s)neutrinos,
\begin{equation}
\Gamma^{SM}_{N,\tilde{N}}=\frac{y_{eff}^2}{8\pi}M_\pm\,. 
\end{equation}
The right-handed (s)neutrinos decays into the inflaton particles ($\delta\phi$) are parameterized with
\begin{figure}[H]
	\centering
	\begin{tikzpicture}[line width=1 pt, scale=1.5]
\draw[fermion] (0:0)--(-40:1);
\draw[scalarnoarrow] (0:0)--(40:1);
\draw[fermion] (180:1)--(0:0);
\node at (-40:1.2) {$\psi_{\phi}$};
\node at (40:1.2) {$\phi$};
\node at (180:1.2) {$N_i$};
\node at (145:0.25) {$\kappa$};
\end{tikzpicture}
\begin{tikzpicture}[line width=1 pt, scale=1.5]
\draw[scalarnoarrow] (0:0)--(-40:1);
\draw[scalarnoarrow] (0:0)--(40:1);
\draw[scalarnoarrow] (180:1)--(0:0);
\node at (-40:1.2) {$\tilde{N}_j$};
\node at (40:1.2) {$\phi$};
\node at (180:1.2) {$\tilde{N}_i$};
\node at (145:0.3) {$\kappa^2\left\langle \phi \right\rangle $};
\end{tikzpicture}
\begin{tikzpicture}[line width=1 pt, scale=1.5]
\draw[scalarnoarrow] (0:0)--(-40:1);
\draw[scalarnoarrow] (0:0)--(40:1);
\draw[scalarnoarrow] (180:1)--(0:0);
\node at (-40:1.2) {$\phi$};
\node at (40:1.2) {$\phi$};
\node at (180:1.2) {$\tilde{N}_i$};
\node at (145:0.35) {$M_1\kappa$};
\end{tikzpicture}
	\label{fig:FeynA2}
\end{figure}	 
\begin{equation}
\Gamma^{\delta\phi}_{\tilde{N}_i}=\frac{M_1^2\kappa^2}{16\pi} \frac{1}{M_\pm}\left(1-4\frac{m_\phi^2}{M_\pm^2}\right) ^\frac{1}{2}
+\frac{\kappa^4\left<\phi\right>^2  }{16\pi}\frac{1}{M_\pm}\left( 1+\frac{\left(m_\phi^2-M_\mp^2 \right)^2}{M_\pm^4}-2\frac{m_\phi^2+M_\mp^2}{M_\pm^2}\right)^\frac{1}{2},
\end{equation}
and $\Gamma^{N_i}_{\delta\phi}=0$ since we are neglecting the inflatino production. 

The thermal scatterings cross-sections $\left<\sigma v\right>_a^{eff}$ are given by: 
\begin{figure}[H]
	\begin{tikzpicture}[line width=1 pt, scale=1]
	\draw[fermion] (-2,2.0) -- (0,2);
	\draw[fermion] (-2,0.0) -- (0,0);
	\draw[scalarnoarrow] (0,2) -- (0,0);
	\draw[fermion] (0,2) -- (2,2);
	\draw[fermion] (0,0) -- (2,0);
	\node at (-1.8,2.2) {$N_i$};
	\node at (-1.8,0.2) {$N_i$};
	\node at (1.8,2.2) {$L$};
	\node at (1.8,0.2) {$L$};
	\node at (0.3,1) {$H_u$};
	\end{tikzpicture}
	\begin{tikzpicture}[line width=1 pt, scale=1]
	\draw[fermion] (-2,2.0) -- (0,2);
	\draw[fermionbar] (-2,0.0) -- (0,0);
	\draw[fermion] (0,2) -- (0,0);
	\draw[scalarnoarrow] (0,2) -- (2,2);
	\draw[scalarnoarrow] (0,0) -- (2,0);
	\node at (-1.8,2.2) {$N_i$};
	\node at (-1.8,0.2) {$N_i$};
	\node at (1.8,2.2) {$H_u$};
	\node at (1.8,0.2) {$H_u$};
	\node at (0.3,1) {$L$};
	\end{tikzpicture}
	\begin{tikzpicture}[line width=1 pt, scale=1]
	\draw[fermion] (-2,2.0) -- (0,2);
	\draw[fermion] (-2,0.0) -- (0,0);
	\draw[scalarnoarrow] (0,2) -- (0,0);
	\draw[fermion] (0,2) -- (2,2);
	\draw[fermion] (0,0) -- (2,0);
	\node at (-1.8,2.2) {$N_i$};
	\node at (-1.8,0.2) {$N_i$};
	\node at (1.8,2.2) {$\tilde{H}_u$};
	\node at (1.8,0.2) {$\tilde{H}_u$};
	\node at (0.3,1) {$\tilde{L}$};
	\end{tikzpicture}
	\begin{tikzpicture}[line width=1 pt, scale=1]
	\draw[fermion] (-2,2.0) -- (0,2);
	\draw[fermionbar] (-2,0.0) -- (0,0);
	\draw[fermion] (0,2) -- (0,0);
	\draw[scalarnoarrow] (0,2) -- (2,2);
	\draw[scalarnoarrow] (0,0) -- (2,0);
	\node at (-1.8,2.2) {$N_i$};
	\node at (-1.8,0.2) {$N_i$};
	\node at (1.8,2.2) {$\tilde{L}$};
	\node at (1.8,0.2) {$\tilde{L}$};
	\node at (0.3,1) {$\tilde{H}_u$};
	\end{tikzpicture}
\end{figure}

\begin{equation}
\left<\sigma v\right>^{N_i}_{eff}\simeq\frac{y_{eff}^4}{16\pi E_{N_i}^2};
\end{equation}

\begin{figure}[H]
	\centering
	\begin{tikzpicture}[line width=1 pt, scale=1]
	\draw[scalarnoarrow] (-2,2.0) -- (0,2);
	\draw[scalarnoarrow] (-2,0.0) -- (0,0);
	\draw[fermionbar] (0,2) -- (0,0);
	\draw[fermion] (0,2) -- (2,2);
	\draw[fermionbar] (0,0) -- (2,0);
	\node at (-1.8,2.2) {$\tilde{N}_i$};
	\node at (-1.8,0.2) {$\tilde{N}_i$};
	\node at (1.8,2.2) {$L$};
	\node at (1.8,0.2) {$L$};
	\node at (0.3,1) {$\tilde{H}_u$};
	\end{tikzpicture}
	\quad
	\begin{tikzpicture}[line width=1 pt, scale=1]
	\draw[scalarnoarrow] (-2,2.0) -- (0,2);
	\draw[scalarnoarrow] (-2,0.0) -- (0,0);
	\draw[fermionbar] (0,2) -- (0,0);
	\draw[fermion] (0,2) -- (2,2);
	\draw[fermionbar] (0,0) -- (2,0);
	\node at (-1.8,2.2) {$\tilde{N}_i$};
	\node at (-1.8,0.2) {$\tilde{N}_i$};
	\node at (1.8,2.2) {$\tilde{H}_u$};
	\node at (1.8,0.2) {$\tilde{H}_u$};
	\node at (0.3,1) {$L$};
	\end{tikzpicture}
	\quad\begin{tikzpicture} [line width=1 pt, scale=1.5]
	\draw[scalarnoarrow] (-140:1)--(0,0);
	\draw[scalarnoarrow] (140:1)--(0,0);
	\draw[scalarnoarrow] (-40:1)--(0,0);
	\node at (-125:0.7) {$\tilde{N}_i$};
	\node at (125:0.8) {$\tilde{N}_i$};
	\draw[scalarnoarrow] (0,0)--(40:1);
	\node at (-52:0.7) {$\tilde{L}$};
	\node at (52:0.8) {$\tilde{L}$};	
	\end{tikzpicture}
	\quad \begin{tikzpicture} [line width=1 pt, scale=1.5]
	\draw[scalarnoarrow] (-140:1)--(0,0);
	\draw[scalarnoarrow] (140:1)--(0,0);
	\draw[scalarnoarrow] (-40:1)--(0,0);
	\node at (-125:0.7) {$\tilde{N}_i$};
	\node at (125:0.8) {$\tilde{N}_i$};
	\draw[scalarnoarrow] (0,0)--(40:1);
	\node at (-52:0.7) {$\tilde{H}_u$};
	\node at (52:0.8) {$\tilde{H}_u$};	
	\end{tikzpicture}
\end{figure}

\begin{flalign}
\left<\sigma v\right>^{\tilde{N}_i}_{eff}\simeq\frac{y_{eff}^4}{16\pi E_{\tilde{N}_i}^2};
\end{flalign}

\begin{figure}[H]
	\centering
	\begin{tikzpicture}[line width=1 pt, scale=1]
	\draw[scalarnoarrow] (-1,2.3) -- (0,2);
	\draw[fermion] (0,2) -- (2,2);
	\draw[scalarnoarrow] (2,2) -- (3,2.3);
	\node at (-0.9,2.5) {$ \phi$};
	\node at (-0.9,0) {$\phi$};
	\node at (-0.5,1) {$N_i$};
	
	\draw[scalarnoarrow] (-1,-.3) -- (0,0);
	\draw[fermion] (2,0) -- (0,0);
	\draw[scalarnoarrow] (2,0) -- (3,-.3);
	\node at (2.9,2.5) {$H_u$};
	\node at (2.9,0) {$H_u$};
	\node at (1,2.4) {$N_i$};
	\draw[fermion] (0,0) -- (0,2);
	\draw[fermion] (2,2) -- (2,0);
	\node at (1,0.3) {$N_i$};
	\node at (2.3,1) {$L$};
	\end{tikzpicture}
	\begin{tikzpicture}[line width=1 pt, scale=1]
	\draw[scalarnoarrow] (-1,2.3) -- (0,2);
	\draw[fermion] (0,2) -- (2,2);
	\draw[scalarnoarrow] (2,2) -- (3,2.3);
	\node at (-0.9,2.5) {$ \phi$};
	\node at (-0.9,0) {$\phi$};
	\node at (-0.5,1) {$N_i$};
	
	\draw[fermion] (-1,-.3) -- (0,0);
	\draw[fermion] (2,0) -- (0,0);
	\draw[fermion] (2,0) -- (3,-.3);
	\node at (2.9,2.5) {$L$};
	\node at (2.9,0) {$L$};
	\node at (1,2.4) {$N_i$};
	\draw[fermion] (0,0) -- (0,2);
	\draw[scalarnoarrow] (2,2) -- (2,0);
	\node at (1,0.3) {$N_i$};
	\node at (2.3,1) {$H_u$};
	\end{tikzpicture}
	\begin{tikzpicture}[line width=1 pt, scale=1]
	\draw[scalarnoarrow] (-1,2.3) -- (0,2);
	\draw[scalarnoarrow] (0,2) -- (2,2);
	\draw[scalarnoarrow] (2,2) -- (3,2.3);
	\node at (-0.9,2.5) {$\phi$};
	\node at (-0.9,0){$\phi$};
	\node at (1,2.4) {$\tilde{N}_{i}$};
	
	\draw[scalarnoarrow] (-1,-.3) -- (0,0);
	\draw[scalarnoarrow] (2,0) -- (0,0);
	\draw[scalarnoarrow] (2,0) -- (3,-.3);
	\node at (2.9,2.5)  {$H_u$};
	\node at (2.9,0) {$H_u$};
	\node at (1,0.3) {$\tilde{N}_{i}$};
	\draw[scalarnoarrow] (0,0) -- (0,2);
	\draw[scalarnoarrow] (2,2) -- (2,0);
	\node at (-0.5,1) {$\tilde{N}_{i}$};
	\node at (2.3,1) {$\tilde{L}$};
	\end{tikzpicture}
	\begin{tikzpicture} [line width=1 pt, scale=1.5]
	\draw[scalarnoarrow] (-140:1)--(0,0);
	\draw[scalarnoarrow] (140:1)--(0,0);
	\draw[scalarnoarrow] (-40:1)--(0,0);
	\node at (-138:1.2) {$\phi$};
	\node at (138:1.2) {$\phi$};
	\draw[scalarnoarrow] (0,0)--(40:1);
	\node at (-42:1.2) {$\tilde{L}$};
	\node at (42:1.2) {$H_u$};	
	\end{tikzpicture}

\end{figure}
\begin{equation}
  \left<\sigma v\right>^{\delta\phi}_{eff}\simeq\frac{k^2y_{eff}^2}{16\pi E_\phi^2}+ \frac{h^4y_{eff}^4}{4\pi^5 E_\phi^2}. \label{sigmadphi}
\end{equation}
The evaporation cross-sections of the inflaton condensate through scattering with the (s)neutrinos are given by

\begin{figure}[H]
	\centering
	\begin{tikzpicture} [line width=1 pt, scale=1.5]
	\draw[fermion] (-140:1)--(0,0);
	\draw[scalarnoarrow] (140:1)--(0,0);
	\draw[fermion] (0:0)--(1,0);
	\node at (-138:1.2) {$N_{i}$};
	\node at (138:1.2) {$\left\langle \phi\right\rangle $};
	\node at (.5,.2) {$N_{i}$};	
	\node at (180:0.25) {$h$};
	\begin{scope}[shift={(1,0)}]
	\draw[fermion] (0,0)--(-40:1);
	\draw[scalarnoarrow] (0,0)--(40:1);
	\node at (-42:1.2) {$N_{i}$};
	\node at (42:1.2) {$\phi$};	
	\node at (0:0.25) {$h$};
	\end{scope}
	\end{tikzpicture}\quad
	\begin{tikzpicture} [line width=1 pt, scale=1.5]
	\draw[scalarnoarrow] (-140:1)--(0,0);
	\draw[scalarnoarrow] (140:1)--(0,0);
	\draw[scalarnoarrow] (-40:1)--(0,0);
	\node at (-138:1.2) {$\tilde{N}_i$};
	\node at (138:1.2) {$\left\langle \phi\right\rangle $};
	\draw[scalarnoarrow] (0,0)--(40:1);
	\node at (-42:1.2) {$\tilde{N}_i$};
	\node at (42:1.2) {$\phi$};	
	\node at (90:0.25) {$h^2$};
	\node at (-90:0.25) {$\kappa^2$};
	\end{tikzpicture}
	\end{figure}
\begin{flalign}
\left<\sigma v\right>^{\tilde{N}_i}_{evap}=\frac{ \kappa ^4+4 \ h^4}{64\pi E_{\tilde{N}_i}^2}\,,\qquad\qquad\left<\sigma v\right>^{N_i}_{evap}=\frac{h^4}{16\pi E_{N_i}^2}.
\end{flalign} 
With the SM particles 
\begin{figure}[H]
	\centering
	\begin{tikzpicture}[line width=1 pt, scale=1]
	\draw[scalarnoarrow] (-1,2.3) -- (0,2);
	\draw[fermion] (0,2) -- (2,2);
	\draw[scalarnoarrow] (2,2) -- (3,2.3);
	\node at (-0.9,2.5) {$\left\langle \phi\right\rangle $};
	\node at (2.9,2.5) {$\phi$};
	\node at (1,2.4) {$N_i$};
	
	\draw[scalarnoarrow] (-1,-.3) -- (0,0);
	\draw[fermion] (2,0) -- (0,0);
	\draw[scalarnoarrow] (2,0) -- (3,-.3);
	\node at (-0.9,0) {$H_u$};
	\node at (2.9,0) {$H_u$};
	\node at (1,0.3) {$L$};
	\draw[fermion] (0,0) -- (0,2);
	\draw[fermion] (2,2) -- (2,0);
	\node at (-0.5,1) {$N_i$};
	\node at (2.5,1) {$N_i$};
	\end{tikzpicture}
	\begin{tikzpicture}[line width=1 pt, scale=1]
	\draw[scalarnoarrow] (-1,2.3) -- (0,2);
	\draw[scalarnoarrow] (0,2) -- (2,2);
	\draw[scalarnoarrow] (2,2) -- (3,2.3);
	\node at (-0.9,2.5) {$\left\langle \phi\right\rangle $};
	\node at (2.9,2.5) {$\phi$};
	\node at (1,2.4) {$\tilde{N}_{i}$};
	
	\draw[scalarnoarrow] (-1,-.3) -- (0,0);
	\draw[scalarnoarrow] (2,0) -- (0,0);
	\draw[scalarnoarrow] (2,0) -- (3,-.3);
	\node at (-0.9,0) {$H_u$};
	\node at (2.9,0) {$H_u$};
	\node at (1,0.2) {$\tilde{L}$};
	\draw[scalarnoarrow] (0,0) -- (0,2);
	\draw[scalarnoarrow] (2,2) -- (2,0);
	\node at (-0.5,1) {$\tilde{N}_{i}$};
	\node at (2.5,1) {$\tilde{N}_{i}$};
	\end{tikzpicture}
	\begin{tikzpicture} [line width=1 pt, scale=1.5]
	\draw[scalarnoarrow] (-140:1)--(0,0);
	\draw[scalarnoarrow] (140:1)--(0,0);
	\draw[scalarnoarrow] (-40:1)--(0,0);
	\node at (-138:1.2) {$H_u$};
	\node at (138:1.2) {$\left\langle \phi\right\rangle $};
	\draw[scalarnoarrow] (0,0)--(40:1);
	\node at (-42:1.2) {$\tilde{L}$};
	\node at (42:1.2) {$\phi$};	
	\end{tikzpicture}
	\label{fig:Feyn3}
\end{figure}
\begin{equation}
\sigma_{\phi H_u}\simeq\frac{k^2y_{eff}^2}{16\pi (3T)^2}+ \frac{h^4y_{eff}^4}{256\pi^5 (3T)^2}\left\{
\begin{array}{ll}
64 & M_1>3T\,, \\16
\left|\frac{ \left(M_1^2-9T^2\right)}{9T^2} \log \left(\frac{M_1^2-9T^2}{M_1^2}\right)\right|^2  & M_1<3T\,.
\end{array} 
\right.
\end{equation}

A detailed calculation of the amplitude for the evaporation processes with the SM dof is given in the next section, Appendix \ref{Evapcalc}. 



%

\section{Evaporation with SM loop calculation}\label{Evapcalc}
On the analysis of the evaporation processes during reheating, namely the scatterings with the standard model particles, we made a simplifying assumption on the effective amplitudes of such effects. We now dwell on these interactions, studying and pointing the relevant processes. Recall the super potential 
\begin{equation}
	W=\frac{1}{2\sqrt{2}}\kappa^2\Phi^2\left(\mathrm{N}_1+ \mathrm{N}_2\right)+ \frac{h}{2}\Phi(\mathrm{N}_2^2-\mathrm{N}_1^2)+\frac{M_1}{2}(\mathrm{N}_1^2+\mathrm{N}_2^2)+y\mathrm{H_uL}\left(\mathrm{N}_1+ \mathrm{N}_2\right).
\end{equation}
From here we can extract the relevant scalar potential and the fermion Yukawa interactions with these bosonic fields 
\begin{flalign}
	&V= V(\phi,\tilde{N}_1,\tilde{N}_2)+y^2\tilde{L}^2(\tilde{N}_1+\tilde{N}_2)^2+ y^2 H_u^2[2\tilde{L}^2+(\tilde{N}_1+\tilde{N}_2)^2]+yM_1 H_u\tilde{L}(\tilde{N}_1+\tilde{N}_2)\\+2yh H_u \tilde{L}\phi&(\tilde{N}_1+\tilde{N}_2)+\sqrt{2}y\kappa H_u \tilde{L}\phi^2 \nonumber\\
	&{\cal{L}} ={\cal{L}}_{kin}+{\cal{L}}_{Yuk}(\phi,\tilde{N}_1,\tilde{N}_2)+y(\tilde{N}_1+\tilde{N}_2)\tilde{H}_uL^c+y\tilde{L}\tilde{H}_u(N_1^c+N_2^c)+y H_u L(N_1^c+N_2^c)+V
	\label{Lagrangian SUSY_evap}	
\end{flalign}
\begin{figure}[H]
	\centering
	
	\begin{tikzpicture}[line width=1 pt, scale=1]
	\draw[scalarnoarrow] (-1,2.3) -- (0,2);
	\draw[fermion] (0,2) -- (2,2);
	\draw[scalarnoarrow] (2,2) -- (3,2.3);
	\node at (-0.9,2.5) {$\left\langle \phi\right\rangle $};
	\node at (2.9,2.5) {$\phi$};
	\node at (1,2.4) {$N_i$};
	
	\draw[scalarnoarrow] (-1,-.3) -- (0,0);
	\draw[fermion] (2,0) -- (0,0);
	\draw[scalarnoarrow] (2,0) -- (3,-.3);
	\node at (-0.9,0) {$H_u$};
	\node at (2.9,0) {$H_u$};
	\node at (1,0.3) {$\nu_{\ell}(e_{\ell})$};
	\draw[fermion] (0,0) -- (0,2);
	\draw[fermion] (2,2) -- (2,0);
	\node at (-0.5,1) {$N_i$};
	\node at (2.5,1) {$N_i$};
	\end{tikzpicture}
	\begin{tikzpicture}[line width=1 pt, scale=1]
	\draw[scalarnoarrow] (-1,2.3) -- (0,2);
	\draw[fermion] (0,2) -- (2,2);
	\draw[scalarnoarrow] (2,2) -- (3,2.3);
	\node at (-0.9,2.5) {$\left\langle \phi\right\rangle $};
	\node at (2.9,2.5) {$\phi$};
	\node at (1,2.4) {$N_i$};
	
	\draw[scalarnoarrow] (-1,-.3) -- (0,0);
	\draw[fermion] (2,0) -- (0,0);
	\draw[scalarnoarrow] (2,0) -- (3,-.3);
	\node at (-0.9,0) {$\tilde{\nu}_{\ell}(\tilde{e}_{\ell})$};
	\node at (2.9,0) {$\tilde{\nu}_{\ell}(\tilde{e}_{\ell})$};
	\node at (1,0.3) {$\tilde{H}_u$};
	\draw[fermion] (0,0) -- (0,2);
	\draw[fermion] (2,2) -- (2,0);
	\node at (-0.5,1) {$N_i$};
	\node at (2.5,1) {$N_i$};
	\end{tikzpicture}
	\begin{tikzpicture}[line width=1 pt, scale=1]
	\draw[scalarnoarrow] (-1,2.3) -- (0,2);
	\draw[fermion] (0,2) -- (2,2);
	\draw[scalarnoarrow] (2,2) -- (3,2.3);
	\node at (-0.9,2.5) {$\left\langle \phi\right\rangle $};
	\node at (2.9,2.5) {$\phi$};
	\node at (1,2.4) {$N_i$};
	
	\draw[fermion] (-1,-.3) -- (0,0);
	\draw[scalarnoarrow] (2,0) -- (0,0);
	\draw[fermion] (2,0) -- (3,-.3);
	\node at (-0.9,0) {$\nu_{\ell}$};
	\node at (2.9,0) {$\nu_{\ell}$};
	\node at (1,0.2) {$H_u$};
	\draw[fermion] (0,0) -- (0,2);
	\draw[fermion] (2,2) -- (2,0);
	\node at (-0.5,1) {$N_i$};
	\node at (2.5,1) {$N_i$};	
	
	\end{tikzpicture}
	\begin{tikzpicture}[line width=1 pt, scale=1]
	\draw[scalarnoarrow] (-1,2.3) -- (0,2);
	\draw[fermion] (0,2) -- (2,2);
	\draw[scalarnoarrow] (2,2) -- (3,2.3);
	\node at (-0.9,2.5) {$\left\langle \phi\right\rangle $};
	\node at (2.9,2.5) {$\phi$};
	\node at (1,2.4) {$N_i$};
	
	\draw[fermion] (-1,-.3) -- (0,0);
	\draw[scalarnoarrow] (2,0) -- (0,0);
	\draw[fermion] (2,0) -- (3,-.3);
	\node at (-0.9,0) {$\tilde{H}_u$};
	\node at (2.9,0) {$\tilde{H}_u$};
	\node at (1,0.2) {$\tilde{\nu}_{\ell}(\tilde{e}_{\ell})$};
	\draw[fermion] (0,0) -- (0,2);
	\draw[fermion] (2,2) -- (2,0);
	\node at (-0.5,1) {$N_i$};
	\node at (2.5,1) {$N_i$};	
	\end{tikzpicture}
	\begin{tikzpicture}[line width=1 pt, scale=1]
	\draw[scalarnoarrow] (-1,2.3) -- (0,2);
	\draw[scalarnoarrow] (0,2) -- (2,2);
	\draw[scalarnoarrow] (2,2) -- (3,2.3);
	\node at (-0.9,2.5) {$\left\langle \phi\right\rangle $};
	\node at (2.9,2.5) {$\phi$};
	\node at (1,2.4) {$\tilde{N}_i$};
	
	\draw[fermion] (-1,-.3) -- (0,0);
	\draw[fermion] (0,0) -- (2,0);
	\draw[fermion] (2,0) -- (3,-.3);
	\node at (-0.9,0) {$\tilde{\nu}_{\ell}(\tilde{e}_{\ell})$};
	\node at (2.9,0) {$\tilde{\nu}_{\ell}(\tilde{e}_{\ell})$};
	\node at (1,0.3) {$\tilde{H}_u$};
	\draw[scalarnoarrow] (0,0) -- (0,2);
	\draw[scalarnoarrow] (2,2) -- (2,0);
	\node at (-0.5,1) {$\tilde{N}_i$};
	\node at (2.5,1) {$\tilde{N}_i$};
	\end{tikzpicture}
	\begin{tikzpicture}[line width=1 pt, scale=1]
	\draw[scalarnoarrow] (-1,2.3) -- (0,2);
	\draw[scalarnoarrow] (0,2) -- (2,2);
	\draw[scalarnoarrow] (2,2) -- (3,2.3);
	\node at (-0.9,2.5) {$\left\langle \phi\right\rangle $};
	\node at (2.9,2.5) {$\phi$};
	\node at (1,2.4) {$\tilde{N}_i$};
	
	\draw[fermion] (-1,-.3) -- (0,0);
	\draw[fermion] (0,0) -- (2,0);
	\draw[fermion] (2,0) -- (3,-.3);
	\node at (-0.9,0) {$\tilde{H}_u$};
	\node at (2.9,0) {$\tilde{H}_u$};
	\node at (1,0.3) {$\nu_{\ell}(e_{\ell})$};
	\draw[scalarnoarrow] (0,0) -- (0,2);
	\draw[scalarnoarrow] (2,2) -- (2,0);
	\node at (-0.5,1) {$\tilde{N}_i$};
	\node at (2.5,1) {$\tilde{N}_i$};
	\end{tikzpicture}
	
	\begin{tikzpicture} [line width=1 pt, scale=1.5]
	\draw[scalarnoarrow] (-140:1)--(0,0);
	\draw[scalarnoarrow] (140:1)--(0,0);
	\draw[scalarnoarrow] (-40:1)--(0,0);
	\node at (-138:1.2) {$H_u$};
	\node at (138:1.2) {$\left\langle \phi\right\rangle $};
	\draw[scalarnoarrow] (0,0)--(40:1);
	\node at (-42:1.2) {$\tilde{\nu}_{\ell}(\tilde{e}_{\ell})$};
	\node at (42:1.2) {$\phi$};	
	\end{tikzpicture}
	\begin{tikzpicture}[line width=1 pt, scale=1]
	\draw[scalarnoarrow] (-1,2.3) -- (0,2);
	\draw[scalarnoarrow] (0,2) -- (2,2);
	\draw[scalarnoarrow] (2,2) -- (3,2.3);
	\node at (-0.9,2.5) {$\left\langle \phi\right\rangle $};
	\node at (2.9,2.5) {$\phi$};
	\node at (1,2.4) {$\tilde{N}_i$};
	
	\draw[scalarnoarrow] (-1,-.3) -- (0,0);
	\draw[scalarnoarrow] (2,0) -- (0,0);
	\draw[scalarnoarrow] (2,0) -- (3,-.3);
	\node at (-0.9,0) {$H_u$};
	\node at (2.9,0) {$H_u$};
	\node at (1,0.2) {$\tilde{\nu}_{\ell}(\tilde{e}_{\ell})$};
	\draw[scalarnoarrow] (0,0) -- (0,2);
	\draw[scalarnoarrow] (2,2) -- (2,0);
	\node at (-0.5,1) {$\tilde{N}_i$};
	\node at (2.5,1) {$\tilde{N}_i$};
	\end{tikzpicture}
	\begin{tikzpicture}[line width=1 pt, scale=1]
	\draw[scalarnoarrow] (-1,2.3) -- (0,2);
	\draw[scalarnoarrow] (0,2) -- (2,2);
	\draw[scalarnoarrow] (2,2) -- (3,2.3);
	\node at (-0.9,2.5) {$\left\langle \phi\right\rangle $};
	\node at (2.9,2.5) {$\phi$};
	\node at (1,2.4) {$\tilde{N}_i$};
	
	\draw[scalarnoarrow] (-1,-.3) -- (0,0);
	\draw[scalarnoarrow] (2,0) -- (0,0);
	\draw[scalarnoarrow] (2,0) -- (3,-.3);
	\node at (-0.9,0) {$\tilde{\nu}_{\ell}(\tilde{e}_{\ell})$};
	\node at (2.9,0) {$\tilde{\nu}_{\ell}(\tilde{e}_{\ell})$};
	\node at (1,0.2) {$H_u$};
	\draw[scalarnoarrow] (0,0) -- (0,2);
	\draw[scalarnoarrow] (2,2) -- (2,0);
	\node at (-0.5,1) {$\tilde{N}_i$};
	\node at (2.5,1) {$\tilde{N}_i$};
	\end{tikzpicture}
	\caption{ Evaporation processes with SM particles (S-symmetry related diagrams are also possible)} \label{fig:Feyn5}
\end{figure}
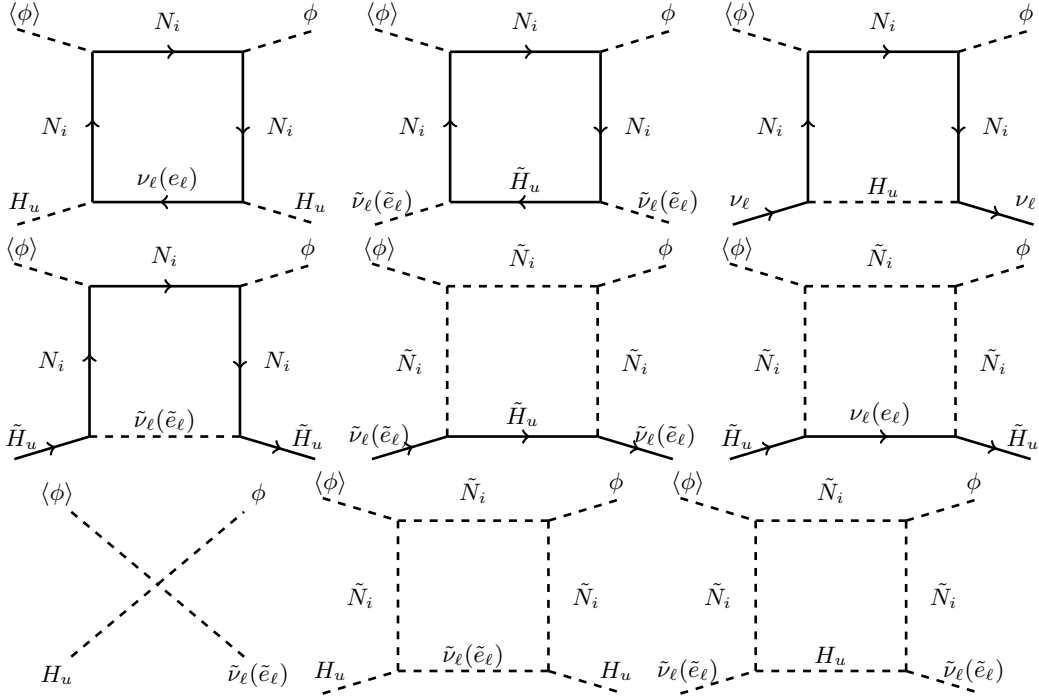
Considering that reheating occurs before the Higgs field acquires a vev and $\left\langle\tilde{N}_i\right\rangle=0$, the processes that may contribute to evaporation with the SM particles are represented in Fig.  \ref{fig:Feyn5}.

The 4-point interaction is proportional to $\kappa^2 \sim 10^{-12}$,  its contribution may be negligible in comparison with the $h$ and $y_{eff}$ for large couplings. However in some regimes, namely with small $y_{eff}$, it can hold the dominant contribution.  Moving to 1-loop interactions, we can reduce our analysis to the calculation of the diagrams with complete fermion and scalar loops. Basically when the external legs are scalar fields. One can see this by looking on the dimension of the effective operators. Having fermion external lines makes the effective operator 5-dimensional, thus it will be suppressed by a mass scale $M_1$.

Let us start with the fermion loop. Since we have massless Higgsinos and seletrons at this stage, we may condense all these process in one calculation. The amplitude becomes
\begin{equation}
	\mathcal{M}_{fermion\ loop}=\int \frac{d^{\mathit{d}}k}{(2\pi )^{\mathit{d}}}\frac{Tr\left[\left(\slashed{k}+M_1\right)\left(\slashed{k}-\slashed{p}_3+M_1\right)\left(\slashed{k}-\slashed{p}_2\right)\left(\slashed{k}+M_1\right)\right]}{\left[k^2-M_1^2\right]^2\left[(k-p_3)^2-M_1^2\right](k-p_2^2)}
\end{equation}
After taking the trace we can reduce it to the analysis of the following integrals
\begin{flalign}
	\mathcal{M}_{fermion\ loop}=4h^2y_{eff}^2\int \frac{d^{\mathit{d}}k}{(2\pi )^{\mathit{d}}} \left[\frac{1}{\left[k^2-M_1^2\right]\left[(k-p_3)^2-M_1^2\right]}+\frac{(p_2-p_3).(k-p_2)}{\left[k^2-M_1^2\right]\left[(k-p_3)^2-M_1^2\right](k-p_2)^2} \right]\\+8h^2y_{eff}^2M_1^2
	\int \frac{d^{\mathit{d}}k}{(2\pi )^{\mathit{d}} }  \left[\frac{2}{\left[k^2-M_1^2\right]^2\left[(k-p_3)^2-M_1^2\right]}+
	\frac{ (2p_2-p_3).(k-p_2)} {\left[k^2-M_1^2\right]^2\left[(k-p_3)^2-M_1^2\right](k-p_2)^2}\right] 
\end{flalign}
represented in order in the diagrams
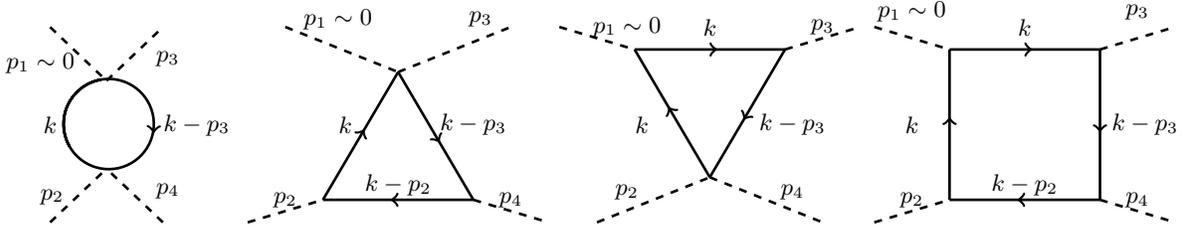
\begin{figure}[H]
	\centering
	\begin{tikzpicture} [line width=1 pt, scale=1.5]
	\draw[scalarnoarrow] (-120:1)--(0,-0.39);
	\draw[scalarnoarrow] (120:1)--(0,0.39);
	\draw[scalarnoarrow] (-60:1)--(0,-0.39);
	\node at (-127:0.8) {$p_2$};
	\node at (137:0.8) {$p_1\sim0$};
	\draw[scalarnoarrow] (0,0.39)--(60:1);
	\node at (-47:0.8) {$p_4$};
	\node at (47:0.8) {$p_3$};	
	\draw[fermionbar] (-0.05,0.4) arc (100:570:.4);
	\node at (-0.5,0) {$k$};
	\node at (0.8,0) {$k-p_3$};
	\end{tikzpicture}	
	\begin{tikzpicture}[line width=1 pt, scale=1]
	\draw[scalarnoarrow] (-0.5,2.3) -- (1,1.7);
	
	\draw[scalarnoarrow] (1,1.7) -- (2.5,2.3);
	\node at (0.2,2.4) {$p_1\sim0 $};
	\node at (2.1,2.4) {$p_3$};

	\draw[scalarnoarrow] (-1,-.3) -- (0,0);
	\draw[fermion] (2,0) -- (0,0);
	\draw[scalarnoarrow] (2,0) -- (3,-.3);
	\node at (-0.5,0) {$p_2$};
	\node at (2.5,0) {$p_4$};
	\node at (1,0.2) {$k-p_2$};
	\draw[fermion] (0,0) -- (1,1.7);
	\draw[fermion] (1,1.7) -- (2,0);
	\node at (0.3,1) {$k$};
	\node at (2,1) {$k-p_3$};
	\end{tikzpicture}
	\begin{tikzpicture}[line width=1 pt, scale=1]
	\draw[scalarnoarrow] (-1,2.3) -- (0,2);
	\draw[fermion] (0,2) -- (2,2);
	\draw[scalarnoarrow] (2,2) -- (3,2.3);
	\node at (-0.1,2.3) {$p_1\sim0 $};
	\node at (2.5,2.3) {$p_3$};
	\node at (1,2.3) {$k$};
	\draw[scalarnoarrow] (-0.5,-.3) -- (1,0.3);
	
	\draw[scalarnoarrow] (1,0.3) -- (2.5,-.3);
	\node at (-0.1,0.1) {$p_2$};
	\node at (2.1,0.1) {$p_4$};

	\draw[fermion] (1,0.3) -- (0,2);
	\draw[fermion] (2,2) -- (1,0.3);
	\node at (0.1,1) {$k$};
	\node at (2.1,1) {$k-p_3$};
	\end{tikzpicture}
	\begin{tikzpicture}[line width=1 pt, scale=1]
	\draw[scalarnoarrow] (-1,2.3) -- (0,2);
	\draw[fermion] (0,2) -- (2,2);
	\draw[scalarnoarrow] (2,2) -- (3,2.3);
	\node at (-0.5,2.5) {$p_1\sim0 $};
	\node at (2.5,2.5) {$p_3$};
	\node at (1,2.3) {$k$};
	\draw[scalarnoarrow] (-1,-.3) -- (0,0);
	\draw[fermion] (2,0) -- (0,0);
	\draw[scalarnoarrow] (2,0) -- (3,-.3);
	\node at (-0.5,0) {$p_2$};
	\node at (2.5,0) {$p_4$};
	\node at (1,0.2) {$k-p_2$};
	\draw[fermion] (0,0) -- (0,2);
	\draw[fermion] (2,2) -- (2,0);
	\node at (-0.5,1) {$k$};
	\node at (2.6,1) {$k-p_3$};
	\end{tikzpicture}
	\caption{Diagram decomposition} \label{fig:Feyn6}
\end{figure}
The solutions are the Passarino-Veltman functions\footnote{Calculations with the Mathematica extension Package-X \cite{Patel:2015tea}}.
\begin{flalign}
	\mathcal{M}_{fermion\ loop} =h^2y_{eff}^2[2\textbf{B}_0(p_2^2;M_1,0)+2\textbf{B}_0(p_3^2;M_1,M1)+4 M_1^2 \textbf{B}_0^{(2,1)}(p_2^2;M_1,0)+8M_1^2\textbf{B}_0^{(2,1)}(p_3^2;M_1,M_1) \nonumber \\+
	4M_1^2(2M_1^2-p_2^2-p_4^2)\textbf{C}^{(2,1,1)}_0(p_2^2,p_4^2,p_3^2;M_1,0,M_1) +(6M_1^2-2p_4^2)\textbf{C}_0(p_2^2,p_4^2,p_3^2;M_1,0,M_1)]
\end{flalign}

The first loop gives a contribution
\begin{flalign}
	\mathcal{M}_{\bigcirc}&=4 h^2y_{eff}^2\textbf{B}_0(p_3^2;M_1,M_1)
\end{flalign}

The triangular loops, in the figure order, give 
\begin{flalign}
	\mathcal{M}_{\Delta}=&2h^2y_{eff}^2\left[\textbf{B}_0(p_2^2;M_1,0)-\textbf{B}_0(p_3^2;M_1,M_1)+(M_1^2-p_4^2)\textbf{C}_0(p_2^2,p_4^2,p_3^2;M_1,0,M_1)\right]\\
	\mathcal{M}_{\nabla}=&16h^2y_{eff}^2M_1^2\textbf{B}^{(2,1)}_0(p_3^2;M_1,M_1)
\end{flalign}
and the square contribution is 
\begin{flalign}
	\mathcal{M}_{\square}=&4h^2y_{eff}^2[M_1^2\textbf{B}^{(2,1)}_0(p_2^2;M_1,0)-2M_1^2\textbf{B}^{(2,1)}_0(p_3^2;M_1,M_1)+M_1^2(2M_1^2-p_2^2-p_4^2)\textbf{C}^{(2,1,1)}_0(p_2^2,p_4^2,p_3^2;M_1,0,M_1)\nonumber\\ &+M_1^2\textbf{C}_0(p_2^2,p_4^2,p_3^2;M_1,0,M_1)]
\end{flalign}

The scattering with the scalar loops gives
\begin{flalign}
	\mathcal{M}_{scalar\ loop}&=h^2y_{eff}^2M_1^2(M_1^2+\left\langle\phi\right\rangle^2)\int \frac{d^{\mathit{d}}k}{(2\pi )^{\mathit{d}}}\frac{1}{\left[k^2-\text{M1}^2\right]^2\left[(k-\text{p3})^2-\text{M1}^2\right](k-\text{p2})^2}\\
	&=h^2y_{eff}^2M_1^2(M_1^2+\left\langle\phi\right\rangle^2)\textbf{C}^{(2,1,1)}_0(p_2^2,p_4^2,p_3^2;M_1,0,M_1)
\end{flalign}

The expansion of these functions, mainly $\textbf{C}_0$, is lengthy and written with several dilogarithms.  We are interested in exploring this results when the external momentum is given by a thermal distribution and in comparing these amplitudes for different ratios of $M_1$ and $T$. For the sake of simplicity, avoiding a careful, detailed and troublesome integration of the Boltzmann distribution, we take $p_i$, with $i=2,3,4$, to be order $3T$, without ever compromising the kinematical conditions, for instance $\kappa^2(p_2^2,p_3^2,p_4^2)>0$.
Collecting all the amplitudes and doing a numerical calculation one sees that $\mathcal{M}\simeq 8h^2y_{eff}^2$ for $M_1>3T$, and for $3T>M_1$ we have $\textbf{B}_0(p_2^2;M_1,0)$ as the dominant contribution, thus
\begin{equation}	
	\mathcal{M}\simeq 4 h^2y_{eff}^2\frac{ \left(M_1^2-9T^2\right)}{9T^2} \log \left(\frac{M_1^2-9T^2}{M_1^2}\right)\,.
\end{equation}

In our numerical analysis of the Boltzmann equations we used an interpolation of these functions to describe the full amplitude
\begin{equation}
	\left|\mathcal{M}\right|^2= 16h^4y_{eff}^4\left|\frac{ \left(M_1^2-9T^2\right)}{9T^2} \log \left(\frac{M_1^2-9T^2}{M_1^2}\right)\right|^2 \Theta\left(\frac{M_1}{3T}\right)+64h^4y_{eff}^4\Theta\left(\frac{3T}{M_1}\right) 
\end{equation}
where $\Theta\left(x\right)=\frac{1}{2}\left(1+\tanh\left[(x-1)\delta\right]\right)$, and $\delta\simeq10$.

In the scalar loops, besides the $h$ and $y_{eff}$ factors in the effective coupling, we will have $M_1^2\left\langle\phi\right\rangle^2$ and  $M_1^4$ proportionality, coming from the scalar potential, that when $M_1<3T$ makes these amplitudes drop very fast as we increase $T$. When $M_1>3T$ by solving the loop integrals one can see numerically that scalar contribution is constant, as a function of $T$, but some orders of magnitude smaller than the fermion counterpart. We therefore neglect their contributions for these decays.   

\acknowledgments

This work has been partially supported by MICINN (PID2019-105943GB-I00/AEI/10.13039/501100011033) and ``Junta de Andaluc\'ia" grants P18-FR-4314 and A-FQM-211-UGR18. ATM is supported by FCT grant SFRH/BD/144803/2019.


\begin{thebibliography}{99}

\bibitem{Manso:2018cba}
A.~Torres Manso and J.~G.~Rosa,
``$\nu$-inflaton dark matter",
JHEP {\bf 1902} (2019) 020
[arXiv:1811.02302 [hep-ph]].

\bibitem{Tyson:1998vp} 
J.~A.~Tyson, G.~P.~Kochanski and I.~P.~Dell'Antonio,
``Detailed mass map of CL0024+1654 from strong lensing'',
Astrophys.\ J.\  {\bf 498} (1998) L107 
[astro-ph/9801193].


\bibitem{Dahle:2007wf} 
H.~Dahle,
``A compilation of weak gravitational lensing studies of clusters of galaxies",
astro-ph/0701598.


\bibitem{Paczynski:1985jf} 
B.~Paczynski,
``Gravitational microlensing by the galactic halo",
Astrophys.\ J.\  {\bf 304} (1986) 1.


\bibitem{Taoso:2007qk} 
M.~Taoso, G.~Bertone and A.~Masiero,
``Dark Matter Candidates: A Ten-Point Test",
JCAP {\bf 0803}  (2008) 022
[arXiv:0711.4996 [astro-ph]].

\bibitem{Guth:1980zm} 
A.~H.~Guth,
``The Inflationary Universe: A Possible Solution to the Horizon and Flatness Problems",
Phys.\ Rev.\ D {\bf 23} (1981) 347
[Adv.\ Ser.\ Astrophys.\ Cosmol.\  {\bf 3}, 139 (1987)].


\bibitem{Linde:1981mu} 
A.~D.~Linde,
``A New Inflationary Universe Scenario: A Possible Solution of the Horizon, Flatness, Homogeneity, Isotropy and Primordial Monopole Problems",
Phys.\ Lett.\  {\bf 108B} (1982)  389
[Adv.\ Ser.\ Astrophys.\ Cosmol.\  {\bf 3}, 149 (1987)].


\bibitem{Albrecht:1982wi} 
A.~Albrecht and P.~J.~Steinhardt,
``Cosmology for Grand Unified Theories with Radiatively Induced Symmetry Breaking",
Phys.\ Rev.\ Lett.\  {\bf 48} (1982) 1220
[Adv.\ Ser.\ Astrophys.\ Cosmol.\  {\bf 3}, 158 (1987)].


\bibitem{Fields:2014uja}
B.~D.~Fields, P.~Molaro and S.~Sarkar,
``Big-Bang Nucleosynthesis",
Chin. Phys. C \textbf{38} (2014) 339 
[arXiv:1412.1408 [astro-ph.CO]]; 
P.A. Zyla et al. (Particle Data Group), Prog. Theor. Exp. Phys. 2020, 083C01 (2020).

\bibitem{Kofman:1997yn}
L.~Kofman, A.~D.~Linde and A.~A.~Starobinsky,
``Towards the theory of reheating after inflation",
Phys. Rev. D \textbf{56} (1997) 3258
[arXiv:hep-ph/9704452 [hep-ph]].

\bibitem{Liddle:2006qz} 
A.~R.~Liddle and L.~A.~Urena-Lopez,
``Inflation, dark matter and dark energy in the string landscape",
Phys. Rev. Lett. \textbf{97} (2006) 161301
[arXiv:astro-ph/0605205 [astro-ph]].

\bibitem{Panotopoulos:2007ri}  G.~Panotopoulos,
``A Brief note on how to unify dark matter, dark energy, and inflation",
Phys.\ Rev.\ D {\bf 75} (2007) 127301 
[arXiv:0706.2237 [hep-ph]].


\bibitem{Cardenas:2007xh} 
V.~H.~Cardenas,
``Inflation, Reheating and Dark Matter",
Phys.\ Rev.\ D {\bf 75} (2007) 083512 
[astro-ph/0701624].

\bibitem{Liddle:2008bm} 
A.~R.~Liddle, C.~Pahud and L.~A.~Urena-Lopez,
``Triple unification of inflation, dark matter, and dark energy using a single field",
Phys.\ Rev.\ D {\bf 77} (2008) 121301
[arXiv:0804.0869 [astro-ph]].


\bibitem{Bose:2009kc} 
N.~Bose and A.~S.~Majumdar,
``Unified Model of k-Inflation, Dark Matter and Dark Energy",
Phys.\ Rev.\ D {\bf 80} (2009) 103508
[arXiv:0907.2330 [astro-ph.CO]].

\bibitem{DeSantiago:2011qb} 
J.~De-Santiago and J.~L.~Cervantes-Cota,
``Generalizing a Unified Model of Dark Matter, Dark Energy, and Inflation with Non Canonical Kinetic Term",
Phys.\ Rev.\ D {\bf 83} (2011) 063502
[arXiv:1102.1777 [astro-ph.CO]].

\bibitem{Bastero-Gil:2015lga}
M.~Bastero-Gil, R.~Cerezo and J.~G.~Rosa,
``Inflaton dark matter from incomplete decay",
Phys.\ Rev.\ D {\bf 93} (2016)  103531
[arXiv:1501.05539 [hep-ph]].



\bibitem{Daido:2017wwb}  
R.~Daido, F.~Takahashi and W.~Yin,
``The ALP miracle: unified inflaton and dark matter",
JCAP {\bf 1705}  (2017) 044
[arXiv:1702.03284 [hep-ph]].

\bibitem{Daido:2017tbr} 
R.~Daido, F.~Takahashi and W.~Yin,
``The ALP miracle revisited",
JHEP {\bf 1802} (2018) 104 
[arXiv:1710.11107 [hep-ph]].

\bibitem{Lerner:2009xg}  
R.~N.~Lerner and J.~McDonald,
``Gauge singlet scalar as inflaton and thermal relic dark matter",
Phys.\ Rev.\ D {\bf 80} (2009) 123507 
[arXiv:0909.0520 [hep-ph]].


\bibitem{Okada:2010jd} 
N.~Okada and Q.~Shafi,
``WIMP Dark Matter Inflation with Observable Gravity Waves",
Phys.\ Rev.\ D {\bf 84} (2011) 043533 
[arXiv:1007.1672 [hep-ph]].


\bibitem{delaMacorra:2012sb}  
A.~de la Macorra,
``Dark Matter from the Inflaton Field",
Astropart. Phys. \textbf{35} (2012) 478
[arXiv:1201.6302 [astro-ph.CO]].


\bibitem{Khoze:2013uia} 
V.~V.~Khoze,
``Inflation and Dark Matter in the Higgs Portal of Classically Scale Invariant Standard Model",
JHEP {\bf 1311} (2013) 215 
[arXiv:1308.6338 [hep-ph]].

\bibitem{Kahlhoefer:2015jma} 
F.~Kahlhoefer and J.~McDonald,
``WIMP Dark Matter and Unitarity-Conserving Inflation via a Gauge Singlet Scalar",
JCAP \textbf{11} (2015) 015
[arXiv:1507.03600 [astro-ph.CO]].

\bibitem{Choubey:2017hsq} 
S.~Choubey and A.~Kumar,
``Inflation and Dark Matter in the Inert Doublet Model",
JHEP \textbf{11} (2017) 080
[arXiv:1707.06587 [hep-ph]].

\bibitem{Hooper:2018buz}  
D.~Hooper, G.~Krnjaic, A.~J.~Long and S.~D.~Mcdermott,
``Can the Inflaton Also Be a Weakly Interacting Massive Particle?",
Phys. Rev. Lett. \textbf{122} (2019) 091802
[arXiv:1807.03308 [hep-ph]].

\bibitem{Borah:2018rca}
D.~Borah, P.~S.~B.~Dev and A.~Kumar,
``TeV scale leptogenesis, inflaton dark matter and neutrino mass in a scotogenic model",
Phys. Rev. D \textbf{99} (2019) 055012
[arXiv:1810.03645 [hep-ph]].



\bibitem{Tenkanen:2016twd}  
T.~Tenkanen,
``Feebly Interacting Dark Matter Particle as the Inflaton",
JHEP {\bf 1609} (2016) 049 
[arXiv:1607.01379 [hep-ph]].

\bibitem{Cosme:2020mck}
C.~Cosme, M.~Dutra, T.~Ma, Y.~Wu and L.~Yang,
``Neutrino Portal to FIMP Dark Matter with an Early Matter Era'', 
arXiv:2003.01723 [hep-ph].


\bibitem{Levy:2020zfo}
M.~Levy, J.~G.~Rosa and L.~B.~Ventura,
``Warm Inflation, Neutrinos and Dark matter: a minimal extension of the Standard Model'', 
arXiv:2012.03988 [hep-ph].

\bibitem{Kallosh:2013yoa} 
R.~Kallosh, A.~Linde and D.~Roest,
``Superconformal Inflationary $\alpha$-Attractors'',
JHEP {\bf 1311} (2013) 198 
%
[arXiv:1311.0472 [hep-th]].


\bibitem{SUSYDM}
L.~Roszkowski, E.~M.~Sessolo and S.~Trojanowski,
``WIMP dark matter candidates and searches\textemdash{}current status and future prospects'', 
Rept. Prog. Phys. \textbf{81} (2018) 066201
[arXiv:1707.06277 [hep-ph]].

\bibitem{sneutrinoDM}
C.~Arina and N.~Fornengo,
``Sneutrino cold dark matter, a new analysis: Relic abundance and detection rates'', 
JHEP \textbf{11} (2007) 029
[arXiv:0709.4477 [hep-ph]].

\bibitem{lowmassLSP}
R.~Kumar Barman, G.~Belanger and R.~M.~Godbole,
``Status of low mass LSP in SUSY'', 
Eur. Phys. J. ST \textbf{229} (2020) 3159 [arXiv:2010.11674 [hep-ph]].



\bibitem{gravitino1}
M.~Kawasaki and T.~Moroi,
Prog. Theor. Phys. \textbf{93} (1995) 879
[arXiv:hep-ph/9403364 [hep-ph]].


  
\bibitem{gravitino2}
M.~Kawasaki, K.~Kohri, T.~Moroi and A.~Yotsuyanagi,
Phys. Rev. D \textbf{78} (2008) 065011
[arXiv:0804.3745 [hep-ph]].
  
\bibitem{gravitino3} 
M.~Kawasaki, K.~Kohri, T.~Moroi and Y.~Takaesu,
``Revisiting Big-Bang Nucleosynthesis Constraints on Dark-Matter Annihilation'',  Phys. Lett. B \textbf{751} (2015) 246
[arXiv:1509.03665 [hep-ph]].


\bibitem{axinoneutralino1}
H.~Baer, A.~Lessa, S.~Rajagopalan and W.~Sreethawong,
``Mixed axion/neutralino cold dark matter in supersymmetric models'',
JCAP \textbf{06} (2011) 031
[arXiv:1103.5413 [hep-ph]].


\bibitem{axinoneutralino2}
H.~Baer, K.~Y.~Choi, J.~E.~Kim and L.~Roszkowski,
``Dark matter production in the early Universe: beyond the thermal WIMP paradigm'', 
Phys. Rept. \textbf{555} (2015) 1 [arXiv:1407.0017 [hep-ph]].





\bibitem{neutrinomass}
R.~N.~Mohapatra and A.~Y.~Smirnov,
``Neutrino Mass and New Physics,''
Ann.\ Rev.\ Nucl.\ Part.\ Sci.\  {\bf 56} (2006), 569
[hep-ph/0603118];
A.~de Gouvea,
``Neutrino Mass Models,''
Ann.\ Rev.\ Nucl.\ Part.\ Sci.\  {\bf 66} (2016), 197


\bibitem{effpot}
S.~R.~Coleman and E.~J.~Weinberg,
``Radiative Corrections as the Origin of Spontaneous Symmetry Breaking",
Phys. Rev. D \textbf{7} (1973), 1888; 
G.~Gamberini, G.~Ridolfi and F.~Zwirner,
``On Radiative Gauge Symmetry Breaking in the Minimal Supersymmetric Model",
Nucl. Phys. B \textbf{331} (1990), 331. 

\bibitem{Starobinsky:1980te} 
A.~A.~Starobinsky,
``A New Type of Isotropic Cosmological Models Without Singularity",
Phys.\ Lett.\  {\bf 91B} (1980), 99 
[Adv.\ Ser.\ Astrophys.\ Cosmol.\  {\bf 3}, 130 (1987)].


\bibitem{Pallis:2016mvm}
C.~Pallis and N.~Toumbas,
``Starobinsky Inflation: From Non-SUSY To SUGRA Realizations",
Adv. High Energy Phys. \textbf{2017} (2017), 6759267
[arXiv:1612.09202 [hep-ph]].

\bibitem{planck}
Y.~Akrami \textit{et al.} [Planck],
``Planck 2018 results. X. Constraints on inflation",
Astron. Astrophys. \textbf{641} (2020), A10
[arXiv:1807.06211 [astro-ph.CO]].
 
\bibitem{Arcadi:2011ev}
G.~Arcadi and P.~Ullio,
``Accurate estimate of the relic density and the kinetic decoupling in non-thermal dark matter models",
Phys.\ Rev.\ D {\bf 84} (2011), 043520
[arXiv:1104.3591 [hep-ph]].

\bibitem{Drees:2018dsj}
M.~Drees and F.~Hajkarim,
``Neutralino Dark Matter in Scenarios with Early Matter Domination",
JHEP {\bf 1812} (2018), 042
[arXiv:1808.05706 [hep-ph]].


\bibitem{Ichikawa:2008ne} 
K.~Ichikawa, T.~Suyama, T.~Takahashi and M.~Yamaguchi,
``Primordial Curvature Fluctuation and Its Non-Gaussianity in Models with Modulated Reheating",
Phys.\ Rev.\ D {\bf 78} (2008), 063545
[arXiv:0807.3988 [astro-ph]].

\bibitem{WIMPabundance}
G.~Steigman, B.~Dasgupta and J.~F.~Beacom,
``Precise Relic WIMP Abundance and its Impact on Searches for Dark Matter Annihilation",
Phys. Rev. D \textbf{86} (2012), 023506
[arXiv:1204.3622 [hep-ph]].




\bibitem{Hall:2009bx}
L.~J.~Hall, K.~Jedamzik, J.~March-Russell and S.~M.~West,
``Freeze-In Production of FIMP Dark Matter",
JHEP \textbf{03} (2010), 080
[arXiv:0911.1120 [hep-ph]].

\bibitem{Bernal:2017kxu}
N.~Bernal, M.~Heikinheimo, T.~Tenkanen, K.~Tuominen and V.~Vaskonen,
``The Dawn of FIMP Dark Matter: A Review of Models and Constraints",
Int. J. Mod. Phys. A \textbf{32} (2017), 1730023
[arXiv:1706.07442 [hep-ph]].

\bibitem{Aghanim:2018eyx}
N.~Aghanim \textit{et al.} ,
``Planck 2018 results. VI. Cosmological parameters,
Astron. Astrophys. \textbf{641} (2020), A6
doi:10.1051/0004-6361/201833910
[arXiv:1807.06209 [astro-ph.CO]].

\bibitem{leptoreheating}
G.~F.~Giudice, A.~Notari, M.~Raidal, A.~Riotto and A.~Strumia,
Nucl. Phys. B \textbf{685} (2004), 89
[arXiv:hep-ph/0310123 [hep-ph]].

\bibitem{Ibe:2004tg}
M.~Ibe, R.~Kitano, H.~Murayama and T.~Yanagida,
Phys. Rev. D \textbf{70} (2004) 075012
[arXiv:hep-ph/0403198 [hep-ph]].


\bibitem{Patel:2015tea}
H.~H.~Patel,
``Package-X: A Mathematica package for the analytic calculation of one-loop integrals",
Comput.\ Phys.\ Commun.\  {\bf 197} (2015), 276
[arXiv:1503.01469 [hep-ph]].




\end{thebibliography}
\end{document}